\documentclass[a4paper,11pt]{article}
\usepackage{jcappub}
\usepackage{aas_macros}
\usepackage{subcaption}
\usepackage{amsmath,amssymb,amsfonts}
\usepackage{mydefs}
\usepackage{siunitx}
\usepackage{booktabs}
\usepackage{mathtools}
\usepackage{multirow}
\usepackage{placeins}

\begin{document}
\title{Extending the modeling of the anisotropic galaxy power
spectrum to $k = 0.4 \ h\mathrm{Mpc}^{-1}$}

\author[a,b]{Nick Hand,}
\author[a,b,c]{Uro{\v s} Seljak,}
\author[d]{Florian Beutler,}
\author[e]{\& Zvonimir Vlah}

\affiliation[a]{Astronomy Department, University of California, CA 94720, USA}
\affiliation[b]{Lawrence Berkeley National Laboratory, Berkeley, CA 94720, USA}
\affiliation[c]{Physics Department, University of California, CA 94720, USA}
\affiliation[d]{Institute of Cosmology \& Gravitation, Dennis Sciama Building, University of Portsmouth, Portsmouth, PO1 3FX, UK}
\affiliation[e]{Stanford Institute for Theoretical Physics and Department of Physics, Stanford University, Stanford, CA 94306, USA}

\emailAdd{nhand@berkeley.edu}
\emailAdd{useljak@berkeley.edu}
\emailAdd{florian.beutler@port.ac.uk}
\emailAdd{zvlah@stanford.edu}

\date{\today}

\abstract
{
We present a new model for the redshift-space power spectrum of
galaxies and demonstrate its accuracy in modeling the
monopole, quadrupole, and hexadecapole of the galaxy density
field down to scales of $k = 0.4 \ihMpc$.
The model describes the clustering of galaxies
in the context of a halo model and the clustering
of the underlying halos in redshift space using a
combination of Eulerian perturbation theory
and $N$-body simulations. The modeling of
redshift-space distortions is done using
the so-called distribution function approach. The
final model has 13 free parameters, and each parameter is
physically motivated rather than a nuisance parameter, which allows
the use of well-motivated priors.
We account for the Finger-of-God effect
from centrals and both isolated and non-isolated satellites
rather than using a single velocity
dispersion to describe the combined effect. We test and validate
the accuracy of the model on several sets of high-fidelity $N$-body
simulations, as well as realistic mock catalogs designed to simulate
the BOSS DR12 CMASS data set. The suite of simulations covers
a range of cosmologies and galaxy bias models, providing a rigorous
test of the level of theoretical systematics present in the model.
The level of bias in the recovered values of $f \sigma_8$ is found to be
small. When including scales to $k = 0.4 \ h\mathrm{Mpc}^{-1}$, we find
15-30\% gains in the statistical precision of $f \sigma_8$ relative to
$k = 0.2 \ h\mathrm{Mpc}^{-1}$ and a roughly 10-15\% improvement
for the perpendicular Alcock-Paczynski parameter $\alpha_\perp$.
Using the BOSS DR12 CMASS mocks as a benchmark for
comparison, we estimate an uncertainty on $f \sigma_8$ that is
$\sim$10-20\% larger than other similar Fourier-space RSD models in the
literature that use $k \leq 0.2 \ h\mathrm{Mpc}^{-1}$,
suggesting that these models likely have a too-limited parametrization.
}

\maketitle

\section{Introduction}

Galaxy redshift surveys measure the three-dimensional clustering of
galaxies in the Universe, and over the past few decades, they have
provided a wealth of cosmological information
\cite{Davis:1983,Maddox:1990,Tegmark:2004,Cole:2005,Eisenstein:2005,
Anderson:2012,Anderson:2014,Anderson:2014b,Alam:2016}.
In combination with other cosmological
probes, such as observations of the cosmic microwave background, type-Ia
supernova samples, and weak-lensing surveys, analyses of the
large-scale structure (LSS) of the Universe have proven
invaluable in establishing the current cosmological
paradigm, the $\Lambda$CDM model, as well as measuring its parameters
with ever-increasing precision.

Crucial to the success of galaxy surveys has been the ability to precisely
and accurately measure the feature imprinted on the clustering of galaxies
by baryon acoustic oscillations in the early Universe (BAO;
see e.g., \cite{Bassett:2010} for a review). The
BAO signal can be used to provide constraints on the
expansion history of the Universe and infer properties of dark energy
(e.g., \cite{Wagner:2008, Shoji:2009}). The isotropic effect was first
detected in the SDSS \cite{Eisenstein:2005} and the 2dFGRS \cite{Cole:2005},
and more recent measurements of the anisotropic signal, combined
with the Alcock-Paczynski (AP; \cite{Alcock:1979}) effect,
have provided percent-level measurements of the Hubble parameter
$H(z)$ and angular diameter distance $D_a(z)$ \cite{Alam:2016}.
Perhaps most encouragingly, the BAO signal is well-understood
theoretically, with systematic effects on the distance scale
expected to be sub-dominant for future generations of surveys
\cite{Eisenstein:2004,Seo:2005,Angulo:2008,Padmanabhan:2009,Mehta:2011}.

Beyond the BAO signal, additional information is present in the
clustering of galaxies through
what are known as redshift-space distortions (RSD). The peculiar velocities
of galaxies affect their measured redshifts through the Doppler effect, and
in turn, these measured redshifts are used to infer the line-of-sight (LOS)
position of those galaxies. The peculiar velocity field
is sourced by the gravitational potential, and thus, an anisotropic
signal containing information about the rate of structure growth in the
Universe is imprinted on the clustering.
Extracting information from RSD is inherently more difficult than with
BAO, as it requires modeling of the full broadband shape of the clustering
statistic and precise understanding of the anisotropy induced by RSD. The
theoretical task is complicated by the fact that the well-understood, linear
Kaiser model \cite{Kaiser:1987} breaks down on relatively large scales,
with various kinds of nonlinear effects complicating the theoretical modeling
(e.g., \cite{Scoccimarro:2004,Okumura:2011,Jennings:2012,Kwan:2012}).
Of particular importance is the large, nonlinear virial motions of
satellite galaxies within halos, known as the
Finger-of-God (FoG) effect \cite{Jackson:1972}.
Since the statistical precision of clustering measurements is
generally higher on smaller scales, where the effects of nonlinearities
are worse, a direct limit on the amount of useable cosmological
information is imposed due to theoretical uncertainties.

Despite these modeling challenges, RSD analyses have developed
into one of the most popular and powerful cosmological probes today
\cite{Peacock:2001,Hawkins:2003,Tegmark:2006,Guzzo:2008,Yamamoto:2008,Blake:2011,
Beutler:2012,Reid:2012,Samushia:2013,Chuang:2013,Beutler:2014,Reid:2014}.
Constraints on the growth rate of structure through measurements of the
parameter combination $f(z) \sigma_8(z)$ can provide tests of General Relativity
(e.g., \cite{Guzzo:2008}), as well as information about the properties of neutrinos
\cite{Lesgourgues:2006,Beutler:2014} and tighter constraints on the
expansion history through the AP effect (e.g., \cite{Shoji:2009}).
Recent results from Data Release 12 (DR12) of the
Baryon Oscillation Spectroscopic Survey (BOSS)
\cite{Sanchez:2017,Beutler:2017,Grieb:2017,Satpathy:2017} have provided the
tightest constraints to date on the growth rate of structure, with roughly
10\% constraints on $f\sigma_8(z_\mathrm{eff})$ in 3 redshift bins centered at
$z_\mathrm{eff}=0.38$, $0.51$, and $0.61$.

To date, RSD analyses have generally either relied directly on the results of
$N$-body simulations or on perturbative approaches
to model the clustering of galaxies in the quasi-linear and nonlinear
regimes. Both approaches have
their pros and cons. For simulation-based analyses, e.g.,
\cite{Tinker:2006,Hikage:2014,Reid:2014,Guo:2015b}, the simulations represent
the best possible description of nonlinearities, although individual
simulations are expensive to run and, often, the relevant
parameter space cannot be as sufficiently explored as one would like. On
the other hand, modeling techniques relying on perturbation theory (PT), e.g.,
\cite{Beutler:2017,Grieb:2016,Satpathy:2017,Sanchez:2017}, are
relatively fast to compute but will always break down on small enough
scales and fail to capture non-perturbative features, such as the FoG effect
from satellites. In either case, simulations play a crucial role in estimating
the range of scales where a model remains accurate enough to recover cosmological
parameters in an unbiased fashion.

In this work, we present a new model for the redshift-space power spectrum
of galaxies, describing the galaxy clustering in the context of a halo model
\cite{Seljak:2000,Ma:2000,Peacock:2000,Cooray:2002} and
relying on a combination of Eulerian PT and $N$-body simulations to model
the power spectrum of dark matter halos in redshift space.
We use several sets of $N$-body simulations
to validate our model, and we perform cosmological parameter analyses on realistic
BOSS-like mock catalogs to verify both the accuracy and constraining power of the
model. The model relies on the distribution
function approach \cite{Seljak:2011,Okumura:2012,Okumura:2012b,
Vlah:2012,Vlah:2013,Blazek:2014} to map real-space statistics to redshift space.
This formalism is different but complementary from other commonly-used approaches
in RSD analyses, such as the TNS model \cite{Taruya:2010} or the Gaussian
streaming model \cite{Reid:2011}. We build upon the
results presented in \cite{Okumura:2015}, which showed
that the characterization of the redshift-space power spectrum of galaxies
in terms of 1-halo and 2-halo correlations is accurate when compared against
$N$-body simulations. We extend that work by improving the accuracy of
the underlying model for the halo redshift-space power spectrum. The model
is based on the PT results presented in \cite{Vlah:2013}, but uses
simulation-based modeling for key terms. In particular, we develop and
extend the Halo-Zel'dovich Perturbation Theory (HZPT) of \cite{Seljak:2015},
which relies on a combination of linear Lagrangian PT and
simulation-based calibration. A Python software package \texttt{pyRSD} that
implements the model described in this work is publicly
available\footnote{https://github.com/nickhand/pyRSD}.

This paper is organized as follows. Section~\ref{sec:sims}
describes the set of simulations that we use to calibrate our model,
as well as the test suite that we use for independent validation.
We describe the power spectrum estimator, covariance matrix,
and likelihood analysis used to perform parameter estimation
in section~\ref{sec:methods}. In section~\ref{sec:power-spectrum-model},
we detail the power spectrum model, first reviewing the halo model
formalism presented in \cite{Okumura:2015} and then discussing several
new modeling approaches for the redshift-space halo power spectrum.
We assess the accuracy and performance of the model
based on an independent test suite of simulations
in section~\ref{sec:performance}. Finally, we discuss
our results and future prospects in section~\ref{sec:discussion}
and conclude in section~\ref{sec:conclusion}.

\begin{table}[tb]
\centering
\resizebox{\textwidth}{!}
{

  \begin{tabular}{l|c|c|c|c|c|c|c}
  \toprule
  Name & $L_\mathrm{box}$ $[\hMpc]$ & $z_\mathrm{box}$ & $\Omega_m$ & $\Omega_b h^2$
  & $h$ & $n_s$ & $\sigma_8$ \\
  \midrule
   RunPB & 1380 & 0.55 & 0.292 & 0.022 & 0.69 & 0.965 & 0.82 \\
   \midrule
   N-series; Challenge D,E & 2600 & 0.562 & 0.286 & 0.02303 & 0.7 & 0.96 & 0.82 \\
   \midrule
   Challenge A,B,F,G & 2500 & 0.5 & 0.30711 & 0.022045 & 0.6777 & 0.96 & 0.82 \\
   \midrule
   Challenge C & 2500 & 0.441 & 0.27 & 0.02303 & 0.7 & 0.96 & 0.82 \\
   \bottomrule
  \end{tabular}
}
\caption{The cosmological and simulation parameters for the various
$N$-body simulations used in this work.}
\label{tab:sim-params}
\end{table}

\section{Simulations}\label{sec:sims}

We use several sets of $N$-body simulations for both calibrating
and testing the model presented in
this paper. The first set of simulations, described in section
\ref{sec:RunPB}, is used heavily in verifying individual components
of the clustering model. Sections~\ref{sec:nseries} and \ref{sec:lettered-challenge}
describe an independent suite of high-fidelity
simulations that we use to independently verify the accuracy and
precision of the model.
The relevant cosmological and simulation parameters for the mocks
discussed in this section are summarized in table~\ref{tab:sim-params}.

\subsection{RunPB}\label{sec:RunPB}

The main set of simulations used for calibration and testing purposes is the
RunPB $N$-body simulation produced by Martin White with the TreePM N-body code
of \cite{White:2002}. These simulations have been used recently in a
number of analyses \cite{White:2014,Reid:2014,Schmittfull:2015,Schmittfull:2015b}.
The simulation set has 10 realizations of $2048^3$ dark matter particles in a
cubic box of length $L = 1380 \hMpc$. The cosmology is a flat $\Lambda$CDM
model with $\Omega_b h^2$ = 0.022, $\Omega_m$ = 0.292, $n_s$ = 0.965,
$h$ = 0.69 and $\sigma_8 = 0.82$.

For testing and calibration of the modeling of halo clustering, we use
halo catalogs generated using a friends-of-friends (FOF) algorithm with a
linking length of 0.168 times the mean particle separation to
identify halos \cite{Davis:1985}. We consider 8 halo mass bins
(as a function of $M_{\text{fof}}$) across 10 redshift outputs, ranging from
$z=0$ to $z=1$. The redshift outputs considered are:
$z \in \{0, 0.1, 0.25, 0.4, 0.5, 0.55, 0.65, 0.75, 1\}$.
The (overlapping) halo mass bins range from
$\mathrm{log}_{10} M_{\mathrm{fof}} = 12.6$ to
$\mathrm{log}_{10} M_{\mathrm{fof}} = 14.4$
and are described in table~\ref{tab:mass-bins}. For reference,
table~\ref{tab:mass-bins} also gives the linear bias values
at $z=0.55$ and $z=0$. The linear biases for
each halo mass bin are determined from the ratio of the
large-scale halo-matter cross power spectrum to the
matter power spectrum at each redshift output.

\begin{table}[b]
\centering
  \begin{tabular}{l|c|c|c|c|c|c|c|c}
  \toprule
  bin & 1 & 2 & 3 & 4 & 5 & 6 & 7 & 8 \\
  \midrule
  $\log_{10} M_\mathrm{fof}^\mathrm{min}$ & 12.6 & 12.8 & 13.0 & 13.2 & 13.4 & 13.6 & 13.8 & 14.0\\
  \midrule
  $\log_{10} M_\mathrm{fof}^\mathrm{max}$& 13.0 & 13.2 & 13.4 & 13.6 & 13.8 & 14.0 & 14.2 & 14.4 \\
  \midrule
  $b_1(z=0.55)$ & 1.40 & 1.56 & 1.78 & 2.04 & 2.36 & 2.77 & 3.28 & 3.93 \\
  \midrule
  $b_1(z=0)$ & 1.00 & 1.07 & 1.19 & 1.33 & 1.51 & 1.74 & 2.03 & 2.41 \\
  \bottomrule
  \end{tabular}
\caption{The halo mass bins used when comparing results from the
RunPB simulations to theoretical modeling of halo clustering.
For each of the 10 redshift outputs ranging from $z=0$ to $z=1$,
we consider 8 fixed halo mass bins. We give the corresponding
large-scale, linear bias for each bin for two
redshifts, $z=0.55$ and $z=0$.}
\label{tab:mass-bins}
\end{table}

We also rely heavily on a set of galaxy catalogs produced using halo
occupation distribution (HOD) modeling from halo catalogs generated from
the $z=0.55$ RunPB realizations. The halo catalog production and the HOD modeling
is the same as in \cite{Reid:2014}: halos are identified using a spherical
overdensity (SO) algorithm and the HOD parameterization follows \cite{Zheng:2005}.
In \cite{Reid:2014}, the RunPB simulations are denoted as the
\texttt{MedRes} simulations. The HOD parameters used to generate the
galaxy catalog used in this work are $\{\log_{10}M_{\mathrm{min}},
\sigma_{\log_{10}M}, \log_{10}M_1, \alpha,
\log_{10}M_{\mathrm{cut}}\} = \{12.99, 0.308, 14.08, 0.824, 13.20 \}$.
These HOD parameters were chosen to reproduce the clustering of the BOSS
CMASS sample \cite{White:2011}, i.e., a large-scale linear bias of $b_1 \sim 2$ at
$z \sim 0.5$.

\subsection{N-series}\label{sec:nseries}

The N-series cubic boxes are a set of realizations from a
large-volume, high-resolution $N$-body simulation, used as part of
a ``mock challenge`` testing procedure by the BOSS collaboration
in preparation for publishing results as part of DR12 in \cite{Alam:2016}.
Details of this mock challenge can be found in \cite{Tinker:2016}.
Briefly, the N-series suite consists of
seven independent, periodic box realizations with the same cosmology, and
a side length of $L_\mathrm{box} = 2600 \ihMpc$ at a redshift
$z_\mathrm{box} = 0.5$. The cosmology is given by: $\Omega_m = 0.286$,
$\Omega_\Lambda =0.714$, $\sigma_8=0.82$, $n_s =0.96$, and $h=0.7$.
The $N$-body simulation was run using the GADGET2 code \cite{Springel:2005},
with sufficient mass and spatial resolution to resolve the
halos that typical BOSS galaxies occupy.
A single galaxy bias model was assumed, and HOD modeling was used to
populate halos from the seven realizations with galaxies. The parameters
of the HOD were chosen to reproduce the clustering
of the BOSS CMASS sample (i.e., linear bias $b_1 \sim 2$ at $z \sim 0.5$).

An additional set of 84 mock catalogs were generated from the
three orthogonal projections of each of the seven N-series cubic boxes
using the \texttt{make\_survey}
software\footnote{https://github.com/mockFactory/make\_survey}
\cite{White:2014b}. Denoted as the ``cutsky'' mocks,
these mocks have the same angular and radial selection function as
the NGC DR12 CMASS sample \cite{Reid:2016,Alam:2016}. They model the
true geometry, volume, and redshift distribution
of the CMASS NGC sample and provide a realistic simulation of the true BOSS
data set. Each catalog is an independent realization, and these mocks
were also used as part of the DR12 mock challenge.

\subsection{Lettered Challenge Boxes}\label{sec:lettered-challenge}

A second part of the BOSS DR12 mock challenge was performed on a suite of
HOD galaxy samples constructed from a heterogeneous set of
high-resolution $N$-body simulations. There are seven different
HOD galaxy catalogs, constructed from large-volume periodic
simulation boxes with varying cosmologies.
The seven catalogs are labeled A through G. Several of the boxes
are based on the Big MultiDark simulation \cite{Riebe:2013}.
The catalogs are constructed out of simulation boxes with
a range of 3 underlying cosmologies, and boxes with the same cosmology
have varying galaxy bias models, which varies the overall galaxy bias
by $\pm$5\%. The redshift of the boxes ranges from $z=0.441$ to
$z = 0.562$.

These cases were designed to quantify the sensitivity of
RSD models to the specifics of the galaxy bias model over
a reasonable range of cosmologies, testing for any possible
theoretical systematics. The cosmology and relevant
simulation parameters for each of these
boxes is given in table~\ref{tab:sim-params}. A comparison of the
results from the mock challenge in context of the BOSS DR12 results is presented
in \cite{Alam:2016}, and individual Fourier-space clustering results from the
challenge are discussed in \cite{Beutler:2017,Grieb:2017}.

\section{Analysis methods}\label{sec:methods}

In this work, we measure the 2-point clustering of galaxies as characterized
by the power spectrum multipoles, defined in terms of the 2D anisotropic
power spectrum as

\begin{equation}\label{eq:multipole-defintion}
P_\ell(k) = \frac{2\ell + 1}{2} \int_{-1}^{1} d\mu P(k,\mu) \mathcal{L}_\ell(\mu),
\end{equation}
where $\mathcal{L}_\ell$ is the Legendre polynomial of order $\ell$. We estimate
the multipoles from catalogs of discrete galaxies in periodic box $N$-body
simulations and from more realistic, cutsky mock catalogs, which mimic real
survey data. We estimate the continuous galaxy overdensity field using
a Triangular Shaped Cloud interpolation scheme (see e.g., \cite{Hockney:1981})
to assign the galaxy positions to a 3D Cartesian grid.

For both periodic boxes
and cutsky mocks, we employ Fast Fourier Transform (FFT) based estimators
to compute the multipoles. In the case of simulation boxes with
periodic boundary conditions, the line-of-sight is assigned to a specific
box axis and the power spectrum $P(k,\mu)$ can be computed
as the square of the Fourier modes of the overdensity field. The desired
multipoles are then found by computing equation~\ref{eq:multipole-defintion}
as a discrete sum over $P(k,\mu)$. In the case of the cutsky mocks, we employ
the FFT-based estimator described in \cite{Hand:2017}, which modifies
the FFT estimator presented by \cite{Bianchi:2015} and \cite{Scoccimarro:2015}.
Building on the ideas of previous power spectrum estimators
\cite{Feldman:1994,Yamamoto:2006}, this estimator uses a spherical
harmonic decomposition to allow the use of FFTs to compute the
higher-order multipoles, with 5 and 9 FFTs required to compute the
quadrupole and hexadecapole, respectively. When computing FFTs, we ensure
that the grid configuration is such that our desired maximum wavenumber is
not greater than one-half of the Nyquist frequency of the grid, which should
eliminate any aliasing effects on our measured power spectra
(i.e., \cite{Sefusatti:2016}). The measured power spectra are estimated on a
discrete $k$-grid which makes the angular distribution of Fourier modes
irregular. This discreteness effect is especially important at low $k$ and can
be accounted for by modifying equation~\ref{eq:multipole-defintion} as

\begin{equation}
\label{eq:irregular-mu}
P_\ell(k) = \frac{2\ell + 1}{2} \int_{-1}^{1} d\mu P(k,\mu)
			\frac{N_\mathrm{modes}(k,\mu)}{N_\mathrm{bin}(k)}\mathcal{L}_\ell(\mu),
\end{equation}
where the $N(k,\mu)$ gives the total number of modes on the $k$-space grid,
and the normalization is

\begin{equation}
N_\mathrm{bin}(k) = \int_{-1}^{1} d\mu N_\mathrm{modes}(k,\mu).
\end{equation}
When computing the theoretical multipoles with the model in this work, which
we compare to simulation results, we use equation~\ref{eq:irregular-mu}
to account for the discreteness effect present in the simulation multipoles.
This procedure has been shown to sufficiently correct for this effect
\cite{Beutler:2017}. For all power spectrum calculations, we use the
publicly-available software package
\texttt{nbodykit}\footnote{https://github.com/bccp/nbodykit}
\cite{Hand:inprep}, which uses massively parallel implementations of these
estimators for fast calculations optimized to run on high-performance
computing machines.

We use the Markov chain Monte Carlo (MCMC) technique to derive the likelihood
distributions of the model parameters described in detail in
section~\ref{sec:model-params}. We employ a modified version of the Python code
\texttt{emcee}\footnote{https://github.com/dfm/emcee} \cite{Foreman-Mackey:2013}
to explore the relevant model parameter space. The data vector used
in these fits is the concatenation of the monopole, quadrupole,
and hexadecapole,

\begin{equation}
\label{eq:data-vector}
\mathcal{D} = \left[P_0(k), P_2(k), P_4(k) \right],
\end{equation}
where we have measured the multipoles from simulations as previously described.
The inclusion of the hexadecapole $P_4(k)$ has been shown to offer significant
improvements on RSD constraints, i.e., \cite{Beutler:2017,Grieb:2017}. In
all fits, we use a bin spacing in wavenumber of $\Delta k = 0.005 \ihMpc$,
and the maximum wavenumber included in the fits ranges from
$k_\mathrm{max} = 0.2 \ihMpc$ to $k_\mathrm{max} = 0.4 \ihMpc$.
The likelihood fits require an estimate of the covariance matrix, and
we use the theoretical Gaussian covariance for the multipoles in Fourier space
(i.e., \cite{Grieb:2016}). In the case of the cutsky mocks, we properly
account for the redshift distribution and survey volume of the mock catalogs
when computing the expected covariance, using e.g., \cite{Yamamoto:2006}. Our
choice for covariance matrix ignores non-Gaussian contributions produced
by e.g., nonlinear structure growth, and in the case of the cutsky mocks,
correlations induced by the window function due to the survey geometry.
We have tested the impact of this choice for covariance matrix by comparing
the parameter fits obtained when using a covariance matrix derived from a
set of 1000 mock catalogs from the Quick Particle Mesh (QPM; \cite{White:2014b})
simulations. While we do find variations in the best-fit parameters found
when using the simulation-based covariance, the shifts are consistent with
the derived errors.

\section{The power spectrum model}\label{sec:power-spectrum-model}

In this section, we present the model for the anisotropic
clustering of galaxies in Fourier space, as characterized by the broadband,
two-dimensional power spectrum. First, we connect the clustering of galaxies
to the clustering of halos, reviewing the halo model formalism presented
in \cite{Okumura:2015} in \S\ref{sec:galaxy-model}.
We describe our model for the redshift-space halo power spectrum and
the various modeling improvements from past work in \S\ref{sec:halo-model}.
In \S\ref{sec:obs-model}, we discuss
how we account for various observational effects when modeling real galaxy
survey data. Finally, we summarize the complete set of model parameters in
\S\ref{sec:model-params}.

\subsection{Halo model formalism for galaxies}\label{sec:galaxy-model}

Our treatment of the clustering of galaxies is based upon the model presented
in \cite{Okumura:2015}. The clustering of a given galaxy sample is
considered within the context of a halo model
\cite{Seljak:2000,Peacock:2000,Ma:2000,Scoccimarro:2001,Cooray:2002},
which allows one to separately consider contributions to the clustering
arising from galaxies within the same halo and those from separate halos,
known as the 1-halo and 2-halo terms, respectively. This formalism is
ideal when accounting for the effects of satellite galaxies on the
anisotropic power spectrum, where the radial distribution of satellites
induces both 1-halo and 2-halo effects. We describe the relevant model
details from \cite{Okumura:2015}, used in this work, below.

\subsubsection{Galaxy sample decomposition}\label{sec:gal-decomp}

In redshift space, we can decompose contributions to the galaxy overdensity
field $\delta_g^S$ into contributions from central and satellite
galaxies as

\begin{equation}
\delta_g^S(\vk) = (1-f_{s})\delta_c^S(\vk) + f_{s}\delta_s^S(\vk),
\end{equation}
where $f_s = N_s / N_g = 1-N_c/N_g$ is the satellite fraction, $N_c$ and $N_s$
are the numbers of central and satellite galaxies, respectively, and
$N_g = N_c + N_s$ is the total number of galaxies. It follows then that
the power spectrum of the galaxy density field
$(2\pi)^3 P_{gg}^S(\vk) \delta(\vk+\vk')   \equiv\la
\delta_g^S(\vk)\delta_g^{S}(\vk') \ra$ can be expressed as

\begin{equation}
 P^S_{gg}(\vk) = (1-f_{s})^2 P^S_{cc}(\vk)
 				+  2f_{s}(1-f_{s}) P^S_{cs}(\vk)
                + f_{s}^2 P^S_{ss}(\vk),
\end{equation}
where $P^S_{cc}$, $P^S_{cs}$, and $P^S_{ss}$ are the centrals auto power spectrum,
the central-satellite cross power spectrum, and the satellite auto power spectrum
in redshift space, respectively.

To fully separate 1-halo and 2-halo contributions to the power spectrum,
we further decompose the central and satellite galaxy samples.
We decompose the central galaxy density field into those centrals
that do and do not have a satellite galaxy in the same halo, denoted as
types ``A'' and ``B'' centrals, respectively. For the latter type, a 1-halo
contribution will exist due to the central-satellite correlations inside
the same halo. We use a similar decomposition for satellite galaxies, where
we consider satellites that only have a single satellite in a
halo (type ``A'') and those satellites that live in halos with more than
one satellite (type ``B''). The latter type will contribute a 1-halo term
to the power spectrum, due to correlations between multiple satellites in
the same halo.

With these galaxy sample definitions, we can express the central-satellite and
satellite-satellite power spectra in terms of 1-halo and 2-halo correlations. Note
that by construction, a halo can only have a single central galaxy, and thus, the
centrals auto spectrum is a purely 2-halo contribution. The central-satellite cross
power spectrum can be expressed as

\begin{align}
P^S_{cs}(\vk) &= (1-f_{c_B}) P^S_{c_As}(\vk) + f_{c_B} P^S_{c_Bs}(\vk), \nn \\
              &= (1-f_{c_B}) \left [(1-f_{s_B}) P^S_{c_A s_A}
              						+ f_{s_B} P^S_{c_A s_B} \right] +
              f_{c_B} \left [(1-f_{s_B}) P^S_{c_B s_A} + f_{s_B} P^S_{c_B s_B} \right], \label{eq:Pcs}
\end{align}
where $f_{c_B} = N_{c_B} / N_c$ is the fraction of centrals
that have a satellite in the same halo, and $f_{s_B} = N_{s_B} / N_s$
is the fraction of satellites that live in halos with more than one
satellite. Because the sample $c_B$ consists of central galaxies that have
satellite galaxies inside the same halo, the term $P^S_{c_Bs}$
(and similarly, $P^S_{c_B s_A}$ and $P^S_{c_B s_B}$) contains a 1-halo
contribution, so we write it as $P^S_{c_Bs}=P^{S, 1h}_{c_Bs} + P^{S, 2h}_{c_Bs}$. All other
power spectra terms in equation \ref{eq:Pcs} are purely 2-halo contributions.

Similarly, we can express the satellite auto power spectrum as

\begin{equation}
P^S_{ss}(\vk) = (1 - f_{s_B})^2 P^S_{s_A s_A}(\vk) + 2 f_{s_B} (1-f_{s_B}) P^S_{s_As_B}(\vk)
			  + f_{s_B}^2 P^S_{s_Bs_B}(\vk). \label{eq:Pss}
\end{equation}
As in the case of $P^S_{c_B s}$, the term $P^S_{s_Bs_B}$ includes
both 1-halo and 2-halo contributions, which we can express as
$P^S_{s_Bs_B} = P^{S, 1h}_{s_B s_B}+P^{S, 2h}_{s_Bs_B}$. All other terms
in equation \ref{eq:Pss} include only 2-halo contributions.

Combining the terms in equations \ref{eq:Pcs} and \ref{eq:Pss}, the
galaxy power spectrum in redshift space is

\begin{equation}
\label{eq:Pgg-model}
P^S_{gg}(\vk) = P^{S, 1h}_{gg}(\vk) + P^{S, 2h}_{gg}(\vk),
\end{equation}
where the 2-halo contributions are given by

\begin{align}
P^{S, 2h}_{gg}(\vk) &= (1-f_s)^2 P^S_{cc} \nn \\
     &+ 2 f_s(1-f_s) \left \{ (1-f_{c_B})
     \left [(1-f_{s_B}) P^S_{c_A s_A} + f_{s_B} P^S_{c_A s_B} \right] \right\} \nn \\
     &+ 2 f_s (1-f_s) \left \{ f_{c_B}
     \left [(1-f_{s_B}) P^{S,2h}_{c_B s_A} + f_{s_B} P^{S,2h}_{c_B s_B} \right] \right \} \nn \\
     &+ f_s^2  \left [(1 - f_{s_B})^2 P^S_{s_A s_A} + 2 f_{s_B} (1-f_{s_B}) P^S_{s_As_B}
			  + f_{s_B}^2 P^{S,2h}_{s_Bs_B} \right], \label{eq:full-2halo}
\end{align}
and the 1-halo contributions are

\begin{align}
P^{S, 1h}_{gg}(\vk) &= 2 f_s (1-f_s) \left \{ f_{c_B}
     \left [(1-f_{s_B}) P^{S,1h}_{c_B s_A}
     		+ f_{s_B} P^{S,1h}_{c_B s_B} \right] \right \} \nn \\
     &+ f_s^2 f_{s_B}^2 P^{S,2h}_{s_Bs_B}. \label{eq:full-1halo}
\end{align}

\subsubsection{Modeling 1-halo and 2-halo terms in redshift space}

In this subsection, we discuss how we model the 1-halo and 2-halo terms in
redshift space that enter into equations \ref{eq:full-2halo} and
\ref{eq:full-1halo}. Our modeling of these terms in redshift space
largely follows the work presented in \cite{Okumura:2015};
for completeness, we reproduce the relevant results from that work below.

Modeling complications arise due to the effects of the radial distribution of
satellite galaxies inside halos on the galaxy power spectrum. In real space,
correlations between galaxies on small scales give rise to the 1-halo term.
In Fourier space, the 1-halo term manifests as a white noise-like term at low $k$,
with departures from white noise at larger $k$ due to the radial
profile of satellites inside halos. As shown in \cite{Okumura:2015},
the deviations from white noise are small on the scales of interest for
cosmological parameter inference ($k \lesssim 0.4 \ihMpc$). Thus, we treat
all 1-halo terms in real space as independent of wavenumber. As discussed in
section \ref{sec:gal-decomp}, there are two sources of 1-halo terms: 1) the
correlation between the $c_B$ sample of centrals and satellites and 2) the
auto-correlation between the $s_B$ sample of satellites. We denote the
real-space amplitude of these terms as $N_{c_B s}$ and $N_{s_B s_B}$,
respectively.

In redshift space, satellite galaxies are spread out in the radial direction
by their large virial velocities inside halos, an effect known as
Fingers-of-God \cite{Jackson:1972}. Affecting both 1-halo and 2-halo
correlations, the FoG effect is a fully nonlinear process, and it is
not possible to accurately model it using perturbation theory.
Quasi-linear perturbative approaches have been developed which use damping
functions, i.e., a Gaussian or Lorentzian, to model the effect
\cite{Scoccimarro:2004,Taruya:2010,Peacock:1994,Park:1994,Percival:2009}.
In previous studies, the effect is typically modeled with a single
damping factor $G(k\mu;\sigma_v)$, with $\sigma_{v}$ corresponding to
the velocity dispersion of the full galaxy sample. In such a model, the
redshift-space power spectrum of galaxies is modeled as
$P_{gg}^S(k,\mu)=G^2(k\mu;\sigma_v)P_{hh}^S(k,\mu)$,
where $P_{hh}^S$ is the redshift-space halo power spectrum.

We separately model the FoG effect from each of the galaxy subsamples
defined in section \ref{sec:gal-decomp}. As demonstrated in
\cite{Okumura:2015}, the FoG effect on the 1-halo and 2-halo terms
from satellite galaxies can be accurately described using a damping function
and the typical virial velocity associated with the halos hosting the satellites.
The functional form of the damping function used in this work is

\begin{equation}
\label{eq:fog-model}
G(k\mu; \sigma_v) = \left( 1+ k^2\mu^2\sigma_v^2/2 \right)^{-2},
\end{equation}
which has a form slightly modified from the commonly-used forms used in
the literature. As shown in \cite{Okumura:2015}, this damping function can
accurately model the FoG effect from satellites over a wide range of scales,
extending down to $k \sim 0.4 \ihMpc$. The dominant FoG effect
arises from satellites, and we include velocity dispersion parameters for
each of the satellite subsamples,
$\sigma_{v, s_A}$ and $\sigma_{v, s_B}$. Recent clustering analyses,
e.g., \cite{Reid:2014,Guo:2015}, have also found evidence that central galaxies
are not at rest with respect to the halo center, giving rise
to an additional FoG effect (albeit smaller than the effect from satellites).
To properly account for this possibility, we also include
a velocity dispersion parameter associated with central galaxies,
$\sigma_{v,c}$. We assume a single velocity dispersion for both the
$c_A$ and $c_B$ galaxy samples.

There are five power spectra in equation \ref{eq:full-2halo}
that include only 2-halo terms. Using the above FoG modeling, these terms become

\begin{align}
\label{eq:P_cc}
P^s_{cc}(k, \mu) &= G(k\mu;\sigma_{v,c})^2 P^S_{cc,h}(k,\mu), \\
\label{eq:P_cAsA}
P^S_{c_A s_A}(k, \mu) &= G(k\mu;\sigma_{v,c})G(k\mu;\sigma_{v,s_A}) P^S_{c_A s_A, h}(k,\mu), \\
\label{eq:P_cAsB}
P^S_{c_A s_B}(k, \mu) &= G(k\mu;\sigma_{v,c})G(k\mu;\sigma_{v,s_B}) P^S_{c_A s_B, h}(k,\mu), \\
\label{eq:P_sAsA}
P^S_{s_A s_A}(k, \mu) &= G(k\mu;\sigma_{v,s_A})^2 P^S_{s_A s_A, h}(k,\mu), \\
\label{eq:P_sAsB}
P^S_{s_A s_B}(k, \mu) &= G(k\mu;\sigma_{v,s_A})G(k\mu;\sigma_{v,s_B})
						P^S_{s_A s_B, h}(k,\mu),
\end{align}
where $P^S_{XX,h}$ represents the auto power spectrum of halos in which
the galaxies of types $X$ reside and $P^S_{XY,h}$ the cross spectrum of
halos in which galaxies of types $X$ and $Y$ reside. Under the assumption
of linear perturbation theory, $P_{XY,h}^S$ converges to the linear
redshift-space power spectrum originally proposed by \cite{Kaiser:1987},
$P^S_{XY,h}(k,\mu)=(b_{1,X}+f\mu^2)(b_{1,Y}+f\mu^2)P_L(k)$, where $P_L(k)$
is the linear matter power spectrum, and $b_1$ is the linear bias factor
of the specified galaxy sample.

The three power spectra that include both 1-halo and 2-halo terms
can be expressed as

\begin{align}
\label{eq:P_cBsA}
P^S_{c_B s_A}(k, \mu) &= G(k\mu;\sigma_{v,c})G(k\mu;\sigma_{v,s_A})
			\left[ P^S_{c_B s_A, h}(k,\mu) + N_{c_B s} \right],  \\
\label{eq:P_cBsB}
P^S_{c_B s_B}(k, \mu) &= G(k\mu;\sigma_{v,c})G(k\mu;\sigma_{v,s_B})
			\left[ P^S_{c_B s_B, h}(k,\mu) + N_{c_B s} \right],  \\
\label{eq:P_sBsB}
P^S_{s_B s_B}(k, \mu) &= G(k\mu;\sigma_{v,s_B})^2
			\left[ P^S_{s_B s_B, h}(k,\mu) + N_{s_B s_B} \right],
\end{align}
where $N_{c_B s}$ is the 1-halo amplitude due to correlations between
centrals and satellites in the same halo, and $N_{s_B s_B}$ is the 1-halo
amplitude between satellites inside the same halo.

\subsection{Halo clustering in redshift-space}\label{sec:halo-model}

The remaining modeling unknown needed in equations
\ref{eq:P_cc} -- \ref{eq:P_sAsB} and \ref{eq:P_cBsA} -- \ref{eq:P_sBsB}
is the prescription for the redshift-space halo power spectrum, $P_{XY,h}^S(k,\mu)$.
In this section, we describe our model for the halo power spectrum, which relies
on a combination of perturbation theory and simulations. The model is based on
the formalism presented in \cite{Vlah:2013}, with important differences and
improvements discussed below.

\subsubsection{Distribution function model for redshift-space distortions}

Our model for the power spectrum of halos in redshift space relies on
expressing the redshift-space halo density field in terms of moments of
the distribution function (DF); the approach has been developed and tested
in a previous series of papers
\cite{Seljak:2011, Okumura:2012, Okumura:2012b, Vlah:2012,Vlah:2013,Blazek:2014}.
If we consider halo samples $X$ and $Y$, with linear biases $b_{1, X}$
and $b_{1, Y}$, the redshift-space power spectrum in the DF model can
be expressed as a sum over mass-weighted, velocity correlators

\begin{equation}
\label{eq:df-power-expansion}
P^S_{XY, h}(k,\mu)=\sum^\infty_{L=0}\sum^\infty_{L'=0}\frac{(-1)^{L'}}{L!L'!}
		\left(\frac{ik\mu}{\mathcal{H}}\right)^{L+L'} P^{XY,h}_{LL'}(k,\mu),
\end{equation}
where $\mathcal{H} = aH$ is the conformal Hubble parameter, and
$P^{XY,h}_{LL'}$ is the power spectrum of the moments $L$ and $L'$
of the radial halo velocity field, weighted by the halo density field.
These spectra are defined as

\begin{equation}
\label{eq:velocity-correlators}
(2\pi)^3 P^{XY,h}_{LL'}(\vk) \delta_D(\vk+\vk') =
		\la T_\parallel^{X, L}(\vk)T_\parallel^{Y, L'}(\vk') \ra,
\end{equation}
where $T_\parallel^{X,L}(\vk)$ is the Fourier transform of the
corresponding halo velocity moment weighted by halo density,

\begin{equation}
T_\parallel^{X,L}(\vx) = \left[ 1+\delta^h_X(\vx)\right]
					\left ( v_{\parallel,X}^h \right)^L,
\end{equation}
where $\delta^h_X$ and $v_{\parallel,X}^h$ are the halo density and
radial velocity fields for sample $X$, respectively. The velocity correlators
defined in equation \ref{eq:velocity-correlators} have well-defined
physical interpretations; for example, $P^{XX,h}_{00}$ represents the
halo density auto power spectrum of sample $X$, whereas $P^{XX,h}_{01}$
is the cross-correlation of density and radial momentum for halo sample $X$.
The DF approach naturally produces an expansion of $P^S_{XY, h}(k, \mu)$
in even powers  of $\mu$, with a finite number of correlators contributing
at a given power of $\mu$. For this work, we consider terms up to and
including $\mu^4$ order in the expansion of equation \ref{eq:df-power-expansion}.

To evaluate the halo velocity correlators in equation
\ref{eq:df-power-expansion}, we largely follow the results outlined
in \cite{Vlah:2012, Vlah:2013}, where the correlators are evaluated
using Eulerian perturbation theory. However, in order to increase
the overall accuracy of the power spectrum model, our work differs from
the results presented in \cite{Vlah:2013} in several crucial areas.
These differences will be discussed in the subsequent subsections
of this section.

\subsubsection{The modeling of halo bias}\label{sec:halo-bias-model}

The spectra $P^{XY,h}_{LL'}(k,\mu)$ in equation \ref{eq:df-power-expansion}
are defined with respect to the halo field, and a biasing model is
needed to relate them to the correlators of the underlying dark matter
density field. Following the results of \cite{Vlah:2013}, we use a
second-order, nonlocal Eulerian biasing model, where the only nonlocal
term results from the second-order tidal tensor. The nonlinear and nonlocal
biasing contributions have been demonstrated to improve the accuracy of
theoretical models, e.g., \cite{Baldauf:2013, Saito:2014}.

As discussed in \cite{Vlah:2013}, the second-order bias in our biasing
scheme is an effective bias, accounting for several free bias parameters that
enter at the 1-loop level, all with similar scale dependence. The spectra
$P^{XX, h}_{00}$ and $P^{XX, h}_{01}$ have distinct values for this effective bias
parameter, $b_2$, and the biasing model in all higher-order correlators enters through
these two terms. The difference can be understood through effects of the
third-order, nonlocal bias, which appears to be equally important to $b_2$
\cite{McDonald:2009,Saito:2014}.

Our biasing scheme has four bias parameters for each halo sample:
the linear bias $b_1$, the two effective second-order biases,
$b_2^{00}$ and $b_2^{01}$, and the nonlocal tidal bias $b_s$. However,
we treat the higher-order biases as functions of $b_1$, and thus, the
only free bias parameter for each halo sample is the linear bias.
In the case of the local Lagrangian bias model, we can predict the
amplitude of the nonlocal tidal bias in terms of the linear bias
\cite{Chan:2012,Baldauf:2012}

\begin{figure}[tb]
\centering
\includegraphics[width=0.7\textwidth]{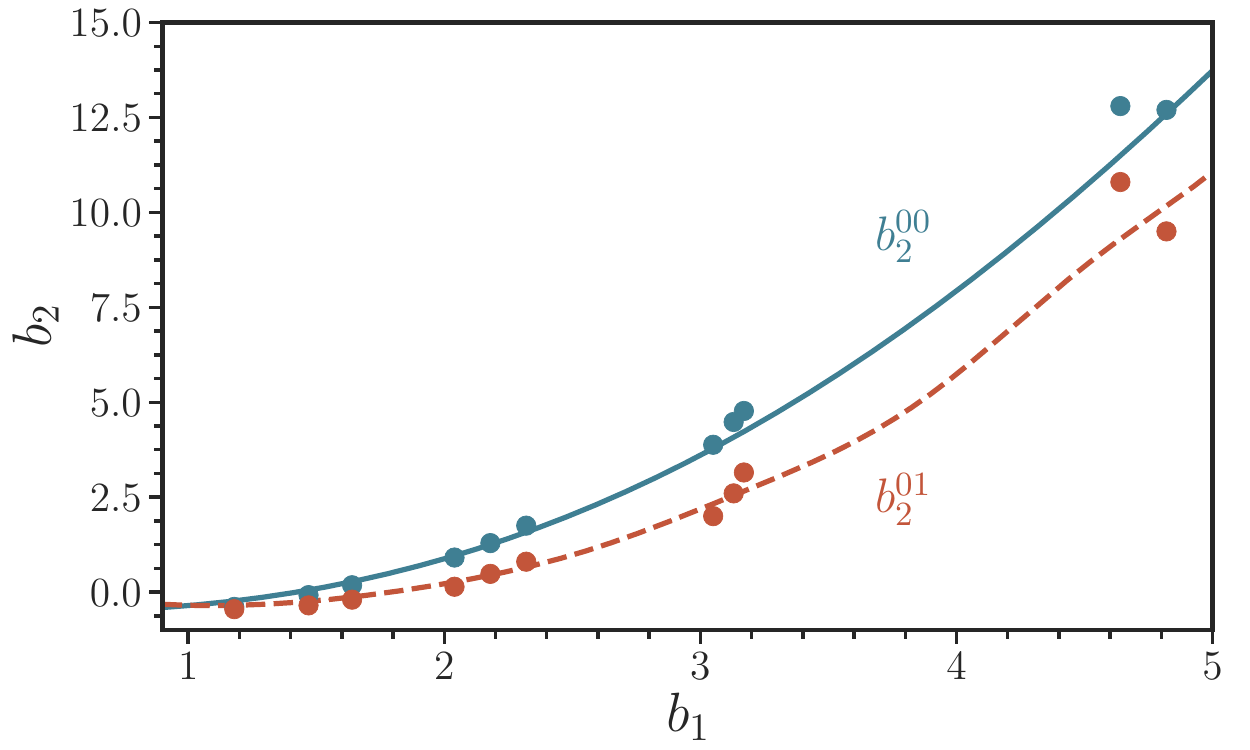}
\caption{The dependence of the second-order nonlinear effective
biases, $b_2^{00}$ (blue, solid) and $b_2^{01}$ (red, dashed), on the
linear bias $b_1$ used in this work, as determined from the RunPB simulations.
For comparison, the best-fit bias parameters from \cite{Vlah:2013} are shown
as circles.}
\label{fig:b2}
\end{figure}

\begin{equation}
b_s = -\frac{2}{7} (b_1 - 1).
\end{equation}
As shown in \cite{Vlah:2013}, the tidal bias does not play a
prominent role in the biasing model, but nonetheless, we include these
terms in our model. The effective biases $b_2^{00}$ and $b_2^{01}$ have
a roughly quadratic dependence on the linear bias $b_1$. Rather than
freely varying these bias parameters, we treat them as a function of $b_1$,
independent of redshift, and use simulations to determine the exact
functional form of this dependence. We use the set of halo mass bins from
the RunPB simulations described in section~\ref{sec:RunPB} and use
Gaussian Process regression (see, e.g., \cite{Rasmussen:2006}) to
predict the functional form of $b_2^{00}(b_1)$ and $b_2^{01}(b_1)$.
For this purpose, we use the public Gaussian Process package
\texttt{george}\footnote{https://github.com/dfm/george}
\cite{Ambikasaran:2014}. The predictions for $b_2^{00}$ and
$b_2^{01}$ used in this work, as determined from the RunPB simulations,
are shown in figure~\ref{fig:b2}. We also show the best-fit bias parameters
used in \cite{Vlah:2013} for several redshifts, which are consistent
with the results obtained from the RunPB simulations.

\subsubsection{Improved modeling of dark matter correlators}\label{sec:hzpt}

\begin{figure}[btp]
\centering
\includegraphics[width=\textwidth]{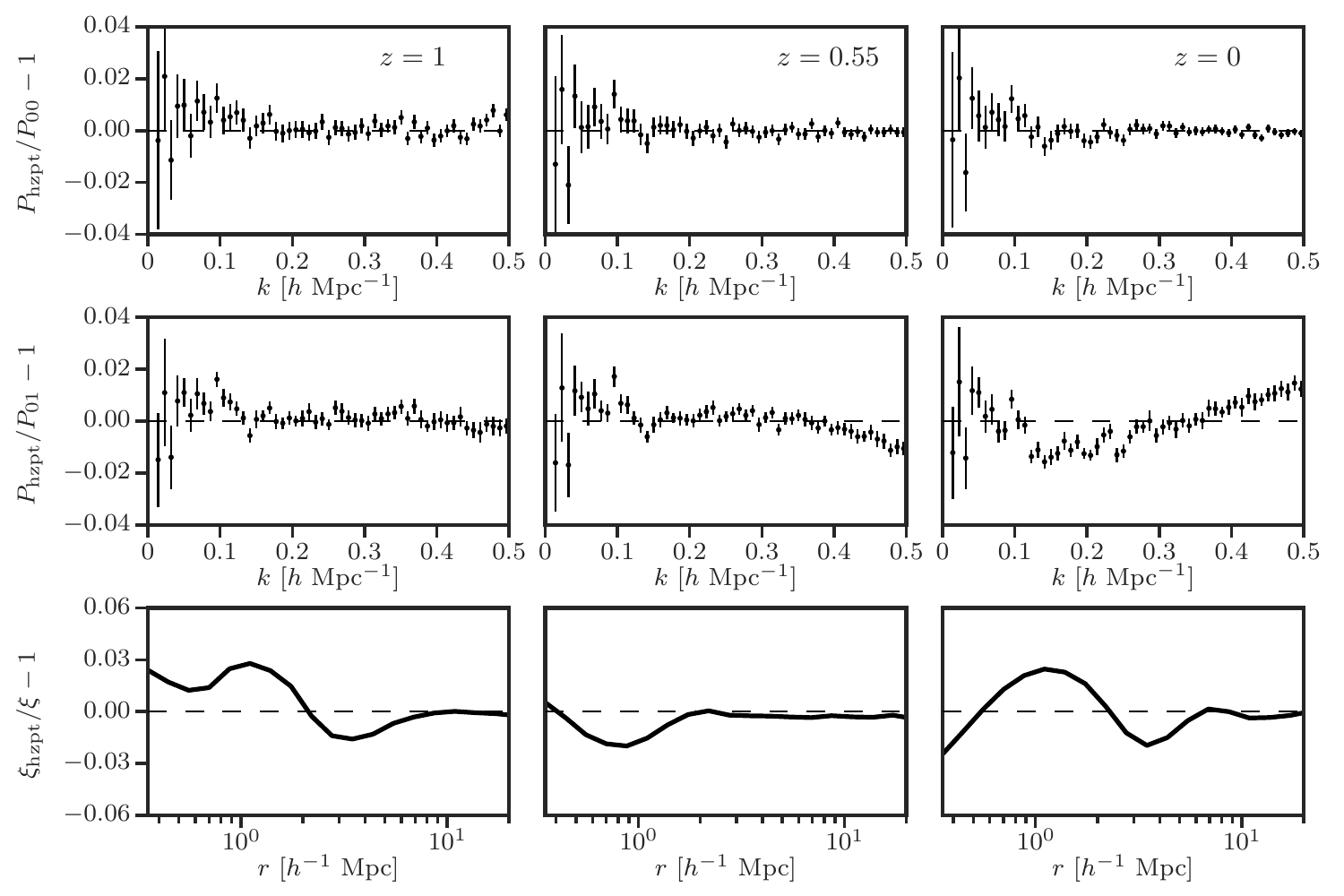}
\caption{The accuracy of the dark matter HZPT modeling results used
in this work, in comparison to the results from the RunPB simulation.
We compare the dark matter power spectrum $P_{00}$ (top),
density -- radial momentum cross-power $P_{01}^S$ (middle), and the small-scale correlation function $\xi_{00}$ (bottom). We give results for
three redshift outputs: $z=1$ (left), $z=0.55$ (center), and
$z=0$ (right). The updated HZPT model parameters are presented in
Appendix \ref{app:hzpt-P00}.}
\label{fig:hzpt-matter}
\end{figure}

We use the Halo-Zel'dovich Perturbation Theory (HZPT)
approach of \cite{Seljak:2015} to model the dark matter density
power spectrum $P_{00}(k)$ and the redshift-space cross-correlation of
dark matter density and radial momentum $P_{01}^S(\vk)$.
This differs from the results presented in \cite{Vlah:2013},
which uses standard perturbation theory (SPT) to evaluate these terms
(which is known to break down at relatively large scales).
Note that the density-momentum cross-correlation can be related to
$P_{00}$ through
$P_{01}^S (\vk) = \mu^2 dP_{00} / d\mathrm{ln}a$ \cite{Vlah:2012},
so only an accurate model for $P_{00}$ is required.
The statistic $P_{01}^S(\vk)$ plays a crucial role in the modeling
of the $\mu^2$ angular dependence of the redshift-space power spectrum.

The HZPT model connects the Zel'dovich approximation
\cite{Zeldovich:1970,White:2014} with a Pad\'e expansion for a 1-halo-like
term that is determined from simulations using simple, physically motivated
parameter scalings. The model for the dark matter
power spectrum has been demonstrated to be accurate to 1\% to $k \sim 1 \ihMpc$
\cite{Seljak:2015}, and the Zel'dovich approximation performs sufficiently well
when modeling BAO, relative to other modeling techniques
\cite{White:2014, Vlah:2015}.

We provide an update to the HZPT results presented in \cite{Seljak:2015},
using the dark matter RunPB simulations detailed
in section \ref{sec:RunPB}. We extend the analysis of \cite{Seljak:2015}
to include measurements of $P_{01}^S(\vk)$, as well as the small-scale
matter correlation function. We also extend the redshift fitting range,
using a set of 10 redshift outputs from the RunPB simulations, ranging
from $z=0$ to $z=1$. We perform a global fit of the amplitude and
redshift-dependence of the 5 parameters in the HZPT model using the
$P_{00}(k)$ and $P_{01}^S(\vk)$ statistics over the range
$k = 0.005-0.5 \ihMpc$, as well as the small-scale correlation
function over the range $r = 0.3 - 25 \ \Mpc/h$. Qualitatively, the results
remain similar to those presented in \cite{Seljak:2015}, but the use of
additional statistics (in particular, the small-scale correlation function)
in the fit does allow some parameter degeneracies to be broken.

We review the HZPT model for $P_{00}$ and $P_{01}^S$ in
appendix~\ref{app:hzpt-P00}, and provide the updated
best-fit model parameters. We also detail the necessary calculation
of $P_{01}^S$ in the Zel'dovich approximation in
appendix~\ref{app:zeldovich}. We show the accuracy of the HZPT model
for the three statistics considered in figure~\ref{fig:hzpt-matter}
for three redshift snapshots, $z=0$, $0.55$, and $1$. It is evident that the
5-parameter HZPT model can provide a consistent picture of the power
spectra to an accuracy of 1-2\% over the range of scales
considered in this work. Furthermore, keeping in mind that the inclusion
of baryonic effects can effect the parameters $R_1, R_{1h}, R_{2h}$ at the
5-10\% level \cite{vanDaalen:2011,Mohammed:2014}, the
model used in this work performs reasonably well at modeling the
notoriously difficult 1-halo to 2-halo regime of the correlation function.

\begin{figure}[btp]
\centering
\includegraphics[width=\textwidth]{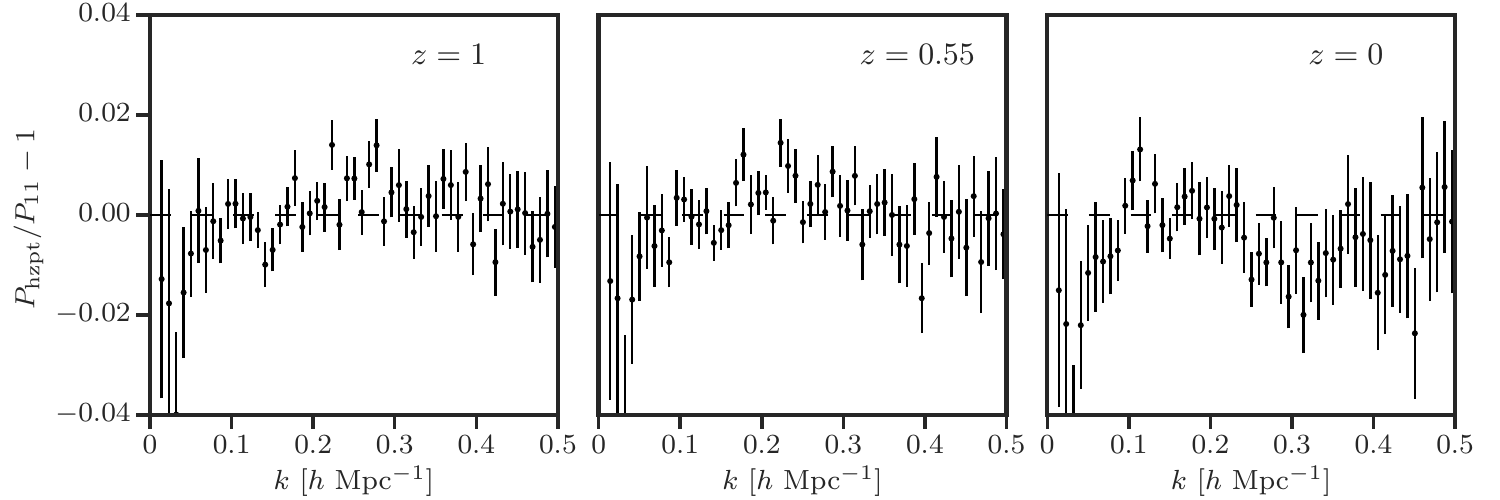}
\caption{The accuracy of the HZPT model for the auto power spectrum of the
dark matter radial momentum $P_{11}^S[\mu^4]$ in comparison to the results from the
RunPB simulations. We give results for
three redshift outputs: $z=1$ (left), $z=0.55$ (center), and $z=0$ (right). The
best-fit HZPT model parameters are presented in Appendix \ref{app:hzpt-P11}.}
\label{fig:hzpt-P11}
\end{figure}

We also extend the HZPT approach to model the dark matter radial
momentum auto power spectrum, $P_{11}^S(\vk)$, which is important for
modeling the $\mu^4$ angular dependence of $P^S_{XY,h}(k,\mu)$. Specifically,
we model the scalar component of $P_{11}^S[\mu^4]$ with the sum of a Zel'dovich
term and Pad\'e expression and the vector contribution using 1-loop SPT
(as was done in \cite{Vlah:2012}). The full model is given by the sum of
the scalar and vector contributions

\begin{align}
P_{11}^S[\mu^4](k) &= P_{11,s}^S[\mu^4](k) + P_{11,v}^S[\mu^4](k), \nn \\
			&= P_{11,s}^\mathrm{zel}(k) + P^{BB}_{11}(k) - f^2 I_{31}(k),
            \label{eq:P11-model}
\end{align}
where the vector contribution $I_{31}(k)$ is defined in \cite{Vlah:2012}, and
we discuss the Zel'dovich term $P_{11,s}^\mathrm{zel}$ in detail in
Appendix \ref{app:zeldovich}. We define the Pad\'e term $P^{BB}_{11}$ and give
the best-fit parameters (fit using the RunPB simulations)
in Appendix \ref{app:hzpt-P11}. Figure~\ref{fig:hzpt-P11} compares the
accuracy of the model in equation~\ref{eq:P11-model} with the results from
the RunPB simulations for three redshift snapshots. The figure shows the model
to be accurate to 1-2\% over the range of scales of interest.

Finally, rather than using the 1-loop SPT expressions for
the dark matter density -- velocity divergence cross power spectrum
$P_{\delta \theta}(k)$ and the velocity divergence auto power spectrum
$P_{\theta \theta}(k)$, we use the fitting formula from \cite{Jennings:2012}.
While the 1-loop SPT expressions for $P_{\delta \theta}$ and $P_{\theta \theta}$
diverge from truth at relatively large scales ($k \sim 0.1 \ihMpc$), the model
of \cite{Jennings:2012} achieves the necessary accuracy over the range of scales
considered in this work.

\subsubsection{Halo stochasticity}

The $\mu^0$ component of the redshift-space halo spectrum, $P^S_{XY,h}(k,\mu)$,
in the DF model is the isotropic, real-space auto spectrum of the halo density
field, $P^{hh}_{00}(k)$. For a complete description of this term, we must
accurately model the contribution from the stochasticity of halos, defined for
two separate halo mass bins ($h$ and $\bar{h}$) as

\begin{equation}
\label{eq:lambda}
\Lambda(k) = P_{00}^{h\bar{h}}(k) - \bar{b}_1(k) P_{00}^{hm}(k) -
		 b_1(k) P_{00}^{\bar{h}m}(k) + b_1(k) \bar{b}_1(k) P_{00}(k),
\end{equation}
where $P_{00}^{hm}$ and $P_{00}^{\bar{h}m}$ are the halo--matter cross
power spectra for the halo mass bins $h$ and $\bar{h}$, respectively, and
$P_{00}$ is the matter power spectrum (modeled using HZPT, as described in
section \ref{sec:hzpt}). The scale-dependent linear bias factors are defined as

\begin{equation}
b_1(k) \equiv \frac{P^{hm}_{00}(k)}{P_{00}(k)}, \ \ \ \
\bar{b}_1(k) \equiv \frac{P^{\bar{h}m}_{00}(k)}{P_{00}(k)}.
\end{equation}

In the Poisson model, the leading-order term of the stochasticity is given by the
Poisson shot noise, $\bar{n}^{-1}$, where $\bar{n}$ is the halo number density.
However, there are significant deviations from this prediction that
have complicated scale dependence. These corrections originate from
two competing effects: first, the halo exclusion effects due to the
finite size of halos and second, the nonlinear clustering of halos
relative to dark matter \cite{Baldauf:2013, Vlah:2013, Baldauf:2016}.
In the $k \rightarrow 0$ limit, the stochasticity behaves close to
white noise, where halo exclusion lowers the stochasticity relative to
the Poisson value and nonlinear clustering leads to a positive contribution.
However, in the high-$k$ limit, the stochasticity must approach the
Poisson value, and these deviations vanish; thus, there exists a
complicated scale dependence that is not well-understood theoretically.

\begin{figure}[btp]
\centering
    \centering
    \begin{subfigure}[t]{\textwidth}
        \centering
        \includegraphics[width=\textwidth]{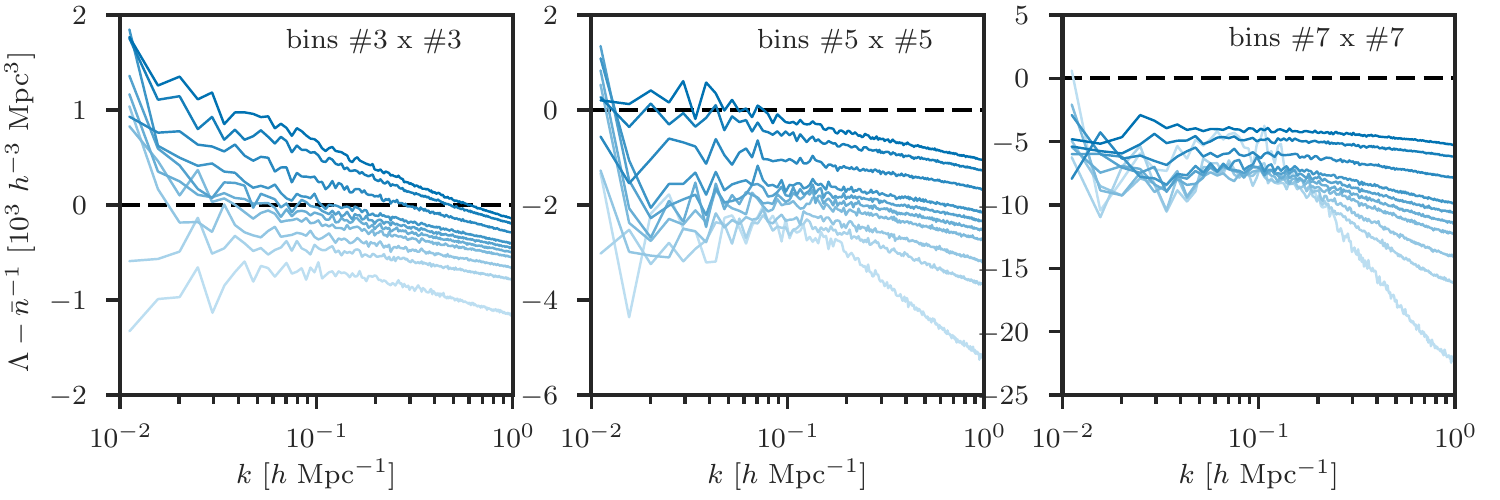}
        \caption{auto halo stochasticity}
    \end{subfigure}
    ~
    \begin{subfigure}[b]{\textwidth}
        \centering
        \includegraphics[width=\textwidth]{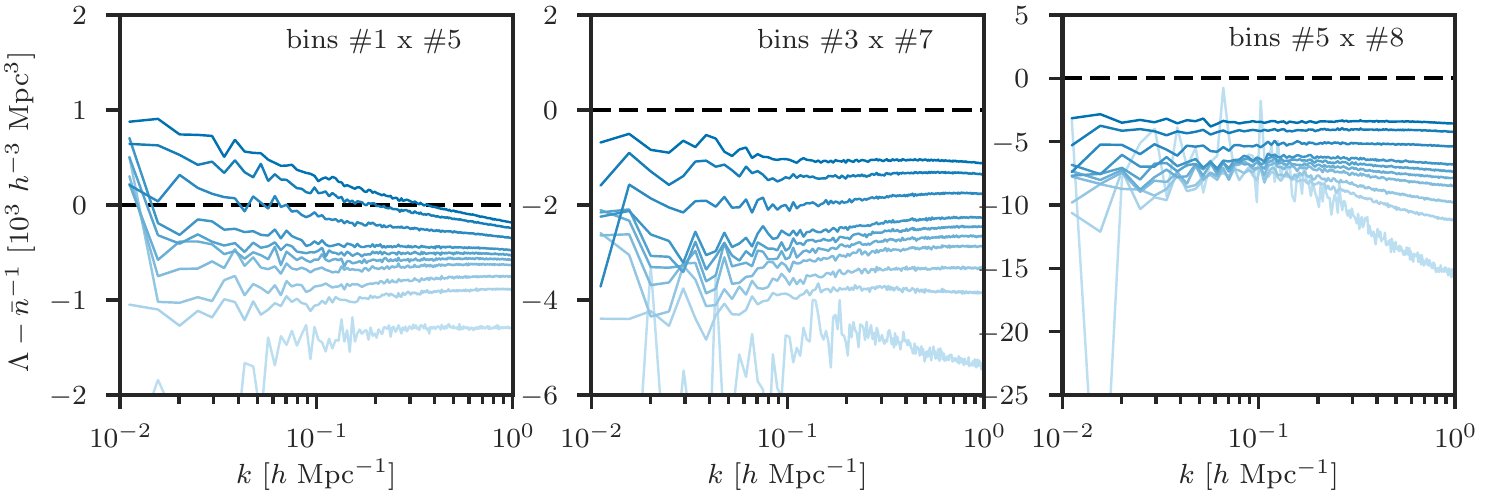}
        \caption{cross halo stochasticity}
    \end{subfigure}
    \caption{The deviation of the halo stochasticity $\Lambda(k)$, as
    defined in equation \ref{eq:lambda}, from the Poisson shot noise,
    for the case of (a) the same halo mass bin and (b) different halo mass bins.
    Results are measured from the RunPB simulations for three separate
    combinations of bins. The average halo mass increases from left to right;
    see Table~\ref{tab:mass-bins} for halo mass bin details. For each subplot,
    we show the results for 10 redshifts, ranging from $z=1$ (dark) to $z=0$ (light).
    Even when the mean halo mass is similar, the scale dependence and
    amplitude in the cases of auto and cross halo stochasticity can differ significantly.}
    \label{fig:stochasticity}
\end{figure}

We use the RunPB simulations at several redshift outputs and the
halo mass bins defined in table~\ref{tab:mass-bins} to investigate the
functional form of the halo stochasticity $\Lambda(k)$
as a function of mass and redshift. In figure~\ref{fig:stochasticity},
we show the deviations of the halo stochasticity from the Poisson shot
noise when considering the same halo mass bin and different halo
mass bins. The trends are consistent with our theoretical understanding: as
the average halo mass increases, the stochasticity becomes sub-Poissonian,
sourced by halo exclusion effects, while positive contributions from
nonlinear biasing become important for lower halo masses. And as halos
grow in time, the exclusion effects become more pronounced at lower redshifts
\cite{Baldauf:2013, Vlah:2013, Baldauf:2016}. However, the scale dependence and
redshift scaling remains non-trivial, and there are significant differences in
the scale dependence and amplitude when considering the cases of auto and cross
mass bins.

The halo stochasticity was studied in simulations in the context of the DF
model in \cite{Vlah:2013}, and the results presented here agree with those
findings. \cite{Vlah:2013} employs a simple model with log scale dependence
to model the auto stochasticity for several mass bins across three
redshifts. We extend those results with finer resolution in both
redshift and halo mass. In an attempt to capture as much complexity as possible,
we treat the halo stochasticity results from the RunPB simulations as a
training set and use Gaussian Process regression to predict the auto
stochasticity $\Lambda(b_1, \sigma_8(z))$ and cross stochasticity
$\Lambda(b_1, \bar{b}_1, \sigma_8(z))$, where we have
parameterized the redshift dependence of the stochasticity using
the value of $\sigma_8$ at each redshift.

\subsubsection{HZPT modeling for the halo-matter cross-correlation}
\label{sec:hzpt-halo-matter}

\begin{figure}[bt]
\centering
\includegraphics[width=\textwidth]{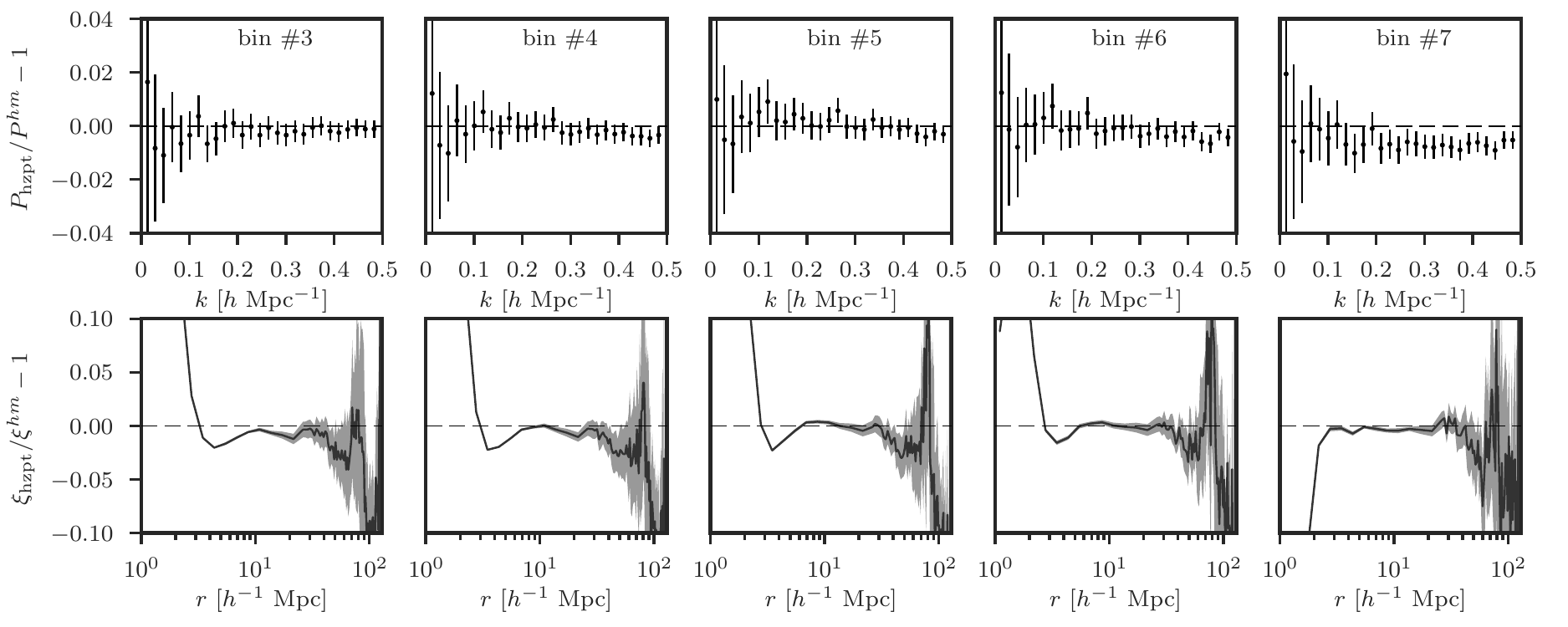}
\caption{The accuracy of the HZPT model used in this work for the halo-matter
cross-correlation, in comparison to the results from the
RunPB simulation. We compare the cross power
spectrum $P_{hm}$ (top) and the correlation function $\xi^{hm}$ (bottom) for
5 halo mass bins at $z = 0.55$ (see table~\ref{tab:mass-bins} for bin details).
We show the measurement uncertainties as error bars for $P^{hm}$ and
as the grey shaded region for $\xi^{hm}$.
The HZPT parameters have been fit using only $P^{hm}(k)$ from
$0.005 \ihMpc < k < 0.5 \ihMpc$. The model is a good description
of $P^{hm}$ on these scales, as well as $\xi^{hm}$ down to
$r \sim 5 \ \Mpc/h$, but fails once entering the 1-halo
regime on small scales.}
\label{fig:hzpt-halo-matter}
\end{figure}

The real-space halo-matter cross correlation $P^{hm}(k)$ plays a
crucial role in accurately modeling the halo auto spectrum using
equation \ref{eq:lambda}. We develop a model for the halo-matter
power spectrum using HZPT and calibrate the model using the suite of
halo mass bins from the RunPB simulations detailed in table~\ref{tab:mass-bins}.
To model the Zel'dovich term of the model, we employ a simple,
linear bias model, such that the full HZPT model is given by

\begin{equation}
P^{hm}(k) = b_1 P_{00}^{\mathrm{zel}}(k)
			+ P_{00}^{BB}(k, A_0, R, R_1, R_{1h}, R_{2h}),
\end{equation}
where $b_1$ is the large-scale, linear bias of the halo field,
$P_{00}^{\mathrm{zel}}$ is the matter density auto spectrum in the
Zel'dovich approximation, and $P_{00}^{BB}$ is a broadband
Pad\'e term, as given by equation \ref{eq:pade_power}.
To fully account for the biased nature of the halo field, the
HZPT model parameters, $\{A_0, R, R_1, R_{1h}, R_{2h}\}$,
now become a function of not only $\sigma_8(z)$ but also the linear
bias $b_1$. We choose a simple power-law functional form for the $b_1$
dependence, which performs well at modeling the bias dependence of $P^{hm}$
over the range of scales of interest in this work. We perform a global
fit across the 8 halo mass bins and 10 redshifts of the RunPB simulations
to determine the best-fit HZPT model parameters. In our parameter fit,
we have included the cross power spectrum $P^{hm}$ on scales
ranging from $k=0.005 \ihMpc$ to $k = 0.5 \ihMpc$. The best-fit
parameters are presented in appendix \ref{app:hzpt-Phm}.

We show the accuracy of the halo-matter HZPT
model in figure~\ref{fig:hzpt-halo-matter} for several halo mass
bins at $z = 0.55$. The trends evident at this redshift are consistent
with the results from the full range of redshifts explored ($z=0-1$).
The model reproduces the cross power spectrum $P^{hm}$ at the
$\sim$2\% level, as well as the cross-correlation $\xi^{hm}$ on
scales $r \gtrsim 5$ Mpc/$h$. However, we see from the
correlation function results on small scales that the model is unable to reproduce
the clustering on scales within the 1-halo regime, where halo profile
details become important. The model
breakdown on these scales is due to the choice to use a power-law
dependence on $b_1$ for the HZPT parameters that are related to
the halo profile. To better describe halo profiles and capture the
effects of nonlinear and nonlocal bias terms, i.e., \cite{Saito:2014},
a more complicated functional form for the linear bias dependence
is required. However, because Fourier-space statistics are the main concern
of this work and the simplified model performs well when modeling the power
spectrum on the scales of interest, we leave the investigation of
improved small-scale modeling to future work.

\subsection{Modeling observational effects}\label{sec:obs-model}

In this section we discuss several details that arise when
modeling data from real galaxy surveys. In section \ref{sec:modeling-AP},
we describe how we account for the geometric distortions that occur
when an inaccurate fiducial cosmology is assumed due to the AP effect.
Section \ref{sec:window-function} discusses how we treat the survey geometry
and window function when modeling realistic ``cutsky'' mocks, which have a
realistic survey geometry imposed.

\subsubsection{The Alcock-Paczynski effect}\label{sec:modeling-AP}

When analyzing data from galaxy surveys, we must transform
observed angular positions and redshifts into physical coordinates,
using a fiducial cosmological model to specify the relation
between the redshift and the LOS distance
(i.e., the Hubble parameter) and between the angular separation
and the distance perpendicular to the LOS
(i.e., the angular diameter distance). If the fiducial
cosmology differs from the true cosmology, an anisotropic,
geometric warping of the clustering signal is introduced.
This distortion, known as the Alcock- Paczynski (AP) effect,
\cite{Alcock:1979} is distinct from RSD and can be used to
measure cosmological parameters. The presence
of the BAO feature at a fixed scale in the power spectrum
helps distinguish the geometric AP effect and the
dynamical RSD anisotropy, thus increasing the
constraining power of full-shape modeling
\cite{Shoji:2009,Ballinger:1996}.

The difference between the assumed and true cosmological models
results in a rescaling of the wavenumbers transverse $k_\perp$ and parallel
$k_\parallel$ to the LOS direction, such that

\begin{equation}
k'_\perp = q_\perp k_\perp \ \mathrm{and}
		\ k'_\parallel = q_\parallel k_\parallel,
\end{equation}
where the primes denote quantities that are observed assuming the
fiducial (and possibly incorrect) cosmology. The two distortion
parameters $q_\perp$ and $q_\parallel$ are given by

\begin{equation}
q_\perp = \frac{D_A(z_\mathrm{eff})}{D'_A(z_\mathrm{eff})}
	\ \mathrm{and} \
	q_\parallel = \frac{H'(z_\mathrm{eff})}{H(z_\mathrm{eff})},
\end{equation}
which are the ratios of the Hubble parameter and angular diameter
distance in the fiducial and true cosmologies at the effective
redshift of the survey. With these definitions, the theoretical
prediction for the multipole power spectrum when including
the AP effect can be expressed as

\begin{equation}
P_\ell(k') = \frac{2\ell+1}{2q_\perp q_\parallel^2}
		\int_{-1}^{1} d\mu \ P^S_{gg} \left [ k(k',\mu'), \mu(\mu') \right]
        \mathcal{L}_\ell(\mu),
\end{equation}
where $\mathcal{L}_\ell$ is the Legendre polynomial of order $\ell$,
and we use the model prediction of equation \ref{eq:Pgg-model} for
$P^S_{gg}[k'(k, \mu), \mu(\mu')]$. The true $(k, \mu)$
can be related to the observed $(k', \mu')$ via

\begin{align}
k(k', \mu') &= \frac{k'}{q_\perp}
		\left [ 1 + (\mu')^2 \left (\frac{1}{F^2}  - 1 \right)
        \right]^{1/2}, \\
\mu(\mu') &= \frac{\mu'}{F}
		\left [1 + (\mu')^2 \left (\frac{1}{F^2}  - 1 \right),
        \right]^{-1/2}
\end{align}
where $F = q_\parallel / q_\perp$. The normalization scaling
of the power spectrum with $q_\perp^{-1} q_\parallel^{-2}$ is
due to the volume distortion between the two different cosmologies.

For comparison with BAO distance analyses, a second set of AP parameters is
usually defined, given by

\begin{align}
\aperp &\equiv
		\frac{D_A(\zeff)}{D'_A(\zeff)}
        \frac{r_d'}{r_d} = q_\perp  \frac{r_d'}{r_d} \\
\apar &\equiv
    		\frac{H'(\zeff)}{H(\zeff)}
             \frac{r_d'}{r_d} = q_\parallel  \frac{r_d'}{r_d},
\end{align}
where we have defined $r_d \equiv r_s(z_d)$ as the sound horizon
scale at the drag redshift $z_d$. BAO measurements are
sensitive to the Hubble parameter and angular diameter distance
relative to the sound horizon scale of a fixed ``template'' cosmology,
and this second set of parameter definitions facilitates comparison
of measurements using different template cosmological models.

\subsubsection{The survey geometry}\label{sec:window-function}

\begin{figure}[tb]
\centering
\includegraphics[]{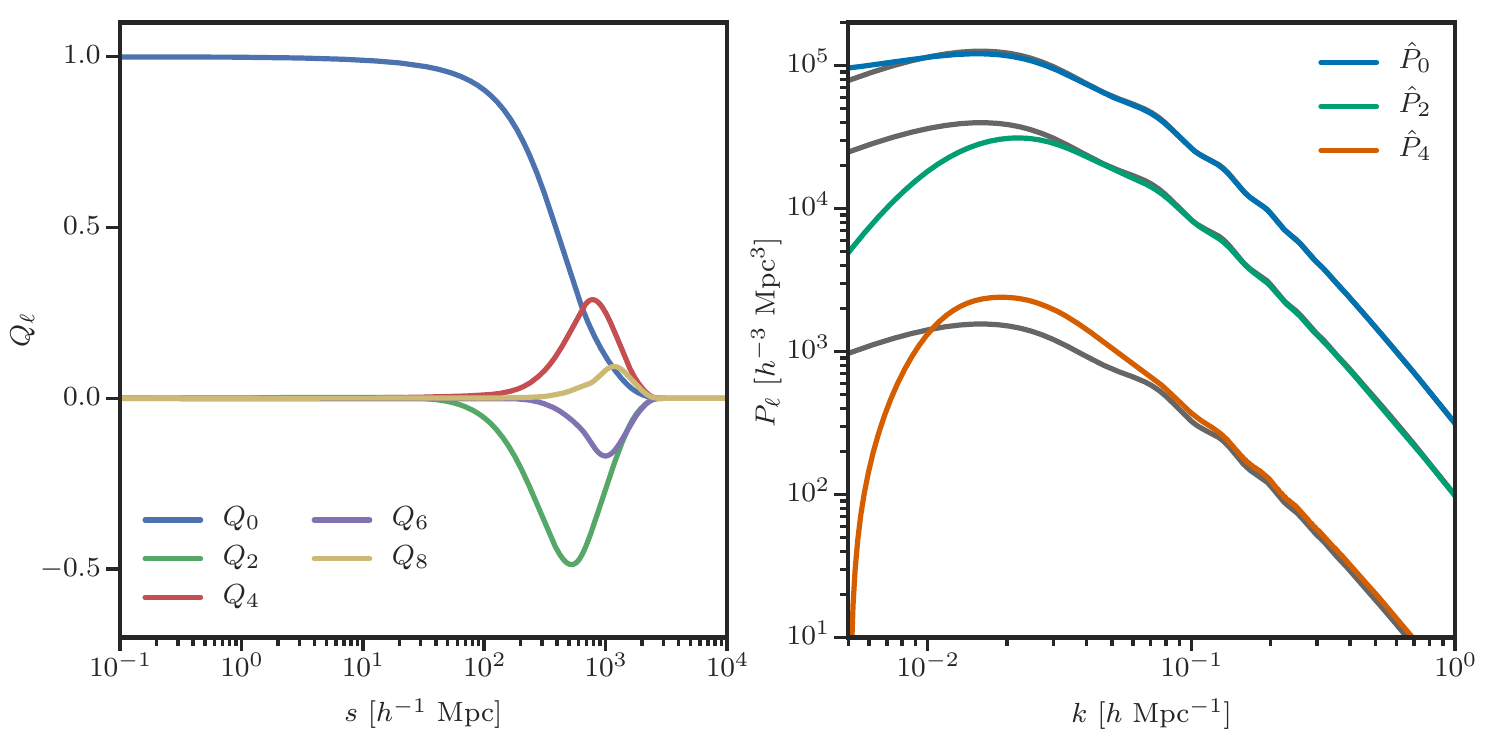}
\caption{The window function multipoles in configuration space (left)
and the effects of the window function on linear Kaiser power
spectrum multipoles (right) for the DR12 CMASS NGC survey geometry.
In the right panel, the solid grey lines show the unconvoled multipoles, while
the colored lines correspond to the model after convolution
with the window function, $\hat{P}_\ell(k)$. The convolution procedure has
large effects at small $k$, and we choose to use
$k_\mathrm{min} = 0.02 \ihMpc$ in our data analysis to minimize these effects.}
\label{fig:window-effects}
\end{figure}

When analyzing cutsky mock catalogs, we must account for
the effects of the survey geometry when comparing our theoretical
model to the measured power spectrum. We do this by convolving
our theoretical model with the survey window function, rather than
trying to remove the effect of the survey geometry from the data itself.
Our window function treatment follows the method first presented in
\cite{Wilson:2017} and used in the analysis of BOSS DR12 data
in \cite{Beutler:2017,Zhao:2017}.

Following \cite{Wilson:2017}, we compute the window function multipoles in
configuration space using a pair counting algorithm and the catalog
of random objects describing the survey geometry. We use the \texttt{Corrfunc}
correlation function code \cite{Sinha:2016} to compute the pair counts of
the random catalog via

\begin{equation}
Q_\ell(s) \propto \int_{-1}^{1} d\mu RR(s, \mu) \mathcal{L}_\ell(\mu)
		\simeq \sum_i RR(s_i, \mu_i) \mathcal{L}_\ell(\mu_i),
\end{equation}
where the normalization is such that $Q_0(s) \rightarrow 1$ for $s \ll 1$.
The resulting multipoles $Q_\ell$ for the BOSS CMASS NGC sample are
shown in the left panel of figure~\ref{fig:window-effects}. The $Q_\ell$ vanish
for scales $\gtrsim 3000 \ \hMpc$, as these are the largest scales in
the volume of the NGC. Note that on small scales, the clustering becomes
isotropic, with the multipoles vanishing. In general, the contribution
of the higher-order multipoles decreases as $\ell$ increases, which guarantees
the convolution converges when including only the first few $Q_\ell$. Here,
we include results up to and including $Q_8$, and have verified that the
inclusion of $Q_{10}$ does not affect our results.

With the measured $Q_\ell$, we compute the convolved theoretical
correlation function multipoles in configuration space as

\begin{align}
\hat{\xi}_0(s) = \xi_0 Q_0 &+ \frac{1}{5} \xi_2 Q_2 + \frac{1}{9} \xi_4 Q_4 + ...\nn \\
\hat{\xi}_2(s) = \xi_0 Q_2 &+ \xi_2 \left[Q_0 + \frac{2}{7}Q_2 + \frac{2}{7}Q_4 \right] \nn \\
		&+ \xi_4 \left[\frac{2}{7}Q_2 \frac{100}{693}Q_4 + \frac{25}{143}Q_6\right] + ...\nn\\
\hat{\xi}_4(s) = \xi_0 Q_4 &+ \xi_2 \left[ \frac{18}{35}Q_2 + \frac{20}{77}Q_4
			+ \frac{45}{143}Q_6\right] \nn\\
            &+ \xi_4 \left[Q_0 + \frac{20}{77}Q_2 + \frac{162}{1001}Q_4 +
            \frac{20}{143}Q_6 + \frac{490}{2431}Q_8 \right] + ....,
\end{align}\label{eq:window-xi-poles}
where $\xi_\ell$ are the theoretical correlation function mutipoles, computed
from the power spectrum multipoles via a 1D Hankel transform, evaluated using
the \texttt{FFTLog} software \cite{Hamilton:2000}. We also perform the transformation
from $\hat{\xi}_\ell(s)$ to $\hat{P}_\ell(k)$ using \texttt{FFTLog}.

The effects of the window function convolution can be seen in the right panel
of figure~\ref{fig:window-effects}, where we illustrate the effects using linear
Kaiser multipoles. The effects are most important on scales
of order the survey size; for the NGC CMASS sample, the window function effects
are only important on scales $k \lesssim 0.05 \ihMpc$. The impact of
the survey geometry increases for the higher-order multipoles, with the anisotropy
of the window function leading to non-trivial effects on our convolved model.
In this work, we use a minimum wavenumber of $k_\mathrm{min} = 0.02 \ihMpc$
when comparing data and theory and have tested that the window function convolution
has minimal impact on our parameter fitting analyses. However, as measurement
errors decrease for future surveys, the window convolution will need to be carefully
tested, given both the constraining power of the $\ell=2$ and $4$ multipoles and
the larger convolution effects.

\subsection{Model parametrization}\label{sec:model-params}

\begin{table}[tb]
\centering
  \begin{tabular}{c|c}
  \toprule
  \multicolumn{2}{c}{Free Parameters} \\
  Name [Unit] & Prior \\
  \midrule
  $\aperp$ & $\mathcal{U}(0.8, 1.2)$ \\
  $\apar$ & $\mathcal{U}(0.8, 1.2)$ \\
  $f$ & $\mathcal{U}(0.6, 1.0)$ \\
  $\sigma_8(\zeff)$ & $\mathcal{U}(0.3, 0.9)$ \\
  $b_{1,c_A}$ & $\mathcal{U}(1.2, 2.5)$ \\
  $f_s$ &  $\mathcal{U}(0, 0.25)$ \\
  $f_{s_B}$ & $\mathcal{U}(0, 1)$ \\
  $\Nsatmult$ & $\mathcal{N}(2.4, 0.1)$ \\
  $\sigma_c$ [$\hMpc$] &  $\mathcal{U}(0, 3)$ \\
  $\sigma_{s_A}$ [$\hMpc$] & $\mathcal{U}(2, 6)$ \\
  $\gamma_{s_A}$ & $\mathcal{N}(1.45, 0.3)$ \\
  $\gamma_{s_B}$ &  $\mathcal{N}(2.05, 0.3)$ \\
  $f^{1h}_{s_B s_B}$ & $\mathcal{N}(4, 1)$ \\
  \bottomrule
  \end{tabular}
  \quad
  \scalebox{0.95}{
  \begin{tabular}{c|c}
  \toprule
  \multicolumn{2}{c}{Constrained Parameters} \\
  Name [Unit] & Constraint Expression \\
  \midrule
  $b_{1,c_B}$ & equation~\ref{eq:b1cB-constraint} \\
  $b_{1,s_A}$ & $\gamma_{s_A} b_{1,c_A}$  \\
  $b_{1,s_B}$ & $\gamma_{s_B} b_{1,c_A}$ \\
  $f_{c_B}$ & equation~\ref{eq:fcB-constraint} \\
  $\sigma_{s_B}$ [$h^{-1}$ Mpc] &
  			$\sigma_{s_A} \left[\sigma_v^\mathrm{model}(b_{1,s_B})
        	/ \sigma_v^\mathrm{model}(b_{1,s_A})\right]$  \\
  $N_{c_Bs}$ [$\powerunit$] & equation~\ref{eq:NcBs-constraint} \\
  $N_{s_Bs_B}$ [$\powerunit$] & equation~\ref{eq:NsBsB-constraint} \\
  \bottomrule
  \end{tabular}
  }
  \caption{The parameter space of our full-shape RSD fits using
  the model described in this work. There are 13 free parameters (left)
  that are varied during the fitting process, with several additional
  parameters subject to constraint expressions (right).
  For all free parameters, we provide the prior used when
  fitting, either a normal prior $\mathcal{N}(\mu, \sigma)$ with
  mean $\mu$ and standard deviation $\sigma$, or a uniform prior
  $\mathcal{U}(a, b)$ with lower bound $a$ and upper bound $b$.
  For a detailed description of the model parameters, see
  section~\ref{sec:model-params}.}
  \label{tab:model-params}
\end{table}

Table~\ref{tab:model-params} gives a summary of the parameters
of the model described in this work. We give both the free parameters
as well as the constrained parameters and the corresponding
constraint expressions. There are 13 free parameters detailed in
table~\ref{tab:model-params}, and these parameters correspond to the
parameter space used in our RSD analyses. The table
also lists the assumed prior distribution for each parameter
used during parameter estimation, which is either a flat (uniform) or
Gaussian prior. We use physically motivated priors when possible and assume
wide, flat priors on all cosmological parameters of interest.
We describe the model parametrization in detail below.

\subsubsection{Cosmology parameters}

The free parameters specifying the cosmology in our model
are the AP distortion parameters ($\apar$ and $\aperp$),
the growth rate $f$, and the amplitude of matter
fluctuations $\sigma_8$, where both $f$ and $\sigma_8$ are
evaluated at the effective redshift of the sample, $\zeff$. During our
fitting procedure, we vary $f$ and $\sigma_8$ independently, although
we only report results for the product $f \sigma_8$, which is the
parameter combination most well-constrained by RSD analyses.
The model requires a linear power spectrum in order to evaluate
several perturbation theory integrals. These integrals are computationally
costly (although see recent advances,
\cite{Schmittfull:2016,Schmittfull:2016b, McEwen:2016}), and for this reason, we do not
vary any cosmological parameters affecting the shape of the linear power spectrum
during parameter estimation. We evaluate the linear power spectrum
using the fiducial cosmology and keep the shape fixed, allowing only the
amplitude to vary through changes in $\sigma_8$.

\subsubsection{Linear bias parameters}

In the most general version of the model discussed in
section~\ref{sec:galaxy-model}, we must specify linear bias parameters
for each of the four galaxy subsamples: $b_{1,c_A}$, $b_{1,c_B}$, $b_{1,s_A}$,
and $b_{1,s_B}$. As discussed in section~\ref{sec:halo-bias-model}, the linear
bias fully predicts the higher-order biasing parameters for a given
sample. When varying the linear bias parameters of the $s_A$ and $s_B$
satellite samples, we enforce the expected ordering of the parameters:
$b_{1,c_A} < b_{1,s_A} < b_{1,s_B}$. We use the relations
$b_{1,s_A} = \gamma_{s_A} b_{1,c_A}$ and $b_{1,s_B} = \gamma_{s_B} b_{1,c_A}$
and choose to vary the parameters $\gamma_{s_A}$ and $\gamma_{s_B}$ instead.
We use relatively wide Gaussian priors for these parameters centered on
their expected fiducial values for a CMASS-like galaxy sample,
$\gamma_{s_A} \sim 1.45$ and $\gamma_{s_B} \sim 2.05$. For the linear
bias of the $c_B$ sample, we use the expected scaling of the bias
with the biases of the satellite samples as given by
equation~\ref{eq:b1cB-constraint} and described in
appendix~\ref{app:b1cB-constraint}.

\subsubsection{Sample fractions, velocity dispersions, and 1-halo amplitudes}

There are three parameters specifying the
fraction of all galaxies that are satellites $f_s$, the fraction
of centrals that live in halos with a satellite $f_{c_B}$, and the
fraction of satellites that live in halos with multiple satellites $f_{s_B}$.
We must also specify the 1-halo amplitudes (assumed to be independent
of $k$) that enter into equations \ref{eq:P_cBsA} - \ref{eq:P_sBsB}. We
denote the 1-halo amplitude due to correlations between centrals
and satellites in the same halo as $N_{c_Bs}$, and between satellites
inside the same halo as $N_{s_B s_B}$. And, finally, we must specify
the velocity dispersion parameters for each galaxy subsample in order
to account for the FoG effect. We include a single
velocity dispersion for centrals, $\sigma_c$, and parameters for each
of the satellite subsamples, $\sigma_{s_A}$ and $\sigma_{s_B}$. Thus,
in the most general case, there are additional 12 model parameters needed
to fully evaluate our model, in addition to the 4 cosmological parameters.

There are dependencies between the parameters previously discussed
that allow us to parametrize the model in terms of alternative parameters
that have well-behaved, physically motivated priors.
In particular, we use the constraints outlined in
appendix~\ref{app:model-constraints} for the relative
fraction of the $c_B$ sample, $f_{c_B}$, in equation~\ref{eq:fcB-constraint},
and for the 1-halo amplitudes, $N_{c_Bs}$ and $N_{c_Bc_B}$,
in equations~\ref{eq:NcBs-constraint} and \ref{eq:NsBsB-constraint}.
In the former case, the constraint allows us to vary the parameter
$\Nsatmult$, which is defined as the mean number of satellite galaxies
in halos with more than one satellite. This parameter is typically
centered on $\Nsatmult \sim 2.4$ for CMASS-like galaxy samples, with little
variation around this center value. For the 1-halo amplitude $N_{s_B s_B}$
we vary a normalization parameter $f^{1h}_{s_Bs_B}$ to account for uncertainty
in the expected value, which should have a value of order unity.

Finally, we do not vary the velocity dispersion of the $s_B$ sample,
$\sigma_{s_B}$, but rather use the physically motivated scaling
with halo mass, $\sigma_v^2 \propto M^{2/3}$, and the halo
bias -- mass relation from \cite{Tinker:2010}. We do not use this model
function $\sigma_v^\mathrm{model}(b_1)$ to predict the absolute amplitude of
$\sigma_{s_B}$, but only the functional form. We always rescale the predicted
value by the current value of $\sigma_{s_A}$
(see table~\ref{tab:model-params} for details).

\section{Performance of the model}\label{sec:performance}

\subsection{RunPB results}\label{sec:runPB-results}

\begin{figure}[!ptb]
\centering
\includegraphics[width=0.85\textwidth]{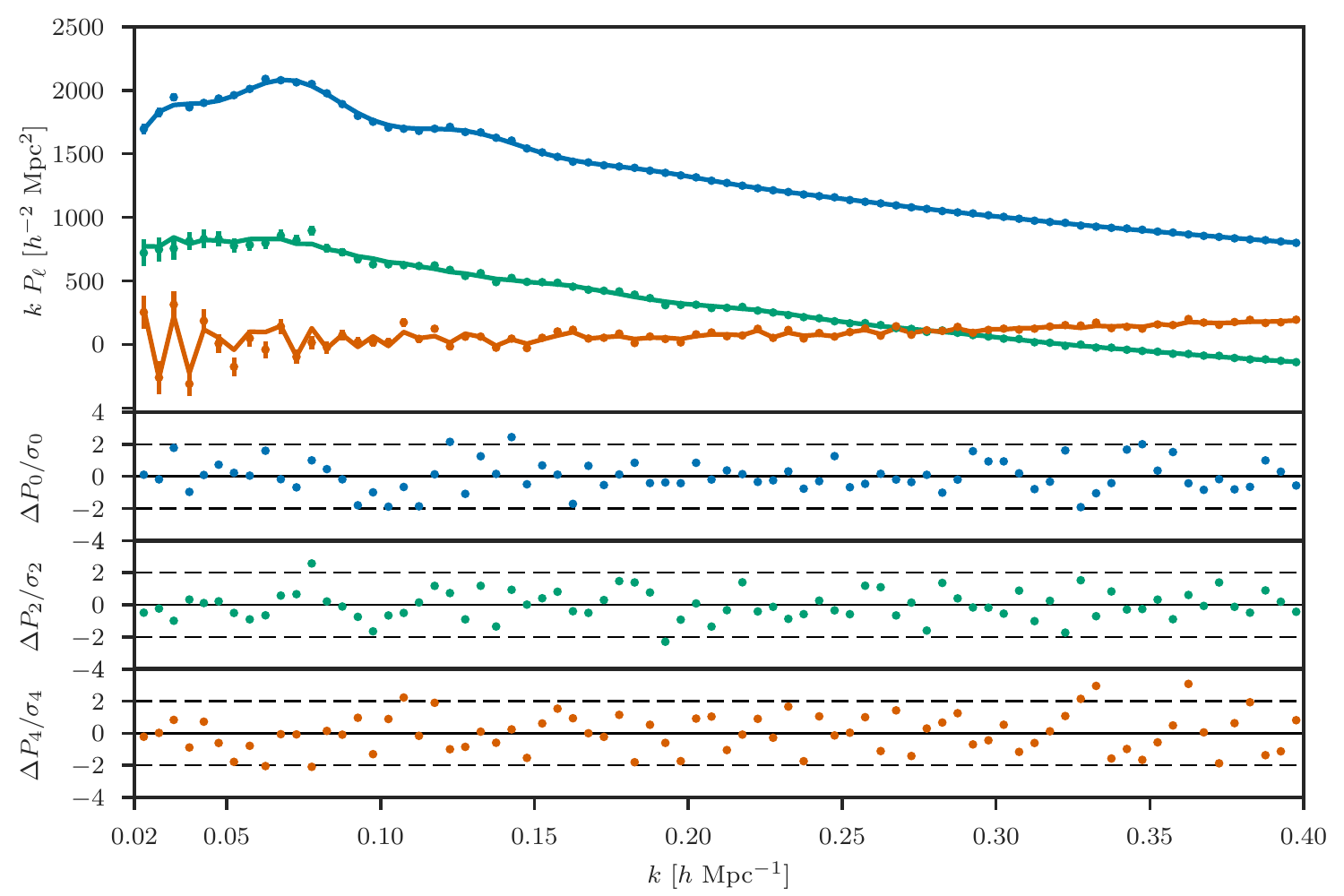}
\caption{The best-fit monopole, quadrupole, and hexadecapole models
(lines) as compared to the measurements (points) from the mean
of 10 RunPB HOD galaxy realizations at $z=0.55$, fit over the
wavenumber range $k = 0.02 - 0.4 \ihMpc$. The lower panels show
the model residuals for each multipole separately. The reduced
chi-squared of the fit to all three multipoles is
$\chi^2_\mathrm{red} = 1.12$. Note that the large variation
from bin to bin in the hexadecapole is due to discrete binning effects.}
\label{fig:runPB-kmax04}
\end{figure}

As a first test of the RSD model described in
section~\ref{sec:power-spectrum-model}, we use a set of HOD galaxy catalogs
constructed from 10 realizations of the $z=0.55$ snapshot of the
RunPB simulation, described previously in section~\ref{sec:RunPB}.
The galaxy catalogs are made by populating halo catalogs according to
a halo occupation distribution with parameters comparable to the BOSS CMASS
sample. The halo catalogs are constructed in the manner described in detail
in \cite{Reid:2014}. Briefly, the halo finder uses the spherical overdensity
implementation of \cite{Tinker:2008}, using an overdensity of
$\Delta_m = 200$ relative to the mean matter density $\rho_m$ to define the
halo virial radius. Central galaxies are not at rest with respect
to the halo center-of-mass; they are assigned a velocity computed
from the halo particles in the densest region of each halo (see \cite{Reid:2014}
for details). Note that the halo catalogs used here are not the same as the
FOF halo catalogs described in section~\ref{sec:RunPB}, which we use
to calibrate certain components of the RSD model.
Differences in the halo finder algorithms
lead to important differences in the clustering of the resulting galaxy catalogs.
While we do not expect RSD fits to these galaxy catalogs to be a
fully-independent validation of the model, they do still provide a useful
test of the accuracy of our model.

Using the model parameterization discussed in section~\ref{sec:model-params},
we fit the mean of the measured monopole, quadrupole, and hexadecapole from
10 realizations at $z = 0.55$ as a function of the maximum wavenumber
included in the fits, $\kmax = [0.2,0.3,0.4] \ihMpc$. The resulting
best-fit model and residuals between measurements and theory are shown
in figure~\ref{fig:runPB-kmax04} for $\kmax = 0.4 \ihMpc$.
We are able to achieve excellent agreement
between the model and simulation multipoles to scales of $k = 0.4 \ihMpc$, well
into the nonlinear clustering regime.

\begin{figure}[!tbp]
\centering
\includegraphics[width=\textwidth]{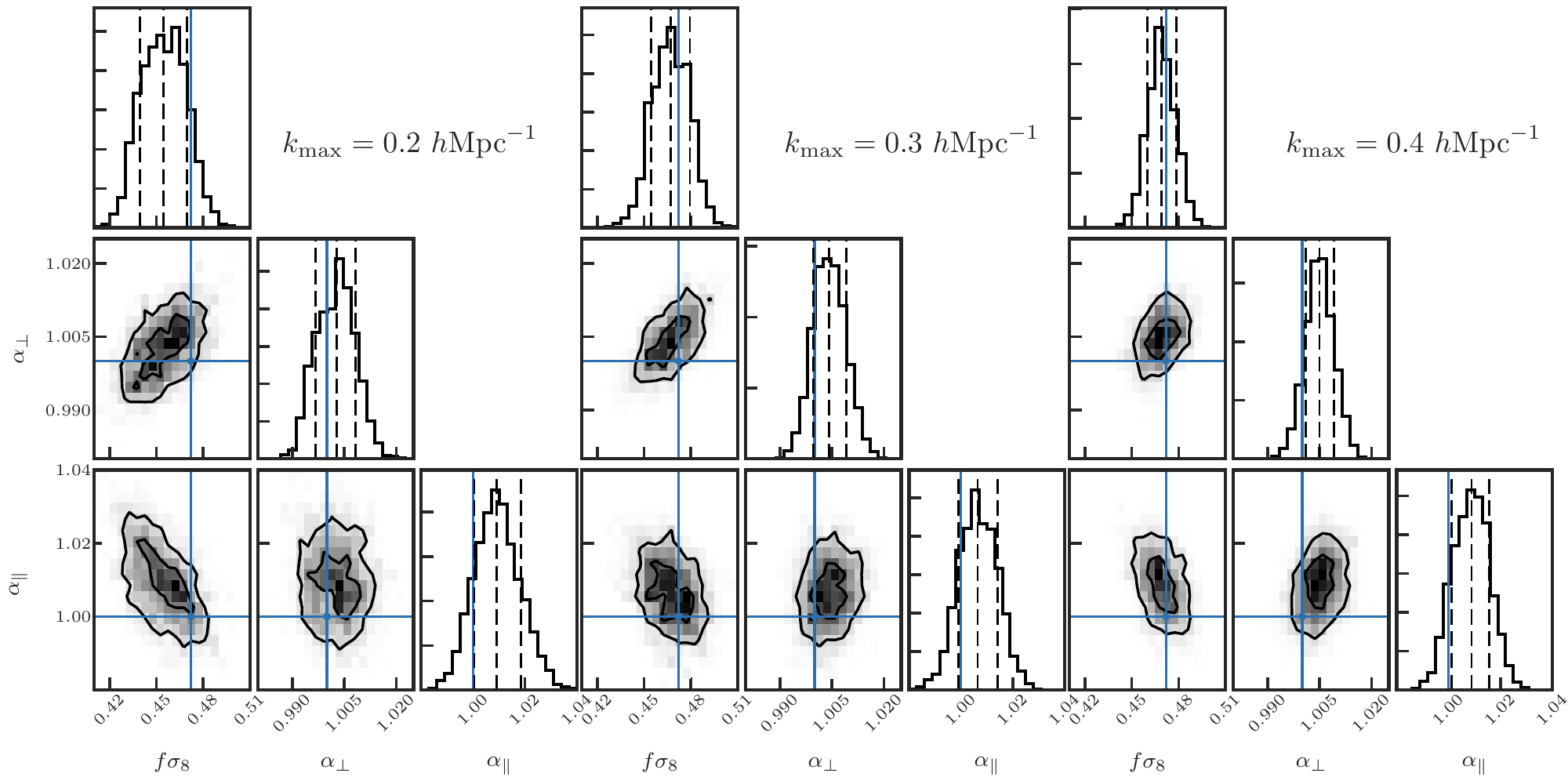}
\caption{The 2D posterior distributions for the cosmology parameters
$f\sigma_8$, $q_\parallel$, and $q_\perp$ obtained from fitting
the mean of 10 RunPB HOD galaxy realizations at $z=0.55$.
We show results when varying the maximum scale included in the fits:
$k_\mathrm{max} = 0.2$ (left), 0.3 (center), 0.4 (right)$\ihMpc$.
The expected parameter values are marked with solid blue lines.}
\label{fig:runPB-2D-cosmo}
\end{figure}

As a function of $\kmax$, we report the mean and $1\sigma$
error for a subset of the model parameters in table~\ref{tab:runPB-kmax-results},
as determined from the posterior distributions obtained via MCMC sampling.
We also show the 2D posterior distributions for $\fsig$, $\apar$,
and $\aperp$ for each $\kmax$ value in figure~\ref{fig:runPB-2D-cosmo}.
As expected, we obtain significant decreases in parameter uncertainties
when including small-scale information in the fits. For the three
cosmology parameters, $\fsig$, $\apar$, and $\aperp$, we find
decreases of 19\%, 18\%, and 18\%, respectively, for $\kmax = 0.3 \ihMpc$
and 38\%, 24\%, and 29\% for $\kmax = 0.4 \ihMpc$, relative to the fit
using $\kmax = 0.2 \ihMpc$. These decreases are roughly consistent
with the expected scaling in the nonlinear regime,
$\sigma \propto \kmax^{-1/2}$ (e.g., \cite{Blazek:2014}).
For the AP parameters, we find more modest decreases in the uncertainty
when ranging from $\kmax = 0.2 \ihMpc$ to $\kmax = 0.4 \ihMpc$. In
particular, extending from $\kmax = 0.3 \ihMpc$ to $\kmax = 0.4 \ihMpc$
offers little improvement in the error on $\apar$. The constraining
power for the AP parameters results from a combination of the BAO signal
and information from the geometric distortion of the full broadband signal.
As nearly all of the information from the BAO signal is present below
$k = 0.2 \ihMpc$, our more modest decreases in uncertainty for the AP
parameters are consistent with our expectations.

We have central/satellite information for each galaxy in the
RunPB catalogs and can assess the accuracy of the halo model decomposition
described in section~\ref{sec:galaxy-model}. As seen in
table~\ref{tab:runPB-kmax-results}, we find
a non-zero velocity dispersion for centrals with an amplitude
$\sigma_c \sim 1 \hMpc$, which is consistent with the expected
amplitude present in the underlying halo catalogs.
The main discrepancy is that our recovered satellite fraction is
significantly higher than the
expected value. This results from the fact that we rely on FOF
halo catalogs for calibration of the halo clustering in the model, while
we are fitting galaxy catalogs created from SO catalogs. The choice
of halo finder alters the clustering on scales around the virial radius.
A FOF halo finder tends to over-merge halos on these scales into a single
halo, whereas a SO finder tends to preserve the multiple smaller halos.
This effect manifests itself as an increase in the model satellite fraction
and is consistent with our fitting results. Thus, we are able to absorb
issues related to these simulation differences into
the satellite fraction parameter.

\begin{table}[!ptb]
\renewcommand{\arraystretch}{1.25}
\centering
\sisetup{round-mode=places, round-precision=3}
\resizebox{\textwidth}{!}
{
\begin{tabular}{c|c|c|c|c}
\toprule
{} &             $k_\mathrm{max} = 0.2$ $h/\mathrm{Mpc}$ &             $k_\mathrm{max} = 0.3$ $h/\mathrm{Mpc}$ &              $k_\mathrm{max} = 0.4$ $h/\mathrm{Mpc}$ & truth \\
\midrule
$f \sigma_8$       &   \num{0.4546353} \numpmerr{0.01487461}{0.01537589} &   \num{0.4672024} \numpmerr{0.01220806}{0.01277618} &  \num{0.4689142} \numpmerr{0.009673357}{0.009075522} & \num{0.472} \\
$f\sigma_8$ [fixed AP] & \num{0.456554949284} \numpmerr{0.00950617885996}{0.00978258252144} & \num{0.465566307306} \numpmerr{0.00855195522308}{0.00834447145462} & \num{0.467769473791} \numpmerr{0.00689902901649}{0.00710368156433} & \num{0.472} \\
$\alpha_\perp$     &   \num{1.00288} \numpmerr{0.005369816}{0.006284306} &   \num{1.00421} \numpmerr{0.004924848}{0.004572193} &   \num{1.004951} \numpmerr{0.004241419}{0.004049029} & \num{1.0} \\
$\alpha_\parallel$ &  \num{1.008968} \numpmerr{0.009548766}{0.008575451} &  \num{1.006492} \numpmerr{0.007784377}{0.007363477} &   \num{1.008892} \numpmerr{0.006819668}{0.007680713} & \num{1.0} \\
$b_1 \sigma_8$     &  \num{1.265523} \numpmerr{0.008938074}{0.008676291} &  \num{1.264623} \numpmerr{0.007589221}{0.008247495} &     \num{1.2681} \numpmerr{0.007574999}{0.008005503} & \num{1.272} \\
$f_s$              &    \num{0.121926} \numpmerr{0.01884478}{0.01842913} &   \num{0.1427805} \numpmerr{0.01262522}{0.01316997} &   \num{0.1433926} \numpmerr{0.007552902}{0.00804628} & \num{0.104} \\
$f_{c_B}$          &   \num{0.1040296} \numpmerr{0.03338925}{0.02977829} &   \num{0.1235227} \numpmerr{0.02246464}{0.02344792} &     \num{0.1219984} \numpmerr{0.01335913}{0.0149048} & \num{0.089} \\
$f_{s_B}$          &      \num{0.437855} \numpmerr{0.2030092}{0.1973982} &     \num{0.4377613} \numpmerr{0.1364825}{0.1234289} &    \num{0.4662209} \numpmerr{0.08145778}{0.07920344} & \num{0.399} \\
$\sigma_c$         &      \num{1.133505} \numpmerr{0.2143937}{0.2383132} &    \num{0.9060378} \numpmerr{0.08815792}{0.1111759} &    \num{0.9297313} \numpmerr{0.06205001}{0.06473397} & -- \\
$\sigma_{s_A}$     &      \num{4.239433} \numpmerr{0.4755335}{0.4125561} &      \num{3.737315} \numpmerr{0.3720634}{0.4643847} &       \num{3.443024} \numpmerr{0.2778931}{0.2701519} & -- \\
\midrule
$\chi^2$/d.o.f.    &                             $113/(108 - 13) = 1.19$ &                             $159/(168 - 13) = 1.03$ &                              $241/(228 - 13) = 1.12$ & \\
\bottomrule
\end{tabular}
}
\caption{Parameter constraints obtained when fitting the 13-parameter RSD model to
$[P_0, P_2, P_4]$, as measured from the mean of the
10 RunPB galaxy catalogs at $z=0.55$. We show results determined as a function
of the maximum wavenumber included in the fits. Parameter posteriors are determined
from MCMC sampling of the likelihood, assuming Gaussian covariance between multipoles.}
\label{tab:runPB-kmax-results}
\end{table}

\begin{figure}[!tbp]
\centering
\includegraphics[width=\textwidth]{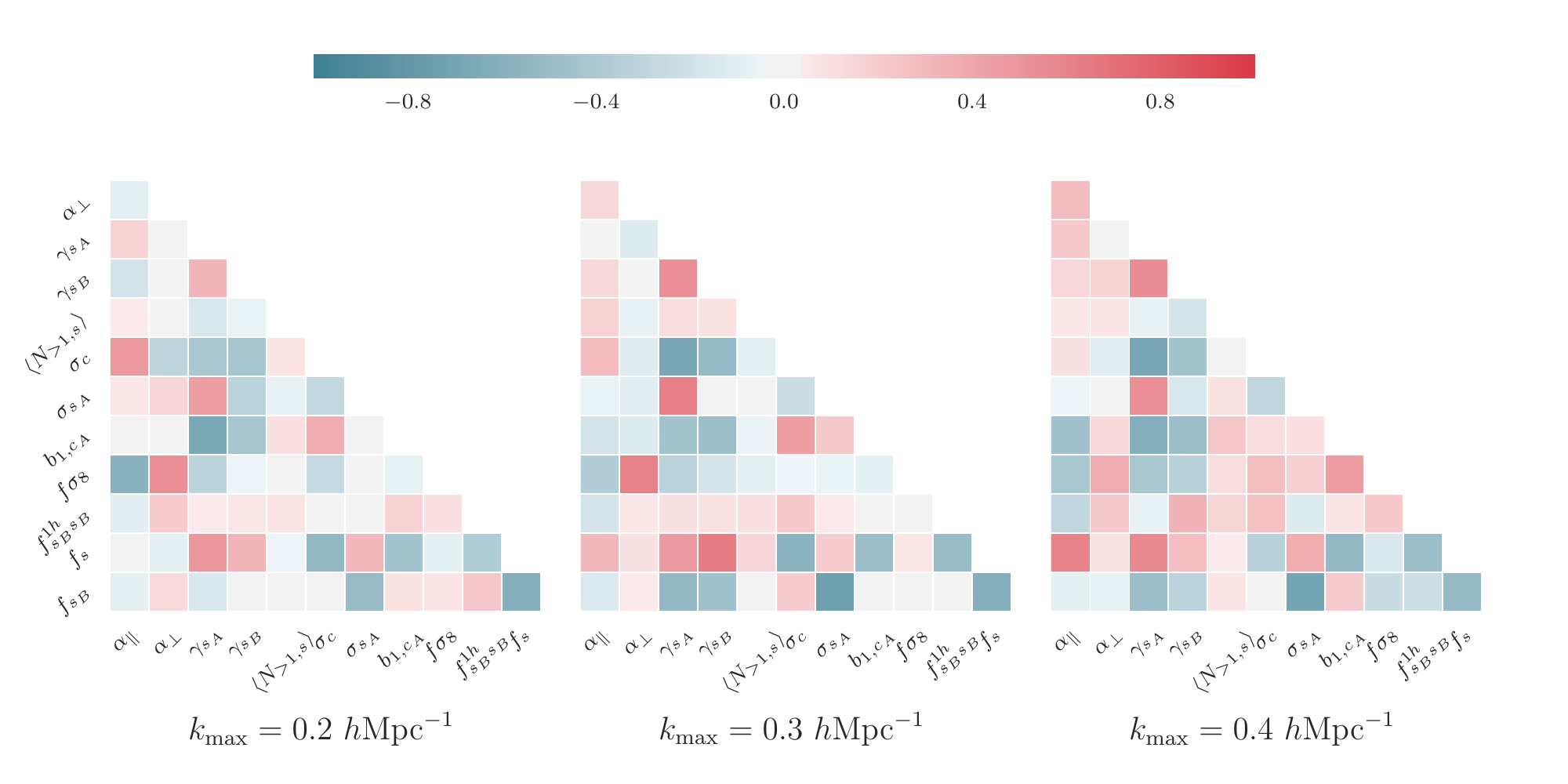}
\caption{Parameter correlations as measured from the posterior
distributions when fitting the mean of the 10 RunPB galaxy catalogs.
We show the correlations as a function of the maximum wavenumber
included in the fit, illustrating the changes in parameter dependencies
when fitting to smaller scales.}
\label{fig:runPB-correlation}
\end{figure}

We analyze the correlations between the posterior distributions
to better understand the constraining power of our 13-parameter model.
We show these parameter correlations for each
fitting range in figure~\ref{fig:runPB-correlation}. As expected, we find
that the main parameter combination measuring the strength of RSD $\fsig$
is most correlated with the AP parameters $\apar$ and $\aperp$, which
measure geometric distortions of the clustering signal. For each of the
$\kmax$ fitting ranges, the correlation matrix for ($\fsig, \aperp, \apar$)
is:

\begin{align}
R^{0.2}[\fsig, \aperp, \apar] & =
	\begin{bmatrix*}[r]
	1.000 & 0.536 & -0.583 \\
    0.536 & 1.000 & -0.094 \\
    -0.583  & -0.094 & 1.000
	\end{bmatrix*}, \\
R^{0.3}[\fsig, \aperp, \apar] & =
	\begin{bmatrix*}[r]
	1.000  & 0.605  & -0.361 \\
    0.605  &  1.000 & 0.133 \\
    -0.361  & 0.133 &  1.000 \\
	\end{bmatrix*}, \\
R^{0.4}[\fsig, \aperp, \apar] & =
	\begin{bmatrix*}[r]
	1.000  & 0.377  &  -0.418 \\
    0.377  & 1.000  & 0.292 \\
    -0.418  & 0.292   & 1.000 \\
	\end{bmatrix*}.
\end{align}

As we extend the maximum wavenumber included in our fits, small-scale
information does help break degeneracies between
$\fsig$ and the AP parameters, reducing the correlation between
$\fsig$ and ($\aperp$, $\apar$). For comparison, \cite{Beutler:2017}
reports a correlation between $\fsig$ and $\aperp$ of 0.503 and
$\fsig$ and $\apar$ of 0.547 for the middle redshift bin for the
combined DR12 BOSS sample, where they have fit [$P_0$, $P_2$] to
$\kmax = 0.15 \ihMpc$ and $P_4$ to $\kmax = 0.1 \ihMpc$. This level of
correlation is similar to our values obtained when fitting to
$\kmax = 0.2 \ihMpc$, however we find a significant reduction in
correlation fitting to $\kmax = 0.4 \ihMpc$. We can assess the
freedom of our RSD modeling using the Fisher formalism, which
predicts a correlation coefficient of unity between $\apar$ and
$\aperp$ in the case where we perfectly understand RSD
\cite{Seo:2003,Seo:2007,Shoji:2009}. In the opposite limit,
we expect $r\sim-0.4$ when only BAO information is used and
RSD information is fully marginalized over. Thus, the correlation
between $\apar$ and $\aperp$ provides a measure of the constraining
power of our RSD parametrization, with the correlation decreasing from
unity as additional freedom is introduced into the RSD model.
Our results are consistent with this expectation, as we find the
correlation increase for large $\kmax$. To model results only to
$\kmax = 0.2 \ihMpc$, our model contains too much freedom, in comparison
to the requirements of modeling to $\kmax = 0.4 \ihMpc$. Again for
comparison, \cite{Beutler:2017} finds a correlation of $r = 0.257$
between $\apar$ and $\aperp$. Thus, our value of $r = 0.292$ indicates
that we are able to recover a similar amount of information using our
RSD model parametrization to $\kmax = 0.4 \ihMpc$.

\subsection{Independent tests on high resolution mocks}

To fully assess the accuracy and precision of our RSD model, we perform
independent tests using two sets of mocks based on high-fidelity, periodic
$N$-body simulations. The first, described in \S \ref{sec:nseries-results}
is a homogenenous set of 21 galaxy catalogs derived from 7 realizations of
a $N$-body simulation with fixed cosmology and bias model. The second,
described in \S \ref{sec:challenge-results}, is a set of 7 heterogenous HOD
galaxy catalogs where both the bias model and underlying cosmology varies from
box to box. For details on the cosmology and simulation parameters for these
mocks, see table~\ref{tab:sim-params}.

\subsubsection{Cubic N-series results}\label{sec:nseries-results}

\begin{figure}[!ptb]
\centering
\includegraphics[width=\textwidth]{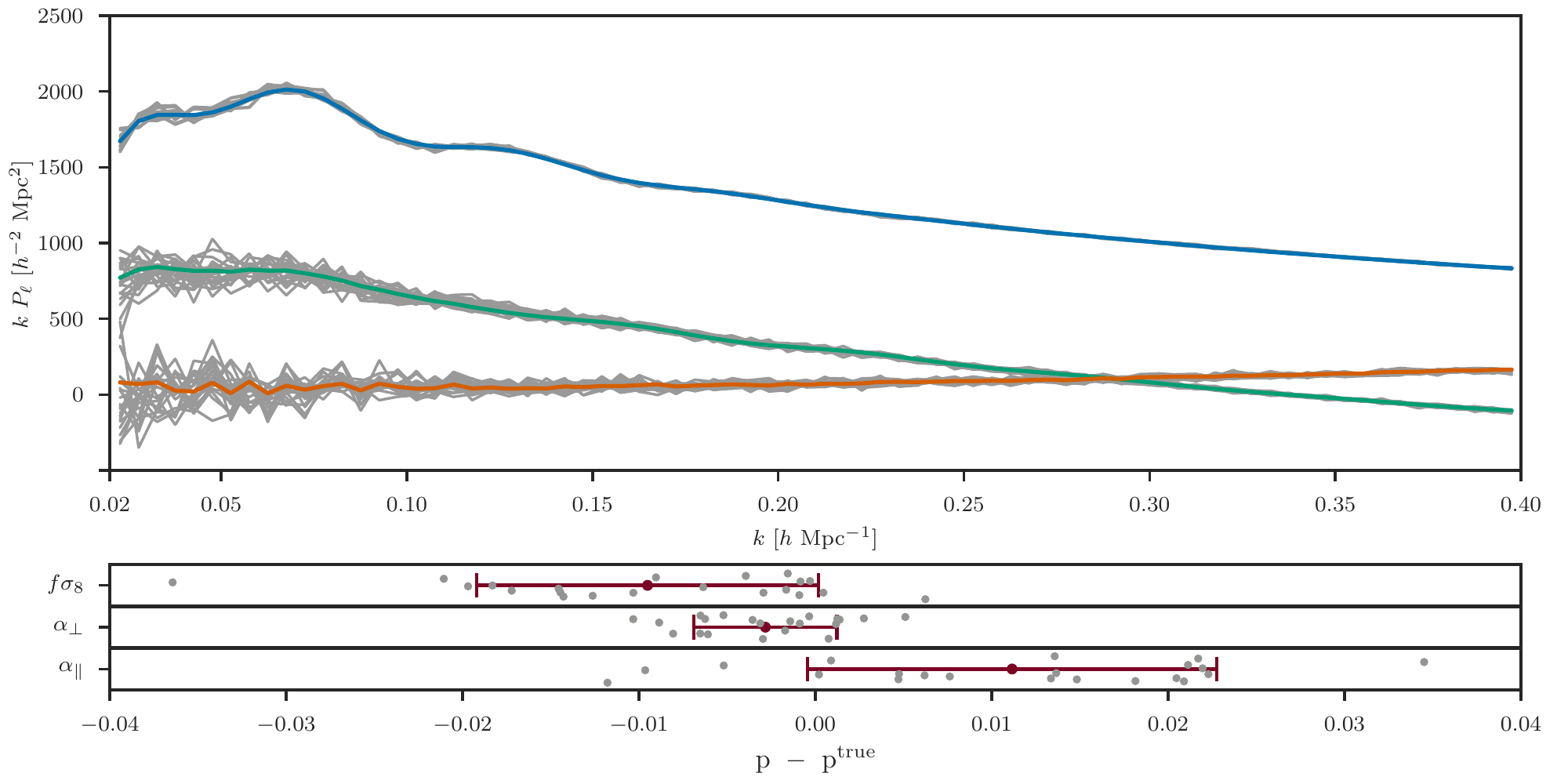}
\caption{The mean of the best-fit monopole, quadrupole, and
hexadecapole models (colored) as compared to the individual
measurements (gray) from the 21 N-series cubic boxes at $z=0.5$, fit over the
wavenumber range $k = 0.02 - 0.4 \ihMpc$. The lower panels show
the scatter in the recovered values for the 3 cosmology parameters,
$\fsig$, $\aperp$, and $\apar$, across the 21 boxes. The error bar
shows the standard deviation of these results (not the error on the mean).}
\label{fig:nseries-kmax04}
\end{figure}

Our first independent tests utilize the cubic N-series simulation, the
large-volume ($L_\mathrm{box} = 2600 \hMpc$)
periodic box simulations described in section~\ref{sec:nseries}. We
perform fits to the monopole, quadrupole, and hexadecapole from 21 HOD
galaxy catalogs, constructed from 7 realizations at $z = 0.5$ and 3 orthogonal
line-of-sight projections per box. The cosmology of these boxes is given in
table~\ref{tab:sim-params}. As in section~\ref{sec:runPB-results}, we perform
fits to the data vector [$P_0, P_2, P_4$] over a range of $\kmax$ values.
The best-fitting parameters for each of the 21
catalogs are obtained by maximum a posterior (MAP) estimation using
the LBFGS algorithm.

Figure~\ref{fig:nseries-kmax04} shows the measured $\ell=0$, $2$, and $4$
multipoles from the individual N-series catalogs, and we have over-plotted
the mean of the best-fitting model from each fit using $\kmax = 0.4 \ihMpc$.
We report the mean (with the expected value subtracted) and standard deviation
for the best-fitting $\fsig$, $\aperp$, and $\apar$ values from the 21 fits
as a function of fitting range in table~\ref{tab:nseries-kmax-results}.
We also include the results for $\fsig$ when holding the AP
parameters fixed to their true values.

We find similar trends in our recovered cosmological parameters for the
N-series boxes as for the RunPB results in \S\ref{sec:runPB-results}.
We obtain good fits to the measured $\ell=0,2,4$ multipoles using our RSD model
up to $\kmax = 0.4 \ihMpc$. However, we do find some evidence for
small systematic biases present in our RSD model, although it is difficult
to properly assess the level of statistical significance with only
seven fully independent realizations (clustering from boxes that
vary only the line-of-sight projection are correlated). When
using $\kmax = 0.4 \ihMpc$, we find that are $\fsig$ value is biased
low by $\Delta \fsig$ = 0.008 and $\apar$ is biased high by $\Delta \apar = 0.011$.
These correspond to $\sim$$0.8\sigma$ and $\sim$$0.9\sigma$ shifts, respectively,
relative to the box-to-box dispersion, as determined by the standard deviation
of the 21 fits. Although it is important to note that, again, with only 7
independent realizations and 21 total fits, the standard deviation across
the fits remains noisy. When fixing the AP parameters
to their true values, we see a relatively large upwards shift in the
mean $\fsig$ value across the fits. As our model prefers a slightly
larger $\apar$ value than expected, when its value is fixed to its
correct value, the recovered value for $\fsig$ shifts upwards,
due to the anti-correlation between the parameters.

\begin{table}[!ptb]
\centering
\sisetup{round-mode=places, round-precision=3}
\begin{tabular}{c|c|c|c|c|c|c|c|c}
\toprule
$k_\mathrm{max}$ & $\Delta \langle \alpha_\parallel \rangle$ & $S_{\alpha_\parallel}$ & $\Delta \langle \alpha_\perp \rangle$ & $S_{\alpha_\perp}$ & $\Delta \langle f \sigma_8 \rangle$ & $S_{f \sigma_8}$ & $\Delta \langle f \sigma_8 \rangle$ & $S_{f \sigma_8}$ \\
$[\ihMpc]$ &  &  &  &  &  &  & \multicolumn{2}{c}{fixed AP} \\
\midrule
$0.2$ & $\num{0.0053911488336}$ & $\num{0.0111166119049}$ & $\num{-0.0020981127675}$ & $\num{0.00512039214005}$ & $\num{-0.0155031905855}$ & $\num{0.0142071321446}$ & $\num{-0.00982062}$ & $\num{0.0127417}$\\
$0.3$ & $\num{0.00740526205368}$ & $\num{0.0107480068305}$ & $\num{-0.00344267578256}$ & $\num{0.00443480424324}$ & $\num{-0.00926522249267}$ & $\num{0.0128783713049}$ & $\num{-0.000110138}$ & $\num{0.011042}$\\
$0.4$ & $\num{0.0111505336712}$ & $\num{0.011605961264}$ & $\num{-0.00283480303861}$ & $\num{0.00405946210705}$ & $\num{-0.0075168011302}$ & $\num{0.00969071762689}$ & $\num{0.00257135}$ & $\num{0.00975301}$\\
\bottomrule
\end{tabular}

\caption{The mean (with expected value subtracted) and standard deviation $S$ of
the best-fitting values for $\fsig$, $\aperp$, and $\apar$ found when
fitting $[P_0, P_2, P_4]$ from the 21 cubic N-series catalogs. Results
are reported as a function of the maximum wavenumber included in
the fit. We also give results for $\fsig$ when holding the AP parameters
fixed to their true values.}
\label{tab:nseries-kmax-results}
\end{table}

\subsubsection{Lettered challenge box results}\label{sec:challenge-results}

\begin{figure}[!ptb]
\centering
\includegraphics[width=\textwidth]{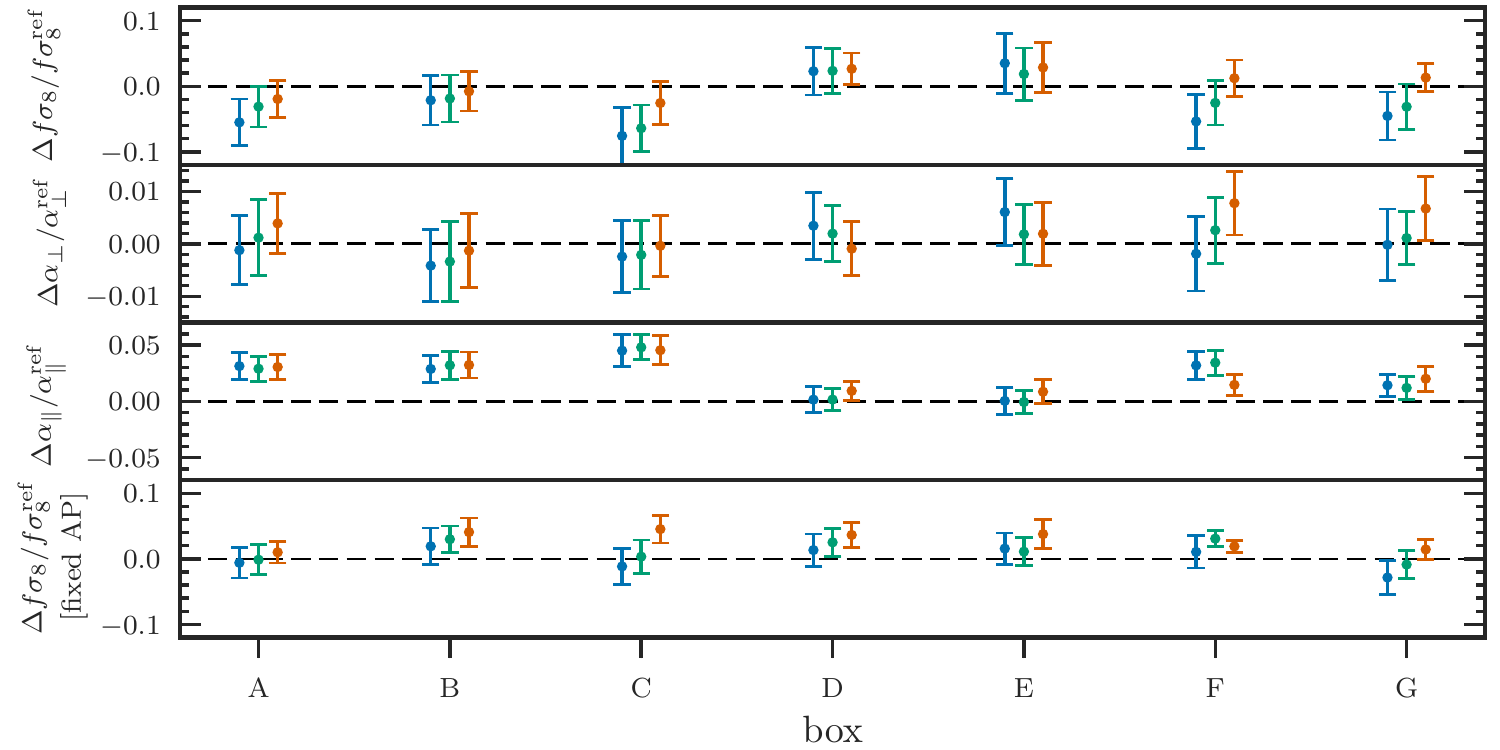}
\caption{The fractional deviation of the best-fitting
$\fsig$, $\aperp$, and $\apar$ values from their true values
for each of the seven lettered challenge boxes. We also
show the deviations for $\fsig$ obtained when the AP parameters
are fixed to their true values. For each box, we show
results obtained using (from left to right)
$\kmax$ = 0.2 (blue), 0.3 (green), and 0.4 (orange)$\ihMpc$.
Error bars show the $1\sigma$ uncertainty as obtained via MCMC
sampling.}
\label{fig:challenge-box-params}
\end{figure}

We perform additional tests of our model using a heterogeneous
set of seven HOD galaxy catalogs, labeled A through G, which
were constructed from high-fidelity cubic $N$-body simulations. These
catalogs are described in detail in section~\ref{sec:lettered-challenge},
and the cosmology and simulation parameters are reported
in table~\ref{tab:sim-params}. The box size for these mocks
is $\sim$2.5$\hGpc$; a single box has roughly 4 times the volume
of the DR12 BOSS CMASS sample and 60\% of the volume of the mean of the
10 RunPB realizations. They were designed to provide stringent stress-tests
of full-shape RSD modeling analyses, and as such, they cover a range of
redshifts ($z=[0.441, 0.5, 0.562]$), $\fsig$ values, and
galaxy bias models. As was done in previous sections, we compute fits to
the monopole, quadrupole, and hexadecapole for each of the seven lettered
challenge boxes, as a function of the maximum wavenumber included in the
fits. We obtain full posterior distributions for each of our 13 model
parameters using MCMC sampling. We report the recovered values for the
cosmological parameters (with the expected value subtracted) and
the $1\sigma$ parameter uncertainties for all seven boxes in
table~\ref{tab:challenge-kmax-results}. Figure~\ref{fig:challenge-box-params}
illustrates the fractional deviation of our recovered cosmology values from
their reference values for each $\kmax$ value.

The recovered values show similar trends as a function of $\kmax$ as the
results from the RunPB and cubic N-series results. We generally find
$\sim$$20 - 30\%$ improvements in the error on $\fsig$ when extending
the fit from $\kmax = 0.2 \ihMpc$ to $\kmax = 0.4 \ihMpc$. We find more
modest decreases in the error for the AP parameters, with little
improvement extending from $\kmax = 0.3 \ihMpc$ to $\kmax = 0.4 \ihMpc$.
Within the expected $1\sigma$ uncertainty of each mock, we
recover $\fsig$ and $\aperp$ values consistent with the truth for all
seven boxes, and the best-fitting values generally remain stable
as a function of $\kmax$. However, the recovered values for $\apar$
show a systematic positive bias for all boxes, relative to the truth, which
can be most easily seen in figure~\ref{fig:challenge-box-params}.
This bias is present for each value of $\kmax$ used.
It is difficult to assess the statistical significance of this potential
bias, as several of the seven mocks are built on the same underlying
$N$-body simulation, which introduces the derived parameters.
Weighting each derived $\apar$ by the inverse uncertainty, we find
a mean positive bias of $\Delta \apar = 0.02$, independent
of $\kmax$. This bias is slightly larger than was found for either the RunPB
or cubic N-series mocks, where both results show a $\sim$0.01 positive
bias in $\apar$.

We also include in figure~\ref{fig:challenge-box-params} and
table~\ref{tab:challenge-kmax-results} the results for $\fsig$
when fixing the AP parameters to their true values. As expected, we see
substantial ($\sim$20-30\%) error decreases since the correlation between $\fsig$
and the AP parameters degrades constraints when $\apar$ and $\aperp$
are allowed to vary. Similar to previous results, we also find a
systematic positive shift in the recovered $\fsig$ values when
holding $\apar$ and $\aperp$ fixed to their true values. This is expected due
to the correlation between $\fsig$ and $\apar$ and the systematic positive
shift found for $\apar$.

\begin{table}[!ptb]
\renewcommand{\arraystretch}{1.25}
\centering
\sisetup{round-mode=places, round-precision=3}
\begin{tabular}{c|c|c|c|c|c}
\toprule
box & $k_\mathrm{max}$ & $\Delta \alpha_\parallel$ & $\Delta \alpha_\perp$ & $\Delta f \sigma_8$ & $\Delta f \sigma_8$ \\
{} & $[\ihMpc]$ &  &  & & fixed AP \\

\midrule
\multirow{3}{*}{A} & $0.2$ & \hphantom{+}\num{0.0313294974701} \numpmerr{0.01176842294}{0.0121401640806} & \num{-0.00122155650179} \numpmerr{0.00643063174933}{0.00671462206836} & \num{-0.0261522293091} \numpmerr{0.0183440019584}{0.0155916928098} & \num{-0.00282991528511} \numpmerr{0.0105482339859}{0.0115848779678}\\
 & $0.3$ & \hphantom{+}\num{0.0290108178539} \numpmerr{0.0112696654642}{0.0108188668489} & \hphantom{+}\num{0.0011659924003} \numpmerr{0.00702679274635}{0.00750759935078} & \num{-0.0147649109364} \numpmerr{0.0152288157461}{0.0141837000847} & \num{-0.000619685649872} \numpmerr{0.0113244354725}{0.0104206204414}\\
& $0.4$ & \hphantom{+}\num{0.0305263666475} \numpmerr{0.00997213304725}{0.0128014389122} & \hphantom{+}\num{0.00388469139248} \numpmerr{0.00559380345009}{0.00584367782808} & \num{-0.00922668576241} \numpmerr{0.0127903989438}{0.0137296617031} & \hphantom{+}\num{0.00476197302341} \numpmerr{0.00819198787212}{0.00727413594723}\\
\midrule
\multirow{3}{*}{B} & $0.2$ & \hphantom{+}\num{0.0287989272924} \numpmerr{0.0116804721197}{0.0127364360142} & \num{-0.00416386939262} \numpmerr{0.00713102064849}{0.00658175555749} & \num{-0.0101383388042} \numpmerr{0.0175936424264}{0.0184044241905} & \hphantom{+}\num{0.00910308361053} \numpmerr{0.013516575098}{0.0131322474793}\\
 & $0.3$ & \hphantom{+}\num{0.0320666794252} \numpmerr{0.0125024460656}{0.0123249228553} & \num{-0.00339107026124} \numpmerr{0.00811761393375}{0.00708038702635} & \num{-0.00887552499771} \numpmerr{0.0164842699515}{0.0176532864571} & \hphantom{+}\num{0.0141946613789} \numpmerr{0.0104447574547}{0.0086717903614}\\
& $0.4$ & \hphantom{+}\num{0.0323007701767} \numpmerr{0.0123008087369}{0.0109549334328} & \num{-0.00131172892385} \numpmerr{0.0069907024046}{0.00703476259573} & \num{-0.00369012951851} \numpmerr{0.0146518051624}{0.0138373374939} & \hphantom{+}\num{0.0193822622299} \numpmerr{0.00964315735826}{0.0109216570854}\\
\midrule
\multirow{3}{*}{C} & $0.2$ & \hphantom{+}\num{0.045094375054} \numpmerr{0.0142246515808}{0.0140230374109} & \num{-0.00243800657761} \numpmerr{0.00731785862282}{0.00651749946428} & \num{-0.0350385015011} \numpmerr{0.0204335153103}{0.0195835530758} & \num{-0.00532528805733} \numpmerr{0.0115975141525}{0.0137221515179}\\
 & $0.3$ & \hphantom{+}\num{0.0480353723861} \numpmerr{0.010639973063}{0.0115125511708} & \num{-0.00211349994283} \numpmerr{0.00608512288534}{0.00695626816428} & \num{-0.0296496157646} \numpmerr{0.0157014727592}{0.0171589553356} & \hphantom{+}\num{0.00157460522652} \numpmerr{0.0110175907612}{0.0125771164894}\\
& $0.4$ & \hphantom{+}\num{0.0454516041961} \numpmerr{0.0131508804274}{0.0125154631005} & \num{-0.000383365617286} \numpmerr{0.00584223682265}{0.00572859916294} & \num{-0.0118462150097} \numpmerr{0.0161693287498}{0.014334499836} & \hphantom{+}\num{0.0210096235275} \numpmerr{0.00972384214401}{0.0100001096725}\\
\midrule
\multirow{3}{*}{D} & $0.2$ & \hphantom{+}\num{0.00150364979395} \numpmerr{0.0119533388002}{0.0109400516867} & \hphantom{+}\num{0.00344387400097} \numpmerr{0.00663030412769}{0.00614385538545} & \hphantom{+}\num{0.0107479381561} \numpmerr{0.015782982111}{0.0181177280719} & \hphantom{+}\num{0.00628396749496} \numpmerr{0.0116500387359}{0.0115258395672}\\
 & $0.3$ & \hphantom{+}\num{0.00156883981863} \numpmerr{0.00981251359886}{0.0102069600959} & \hphantom{+}\num{0.00193707415235} \numpmerr{0.00512366112161}{0.00560043231739} & \hphantom{+}\num{0.0110403585434} \numpmerr{0.0171343386173}{0.0153649747372} & \hphantom{+}\num{0.0118354398012} \numpmerr{0.0104676634073}{0.00926701724529}\\
& $0.4$ & \hphantom{+}\num{0.00931781167063} \numpmerr{0.00880696377074}{0.00831197023028} & \num{-0.000931076777046} \numpmerr{0.00524317912346}{0.00509371060746} & \hphantom{+}\num{0.0125087785721} \numpmerr{0.0111439228058}{0.0115324556828} & \hphantom{+}\num{0.0171369302273} \numpmerr{0.00885191559792}{0.00924065709114}\\
\midrule
\multirow{3}{*}{E} & $0.2$ & \hphantom{+}\num{0.000392393732464} \numpmerr{0.0128686001303}{0.011626849515} & \hphantom{+}\num{0.00604399402862} \numpmerr{0.00600042292417}{0.00679027438044} & \hphantom{+}\num{0.0145524209738} \numpmerr{0.0182068198919}{0.0196305066347} & \hphantom{+}\num{0.00650142788887} \numpmerr{0.0102492747875}{0.00942009687424}\\
 & $0.3$ & \num{-0.000633101278127} \numpmerr{0.0104856327579}{0.010146525684} & \hphantom{+}\num{0.00180933829558} \numpmerr{0.00587431508609}{0.00549394525702} & \hphantom{+}\num{0.00772511124611} \numpmerr{0.0161841511726}{0.0168793201447} & \hphantom{+}\num{0.00458637535572} \numpmerr{0.00923179090023}{0.00867724873447}\\
& $0.4$ & \hphantom{+}\num{0.00851639887903} \numpmerr{0.0106743635318}{0.0109098433885} & \hphantom{+}\num{0.00190491432553} \numpmerr{0.00640916330704}{0.00559075206272} & \hphantom{+}\num{0.0118800020218} \numpmerr{0.0168269276619}{0.0151093006134} & \hphantom{+}\num{0.0156574165821} \numpmerr{0.00950568914413}{0.00870596003611}\\
\midrule
\multirow{3}{*}{F} & $0.2$ & \hphantom{+}\num{0.0319696650322} \numpmerr{0.0130070047704}{0.0117280097091} & \num{-0.00190786579983} \numpmerr{0.00728757366228}{0.00687005315375} & \num{-0.0254867434502} \numpmerr{0.0197129547596}{0.0192892706385} & \hphantom{+}\num{0.00502806305885} \numpmerr{0.0113589465618}{0.0119889378548}\\
 & $0.3$ & \hphantom{+}\num{0.0344000430911} \numpmerr{0.010797153312}{0.0113808930594} & \hphantom{+}\num{0.0025844653544} \numpmerr{0.00627353741598}{0.00634662876117} & \num{-0.0120578169823} \numpmerr{0.0150167942047}{0.017167121172} & \hphantom{+}\num{0.0146441549063} \numpmerr{0.00579063594341}{0.0060808211565}\\
& $0.4$ & \hphantom{+}\num{0.0146365869764} \numpmerr{0.00988840778843}{0.00866331732474} & \hphantom{+}\num{0.00773900712139} \numpmerr{0.00638232904331}{0.00570994836439} & \hphantom{+}\num{0.00579669475555} \numpmerr{0.0137937068939}{0.0127347042968} & \hphantom{+}\num{0.0090447306633} \numpmerr{0.00467213988304}{0.00430336594582}\\
\midrule
\multirow{3}{*}{G} & $0.2$ & \hphantom{+}\num{0.0143034372163} \numpmerr{0.00950768159073}{0.00985080605296} & \num{-0.000195510933935} \numpmerr{0.00662857199046}{0.00701608889602} & \num{-0.0215197116137} \numpmerr{0.0174017399549}{0.0174195021391} & \num{-0.0135231375694} \numpmerr{0.0120331346989}{0.0121001303196}\\
 & $0.3$ & \hphantom{+}\num{0.0119664488414} \numpmerr{0.00909994505744}{0.0110031699098} & \hphantom{+}\num{0.0010944754222} \numpmerr{0.00483371679006}{0.00519439502618} & \num{-0.0148594141006} \numpmerr{0.0154866874218}{0.0174317061901} & \num{-0.00412479639053} \numpmerr{0.0099746286869}{0.0105926394463}\\
& $0.4$ & \hphantom{+}\num{0.0200707702687} \numpmerr{0.0111893469221}{0.0111114650725} & \hphantom{+}\num{0.00671295004199} \numpmerr{0.00611914628635}{0.00613368665229} & \hphantom{+}\num{0.00624599456787} \numpmerr{0.00979697555663}{0.0102607607841} & \hphantom{+}\num{0.00682398080826} \numpmerr{0.00769207035434}{0.00674498081207}\\
\bottomrule
\end{tabular}
\caption{The best-fitting values for $\fsig$, $\aperp$, and $\apar$
obtained when fitting our RSD model to the measured monopole, quadrupole, and
hexadecapole from the 7 lettered challenge boxes. We report
results as a function of the maximum wavenumber included in the fits.
The $1\sigma$ uncertainties obtained via MCMC sampling are also shown.}
\label{tab:challenge-kmax-results}
\end{table}

\subsection{Tests on realistic DR12 BOSS CMASS mocks}\label{sec:ncutsky-results}


Finally, we test our RSD model using BOSS DR12 CMASS mock catalogs,
using the 84 independent, N-series cutsky catalogs
described in \S\ref{sec:nseries}. This set of catalogs offers a chance
to test the performance of our model in a realistic setting with
a large enough number of catalogs to identify systematic biases up to
the level of $\sqrt{84}=9.16$ times smaller than the measurement uncertainty
from a single mock. The 84 N-cutsky catalogs
accurately model the geometry, volume, and redshift distribution
of the DR12 CMASS NGC sample \cite{Reid:2016}. We use the
window function convolution procedure outlined in
section~\ref{sec:window-function} to properly account for effects
of the selection function on the measured power spectrum multipoles.
We measure the monopole, quadrupole, and hexadecapole for each of the
84 catalogs and estimate the best-fitting model parameters using
MAP optimization via the LBFGS algorithm. The power spectra have
been measured using FKP weights with a value of
$P_0 = 10^4$ $[h^{-3} \mathrm{Mpc}^3]$. Similar
to previous fits, we report parameter constraints as a function of
the maximum wavenumber included in the fits. The minimum
wavenumber included in the fits is $k_\mathrm{min} = 0.02 \ihMpc$,
chosen to minimize any large-scale effects of the window function on
our parameter fits.

\begin{figure}[t]
\centering
\includegraphics[width=\textwidth]{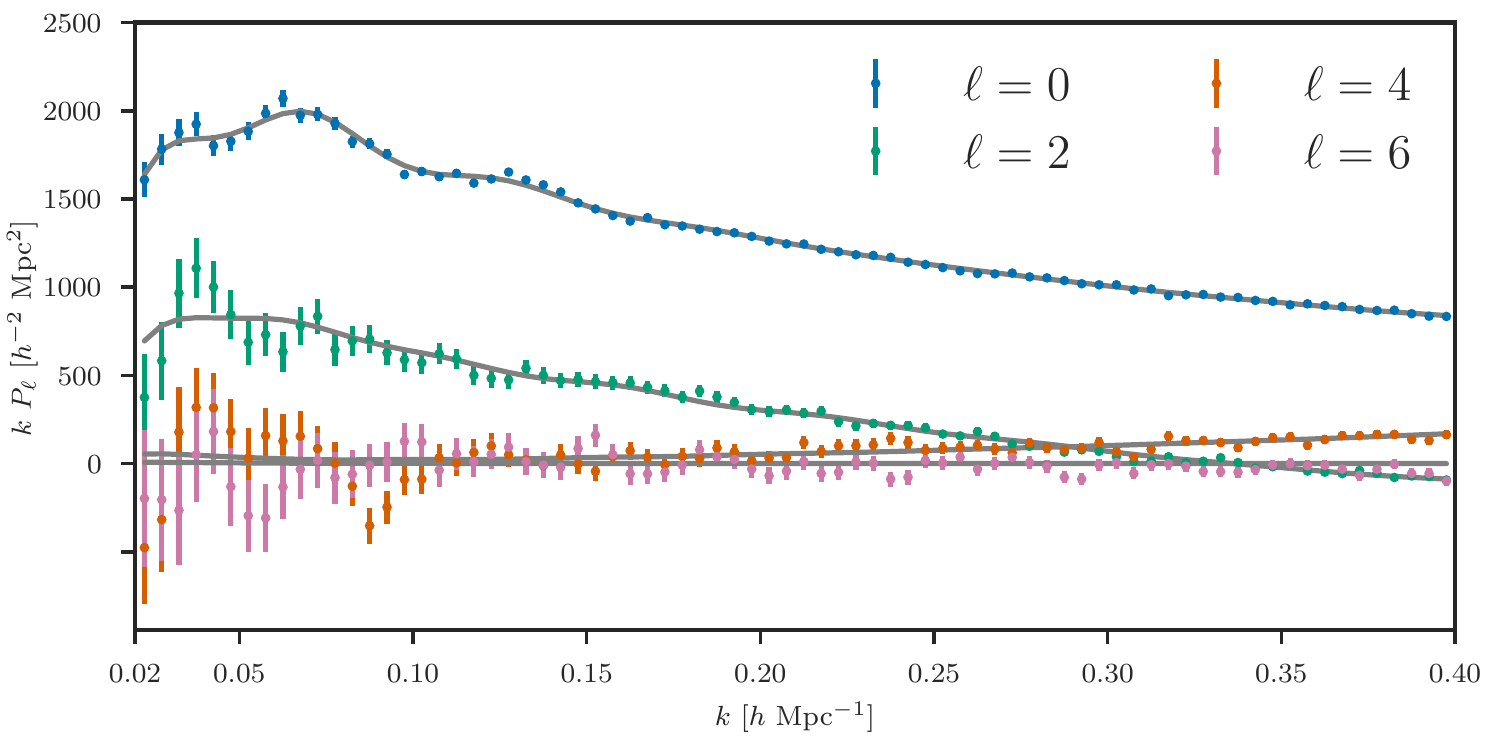}
\caption{
The best-fitting $\ell=0$, $2$, $4$, and $6$ theory (grey lines)
and measurements (points with errors) from a single catalog
of the N-series cutsky test suite, which accurately simulates the BOSS DR12
CMASS data set. The best-fit model has been estimated using the data vector
$[P_0, P_2, P_4]$ while fitting over the wavenumber range
$k = 0.02 - 0.4 \ihMpc$. We also show the tetra-hexadecapole ($\ell=6$)
to illustrate that the model can accurately predict this higher-order multipole
and that it contains little measurable signal. The reduced chi-squared of the
fit for this mock catalog is $\chi^2_\mathrm{red} = 1.01$.
}
\label{fig:ncutsky-poles}
\end{figure}

We plot the best-fitting $\ell=0$, $2$, $4$, and $6$ theoretical
model and the measurements multipoles from a single catalog
of the full 84 N-series cutsky test suite in figure~\ref{fig:ncutsky-poles}.
Here, the best-fit model has been estimated using the data vector
$[P_0, P_2, P_4]$ with $\kmax = 0.4 \ihMpc$, but we also show the
tetra-hexadecapole ($\ell=6$) to illustrate that the model can accurately
predict this higher-order multipole and that it contains little measurable
signal. For this single mock, we find good agreement between theory
and data, with a reduced chi-squared of $\chi^2_\mathrm{red} = 1.01$.

\begin{figure}[!ptb]
\centering
\includegraphics[width=\textwidth]{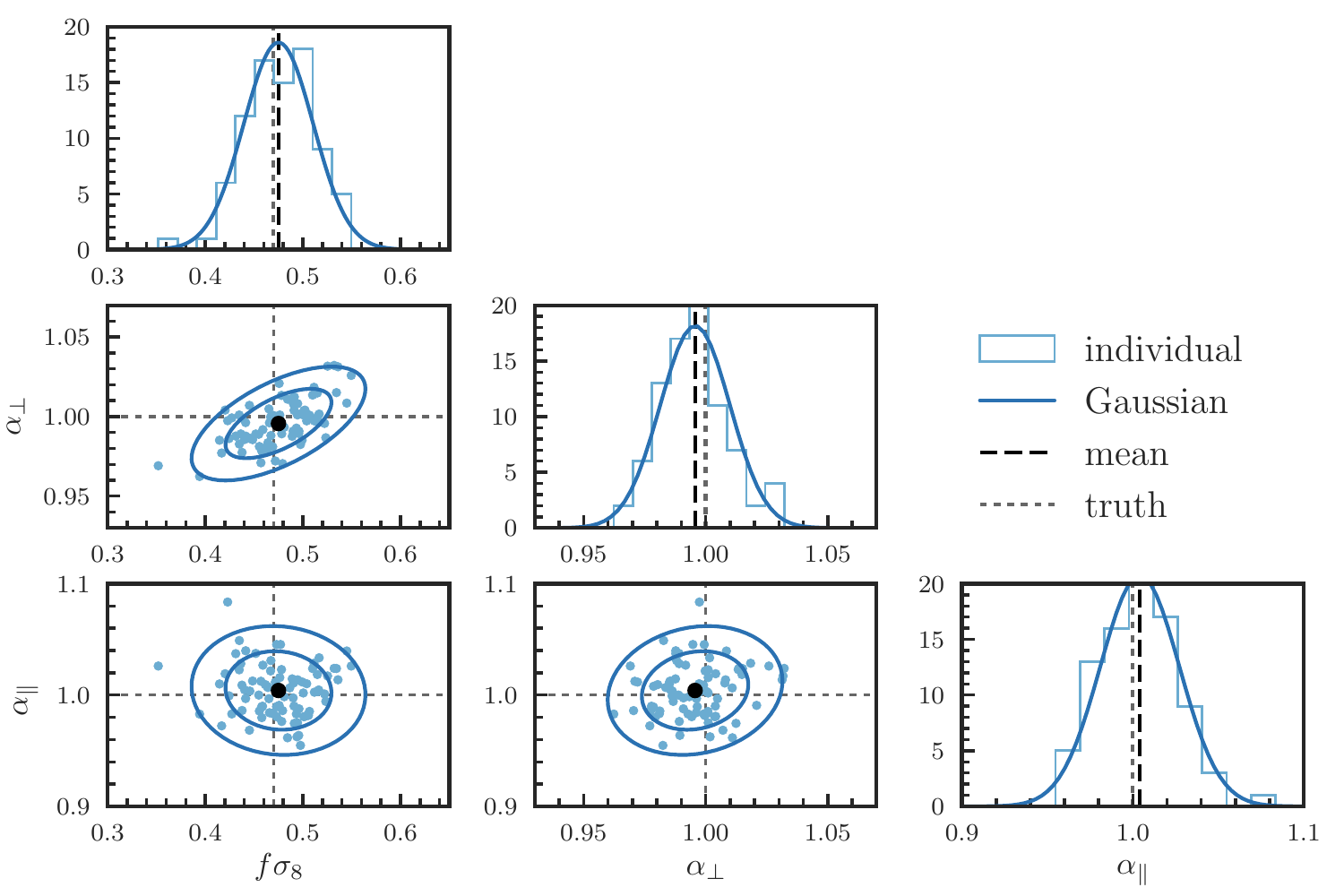}
\caption{The best-fitting $\fsig$, $\aperp$, and $\apar$ parameters
from fitting our RSD model to the measured $[P_0, P_2, P_4]$ multipoles
from the 84 N-series cutsky mocks. We include wavenumbers in the range
$0.02 \ihMpc \le k \le 0.4 \ihMpc$. The diagonal panels show the histogram of
the recovered parameters, with the mean best-fitting parameters indicated as
black dashed lines and the true values as gray dotted lines.
The panels below the diagonal show 2D plots with the 84 individual
best-fitting parameters as blue dots and the mean as a filled circle.
We also show a Gaussian fit to the marginalized parameter distributions
in all panels.}
\label{fig:ncutsky-triangle}
\end{figure}

We give the mean (with the expected value subtracted) and
standard deviation of the best-fitting cosmological parameters from
fits to each of the 84 cutsky mocks in table~\ref{tab:ncutsky-kmax-results}.
We also show the the 1D histograms and 2D correlations of
$\fsig$, $\aperp$, and $\apar$ for the individual fits in
figure~\ref{fig:ncutsky-triangle}, illustrating the constraining
power of our model for these parameters as well as the correlations
between the parameters.
When fitting the monopole, quadrupole, and hexadecapole, we find
good agreement between the mean of the recovered values for
$\fsig$, $\aperp$, and $\apar$. When including scales up to
$\kmax = 0.4 \ihMpc$, we find modest mean biases of
$\Delta \langle \fsig \rangle = 0.005$,
$\Delta \langle \aperp \rangle = -0.004$, and
$\Delta \langle \apar \rangle = 0.004$, which represent
14\%, 28\%, and 17\% of the expected mock-to-mock dispersion of
each parameter, respectively. The statistical precision of the
mean values due to the finite number of catalogs is $84^{-1/2}\simeq0.1$
times the mock-to-mock dispersion. Thus, the results
show evidence for a small bias in the derived
$\aperp$ value and marginal evidence for small biases in $\apar$
and $\fsig$. We also show results in table~\ref{tab:ncutsky-kmax-results}
when fitting only the monopole and quadrupole in order to help
quantify the impact of the hexadecapole on our final constraints.
The mean best-fitting parameters remain consistent with the results
obtained when fitting [$P_0$, $P_2$, $P_4$], and we find the
standard deviation of our best-fitting $\fsig$ values inflates
by roughly 30\%, consistent with the findings of \cite{Beutler:2017}.
When fixing the AP parameters to their true values, we find that
the hexadecapole adds negligible further information to our parameter
constraints.

\FloatBarrier

\begin{table}[t]
\renewcommand{\arraystretch}{1.25}
\centering
\sisetup{round-mode=places, round-precision=3}
\resizebox{\textwidth}{!}
{
\begin{tabular}{c|c|c|c|c|c|c|c|c|c}
\toprule
statistics & $k_\mathrm{max}$ & $\Delta \langle \alpha_\parallel \rangle$ & $S_{\alpha_\parallel}$ & $\Delta \langle \alpha_\perp \rangle$ & $S_{\alpha_\perp}$ & $\Delta \langle f \sigma_8 \rangle$ & $S_{f \sigma_8}$ & $\Delta \langle f \sigma_8 \rangle$ & $S_{f \sigma_8}$ \\
& [$\ihMpc$]  & & & & & &  & \multicolumn{2}{c}{fixed AP} \\
\midrule
\multirow{3}{*}{[$P_0$, $P_2$, $P_4$]} & $0.2$ & $\num{0.00654931581321}$ & $\num{0.0239121287071}$ & $\num{-0.00412064239164}$ & $\num{0.0157353409806}$ & $\num{-0.0198375942146}$ & $\num{0.0408608919404}$ & $\num{-0.00794009234847}$ & $\num{0.0342211376686}$\\
 & $0.3$ & $\num{0.00740194349592}$ & $\num{0.0254278257797}$ & $\num{-0.0050040001996}$ & $\num{0.0152107584096}$ & $\num{-0.00772742168162}$ & $\num{0.0385223062996}$ & $\num{0.00491338453909}$ & $\num{0.0297194302052}$\\
 & $0.4$ & $\num{0.00408223812901}$ & $\num{0.0231472638718}$ & $\num{-0.00433394736816}$ & $\num{0.0143147918214}$ & $\num{0.00502358107403}$ & $\num{0.0356085741637}$ & $\num{0.0127095159665}$ & $\num{0.0272052557811}$\\
\midrule
\multirow{3}{*}{[$P_0$, $P_2$]} & $0.2$ & $\num{-0.00359934399245}$ & $\num{0.0394421337604}$ & $\num{-0.000968386304057}$ & $\num{0.0186425816734}$ & $\num{-0.0142156121899}$ & $\num{0.0521118810106}$ & $\num{-0.0129565429867}$ & $\num{0.0350622747815}$\\
 & $0.3$ & $\num{0.00485771972606}$ & $\num{0.0408386609429}$ & $\num{-0.00445960449609}$ & $\num{0.0190513416908}$ & $\num{-0.00515971498836}$ & $\num{0.0530634183212}$ & $\num{0.00468941555363}$ & $\num{0.0304739709805}$\\
 & $0.4$ & $\num{0.0118872912512}$ & $\num{0.0362612999378}$ & $\num{-0.00837435026334}$ & $\num{0.016040132955}$ & $\num{-0.0104845709297}$ & $\num{0.0401818697017}$ & $\num{0.0066919575272}$ & $\num{0.025023659981}$\\
\bottomrule
\end{tabular}
}
\caption{The mean and standard deviation of the best-fitting values
for $\fsig$, $\aperp$, and $\apar$ from fits
to the 84 N-series cutsky catalogs. Results
are reported as a function of the maximum wavenumber included in
the fit. We show results obtained when including or excluding
the hexadecapole from our fits in order to quantify the influence
of the hexadecapole on our derived constraints.}
\label{tab:ncutsky-kmax-results}
\end{table}

\subsection{Comparison to published models}\label{sec:comparison}

\begin{table}[tbp!]
\renewcommand{\arraystretch}{1.25}
\centering
\sisetup{round-mode=places, round-precision=4}
\resizebox{\textwidth}{!}
{
\begin{tabular}{c|c|c|c|c|c|c|c}
\toprule
{} & $k_\mathrm{max}$ for $[P_0, P_2, P_4]$ & $\Delta \langle \alpha_\parallel \rangle$ & $S_{\alpha_\parallel}$ & $\Delta \langle \alpha_\perp \rangle$ & $S_{\alpha_\perp}$ & $\Delta \langle f \sigma_8 \rangle$ & $S_{f \sigma_8}$\\
\midrule
\cite{Beutler:2017} & [0.15, 0.15, 0.1] & $\num{0.00493269047619}$ & $\num{0.0338038480442}$ & $\num{-0.00139397619048}$ & $\num{0.017998925661}$ & $\num{-0.00485397619048}$ & $\num{0.0374566881589}$ \\
\cite{Grieb:2017} & [0.2, 0.2, 0.2] & $\num{0.00894018206693}$ & $\num{0.0253473318279}$ & $\num{-0.00301058286847}$ & $\num{0.0175468860078}$ & $\num{0.000120513175586}$ & $\num{0.0383068212169}$ \\
\midrule
\multirow{4}{*}{This work} & [0.15, 0.15, 0.1] & $\num{0.000257492539603}$ & $\num{0.0402610784382}$ & $\num{0.00111401469659}$ & $\num{0.0183373204385}$ & $\num{0.0042999941222}$ & $\num{0.0446893980915}$ \\
{} & [0.2, 0.2, 0.2] & $\num{0.00654931581321}$ & $\num{0.0239121287071}$ & $\num{-0.00412064239164}$ & $\num{0.0157353409806}$ & $\num{-0.0198375942146}$ & $\num{0.0408608919404}$ \\
{} & [0.3, 0.3, 0.3] & $\num{0.00740194349592}$ & $\num{0.0254278257797}$ & $\num{-0.0050040001996}$ & $\num{0.0152107584096}$ & $\num{-0.00772742168162}$ & $\num{0.0385223062996}$ \\
{} & [0.4, 0.4, 0.4] & $\num{0.00408223812901}$ & $\num{0.0231472638718}$ & $\num{-0.00433394736816}$ & $\num{0.0143147918214}$ & $\num{0.00502358107403}$ & $\num{0.0356085741637}$ \\
\bottomrule
\end{tabular}
}
\caption{The mean (with expected value subtracted) and
standard deviation of the best-fitting cosmology parameters for the 84 N-cutsky
mocks using the model in this work as well as the Fourier space models
described in \cite{Beutler:2017} and \cite{Grieb:2017}.
Results in all cases were computed using FKP weights
with $P_0 = 10^4 \ h^{-3}\mathrm{Mpc}^3$.}
\label{tab:ncutsky-comparison}
\end{table}

The set of 84 N-series mocks described in the previous section were utilized as
part of the BOSS collaboration's internal RSD modeling tests in preparation for the
DR12 parameter constraint analyses. This enables us to perform a direct
comparison of our model with the main Fourier space RSD models
used in the DR12 consensus results, which are described in the
companion papers in \cite{Beutler:2017} and \cite{Grieb:2017} and the
main DR12 consensus paper in \cite{Alam:2016}. The model used in
\cite{Grieb:2017} was also applied to BOSS DR12 data in configuration space,
with results presented in \cite{Sanchez:2017}. These
analyses differ in a number of ways from ours. In particular,
these models have significantly fewer parameters (7-8 instead of 13) and
use a smaller $\kmax$ value in their fits. We limit our fitting
range to the same as those used in these works and directly compare the
derived parameter constraints for the N-cutsky mocks in
table~\ref{tab:ncutsky-comparison}. For comparison, this table also includes
results from fits using our model that include scales to $\kmax=0.3\ihMpc$ and
$\kmax=0.4\ihMpc$, which goes beyond the scales used in \cite{Grieb:2017} and
\cite{Beutler:2017}. When using comparable fitting ranges, we find that
our model yields a standard deviation for $\fsig$ that is larger by
$\sim$10\% and $\sim$20\% as compared to when using the models of
\cite{Beutler:2017} and \cite{Grieb:2017}, respectively. We find
comparable constraints on $\fsig$ when extending our model to
$\kmax=0.3\ihMpc$ and a modest 5-10\% improvement when using
$\kmax=0.4\ihMpc$.

For the AP parameters, we find a comparable constraint
on $\aperp$ and a slighter worse constraint on $\apar$ as compared
to the model of \cite{Beutler:2017}. We find modest 5\% and 10\% reductions
in the error on $\apar$ and $\aperp$ as compared to the model of
\cite{Grieb:2017}. Extending the fits with our model to $\kmax=0.4\ihMpc$
does not provide much gain for the uncertainty of $\apar$, but we do find
a roughly 20\% reduction in the uncertainty of $\aperp$ as compared to the
models of \cite{Beutler:2017} and \cite{Grieb:2017}. As seen
in the results of \cite{Alam:2016}, the most powerful
method for constraining the AP parameters, and thus $D_A(z)$ and $H(z)$,
remains a BAO-only analysis that takes advantage of the additional statistical
precision gained by the process of density field reconstruction. However,
some additional constraining power can be gained from full-shape RSD analyses
due the AP effect on sub-BAO scales. Here, the extra information provided
by extending the modeling to $\kmax=0.4\ihMpc$ will aid the constraints
on $D_A(z)$ and $H(z)$ and help de-correlate these parameters.

It is also instructive to compare our results for the N-cutsky mocks
to the results published in \cite{Gil-Marin:2016},
which fits the monopole and quadrupole of the DR12 CMASS sample
with $\kmax = 0.24 \ihMpc$. The comparison yields similar conclusions as
previously. In particular, \cite{Gil-Marin:2016} finds errors of
$\sigma_{\fsig} = 0.038$ and $\sigma_{\fsig} = 0.022$ when varying and fixing
the AP parameters, respectively.
These errors are both smaller than the uncertainties derived from our model
by $\sim$30\% when fitting the monopole and quadrupole over similar wavenumber
ranges. The constraints for the AP parameters in \cite{Gil-Marin:2016}
are similarly smaller than those from our model by a comparable amount.

And, finally, it is worth noting that the $\fsig$ constraints using
the model in this work are not competitive with the
2.5\% constraint on $\fsig$ published in \cite{Reid:2014}, which remains
the tightest measurement of $\fsig$ in the literature to date.
With fixed AP parameters, the work used a simulation-based analysis to model
the small-scale correlation function of the DR10 CMASS sample
well into the nonlinear regime, down to scales of $\sim$$0.8 \hMpc$. They relied on
simulations to accurately model the galaxy-halo connection whereas we use the
analytic, halo model decomposition described in
section~\ref{sec:halo-model}. Given the tight constraint on $\fsig$ found by
\cite{Reid:2014}, one might hope that Fourier space
models could be similarly extended into the nonlinear regime and yield comparable
increases in precision. However, while acknowledging the number of differences in the
two analyses, we note that we do not find such large
increases in precision in our measurement of $\fsig$ when including
small-scale information down to $k = 0.4\ihMpc$.


\section{Discussion}\label{sec:discussion}

The results presented in section~\ref{sec:performance} provide tests
of the RSD model presented in this work for a suite of simulations
that span a wide range in both cosmology and galaxy bias models.
Given the measurement uncertainties and the degrees of freedom in our model,
we are able to achieve excellent agreement between the $\ell=0,2,4$
multipoles measured from simulations and our best-fitting theory down to
scales of $k = 0.4 \ihMpc$. To quantify the impact of small-scale
physics on our model, we perform fits for $\kmax=0.2$, $0.3$, and
$0.4 \ihMpc$. The results across the different sets of simulations
indicate a positive systematic shift in
the parallel AP parameter $\apar$ at the level of $0.01-0.02$
that is independent of the $\kmax$ value used. For fits using
$\kmax=0.4\ihMpc$, we find small biases at the level of $\sim$0.005 for
$\fsig$ and $\aperp$. These deviations are small and can be effectively
calibrated with simulations. The amplitude of the shifts is
similar to the level of theoretical systematics present when using
other RSD models in the literature, i.e., \cite{Alam:2016}.
The positive bias in $\apar$ propagates into a small
bias in $\fsig$ when fixing the AP parameters to their expected values, due
to the anti-correlation between $\fsig$ and $\apar$. The exact amplitude of the
bias in $\apar$ can be robustly estimated from a larger set of simulations than
is considered in this work and the best-fitting $\apar$ value modified
accordingly, while accounting for the systematic uncertainty in the error
budget.

A primary goal of this work is to ensure that any model parameters that we
introduce have physically meaningful values and are not just nuisance parameters.
We attempt to capture the complex effects of satellite galaxies on the
clustering signal in redshift space by considering separately the
clustering of isolated satellites and those that live in halos with at least
two satellites. This parametrization leads to a total of 13 model
parameters, significantly more than other Fourier space RSD models in the
literature, i.e., \cite{Beutler:2017,Grieb:2017}, which typically only have
7-8 parameters. In addition to differences in the
treatment of RSD and perturbation theory choices, perhaps the most significant
difference is the use of a single parameter to model the nonlinear
FoG effect of the full galaxy sample, instead of separately modeling the effects
for central and satellite subsamples, as is done in this work. They also
typically float a constant, shot noise parameter, designed to absorb any
potential deficiencies in the model. In some sense, these models are a limit
of the more general parametrization considered in this work and are only valid
over a certain range of scales and galaxy bias values.

As demonstrated in the analysis of \cite{Alam:2016}, the level of
theoretical errors in $\fsig$ measurements from full-shape RSD analyses
ranges from $\sim$25-50\% of the statistical precision for the three redshift
bins considered for the completed BOSS DR12 sample. Another recent analysis
\cite{Beutler:2016b} provides evidence for the possible shortcomings of the
RSD model of \cite{Beutler:2017}. The work extends the modeling to include the
relative velocity effect of baryons and cold dark matter at
decoupling but fails several null tests. The systematics situation is perhaps
even more dire when considering the fact that background cosmology model is
essentially fixed by the Planck results (see Fig. 11 of \cite{Alam:2016}),
indicating that a more relevant test
of systematics should be done with the AP parameters fixed, often resulting in
a $\sim$20-30\% smaller error on $\fsig$. This suggests that RSD analyses
from full-shape modeling are already systematics dominated and will
certainly be so for future galaxy surveys, without subsequent
modeling improvements. While the model presented in this work has its own
shortcomings, one such avenue for improvement is exploring
more physically motivated model descriptions.

As discussed in section~\ref{sec:comparison}, our
parametrization leads to a derived uncertainty of $\fsig$ that is roughly
10-20\% larger than the constraint from the models
of \cite{Beutler:2017,Grieb:2017}, which use fewer parameters.
Each of our parameters has a physical motivation, and we apply reasonable
priors based on these motivations when appropriate. Thus, we
find no clear path to reduce the number of parameters in our model and
do not believe that additional constraining power can be gained through
the use of stronger priors. As such,
RSD models in the literature are likely too-limited in their parametrization,
with the uncertainty on $\fsig$ underestimated by $\sim$10-20\%.
For a galaxy sample such as the BOSS CMASS sample with a satellite
fraction $f_s \sim 0.1$, the clustering is dominated by the 2-halo
correlations of centrals. However, we find the inclusion of parameters to
properly treat the 1\% effects of satellite-satellite correlations to be
crucial to modeling the clustering down to scales of $k \sim 0.4 \ihMpc$.
Using a Fisher analysis, we find similar errors on $\fsig$ as found
by the models of \cite{Beutler:2017,Grieb:2017} for the N-cutsky mocks
(see table~\ref{tab:ncutsky-comparison}) when fixing the relative fraction
of non-isolated satellites $f_{s_B}$ and the central galaxy velocity
dispersion $\sigma_c$. In this case, we only vary a single FoG
velocity dispersion, as is the case for the models of
\cite{Beutler:2017,Grieb:2017}, and fix the $\sim$1\% contribution to the
overall power spectrum from satellites living in halos containing multiple
satellites.

Fully perturbative modeling approaches
cannot accurately capture the effects of nonlinearities, i.e., the FoG
effect from satellites, on small scales, and at some point, the modeling
must become sensitive to the poorly-understood physics of galaxy evolution.
Presently, it is unclear how sensitive cosmological growth of structure measurements
are to such small-scale physics. In particular, assembly bias remains
a worrying potential systematic for galaxy clustering analyses
\cite{Zentner:2014,Zentner:2016}. The most promising avenue for including
small-scale information ($k \sim 0.4 \ihMpc$) in growth of structure
analyses appears to be simulation-based modeling efforts. The most
competitive constraint to date for $\fsig$ published in \cite{Reid:2014} uses a
simulation-based model to describe the correlation function down to scales
of $r \sim 0.8 \hMpc$. In order to achieve the desired accuracy for the RSD
model presented in this work, we also find it necessary to include calibrations
from simulations for key components of the model. The combination of
perturbation theory with simulation-based calibration in our model
likely limits the applicability of the model in comparison to a
fully general, simulation-based approach. An emulator-based approach for
the nonlinear clustering of galaxies in redshift-space using the FastPM
simulation method \cite{Feng:2016} is under active
development.

An alternative approach for maximizing the constraining power
of RSD analyses relies on limiting the effects of satellites on the
modeling. These so-called halo reconstruction methods attempt to modify
the measurement procedure to preferentially exclude satellites galaxies,
thus measuring the clustering of the underlying halo density field, rather
than the galaxy density field \cite{Tegmark:2006,Reid:2009,Okumura:2017}.
The difficulty of these methods remains achieving a transformation accurate
enough such that the added modeling complications from the transformation itself
do not outweigh the benefits gained by removing satellites.
The advantages include limiting the effects of nonlinearities,
which simplifies the modeling and could allow use of models closer to purely
linear theory. Reducing FoG effects raises the overall amplitude of the quadrupole
and boosts the signal-to-noise of the measurement, although removing satellite
galaxies does lower the overall bias, which reduces the constraining power
of a given measurement.

\begin{figure}[tbp!]
\centering
\includegraphics[width=0.8\textwidth]{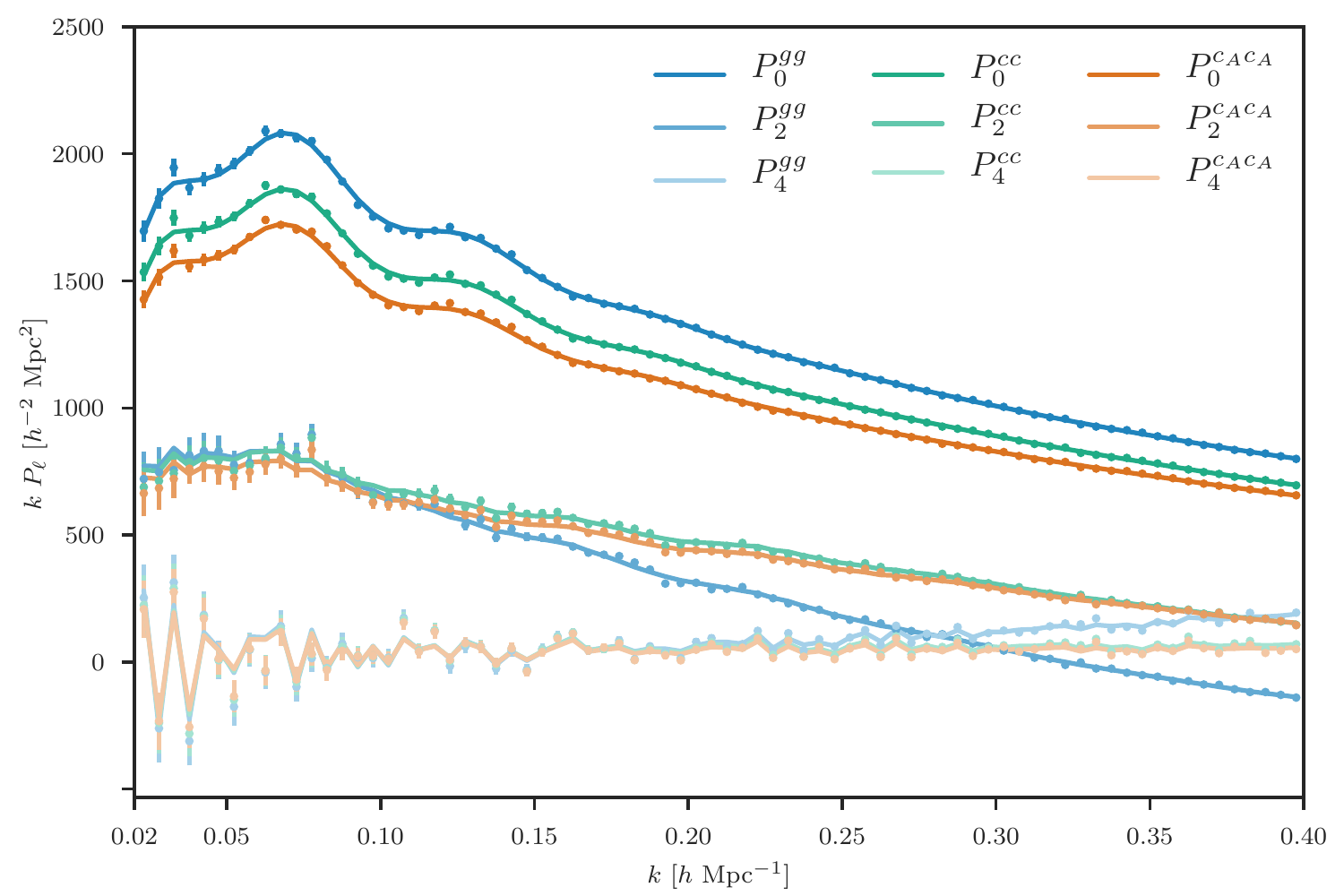}
\caption{The best-fitting model and measured simulation points
for the monopole (darkest shade), quadrupole,
and hexadecapole (lightest shade) from the mean of 10
RunPB galaxy catalogs at $z=0.55$ for all galaxies $P_\ell^{gg}$ (blue),
all centrals $P_\ell^{cc}$ (green), and isolated
centrals with no satellites in the same halo $P_\ell^{c_A c_A}$
(orange). Linear biases for each sample are $b_{1,g} = 2.05$,
$b_{1,c} = 1.93$, and $b_{1,c_A} = 1.84$.
}
\label{fig:runPB-cen-comparison}
\end{figure}

As an illustration, we have compared our
RSD constraints when fitting our 13 parameter model to the
clustering of centrals and type A centrals (isolated centrals that have
no satellites) from the RunPB simulations. We show the best-fit multipoles
for these cases in comparison to the spectra of the full galaxy sample
in figure~\ref{fig:runPB-cen-comparison}. As expected, the small-scale quadrupole
shows a significant reduction in the effects of RSD, and we find a reduction
in the linear bias due to the removal of the highly biased satellites.
Corresponding parameter constraints for $\fsig$, $\aperp$, and $\apar$
are presented in table~\ref{tab:runPB-cen-comparison}. We find the largest
decreases in uncertainty when considering centrals only --
the error on $\fsig$ decreases by 31\%, 19\%, and 15\% when fitting
to $\kmax = 0.2, 0.3$, and $0.4\ihMpc$, respectively. Similarly, we find
decreases of 16\%, 7\%, and 25\% for $\apar$ and 19\%, 10\%, and
0\% for $\aperp$. While we find diminishing benefits to extending
the fitting range from $\kmax = 0.2 \ihMpc$ to $\kmax = 0.4 \ihMpc$,
the constraints using centrals only are in all cases better than using
when using all galaxies. Furthermore, fitting the clustering of only centrals
to $\kmax = 0.2 \ihMpc$ is roughly as competitive in constraining $\fsig$
as fitting the clustering of all galaxies to $\kmax = 0.4 \ihMpc$, and the latter
is significantly more challenging to model than the former. While we
recognize that this is certainly an idealized demonstration, we view
halo reconstruction methods as an important area of future research for both
their constraining power and simplified theoretical modeling.

\begin{table}[tbp!]
\renewcommand{\arraystretch}{1.25}
\centering
\sisetup{round-mode=places, round-precision=4}
\resizebox{\textwidth}{!}
{
\begin{tabular}{c|c|c|c|c}
\toprule
$k_\mathrm{max}$ & {} & all galaxies & centrals only & type A centrals only \\
\midrule
\multirow{3}{*}{$k_\mathrm{max} = 0.2 \ihMpc$} & $\Delta \alpha_\parallel$ & \hphantom{+}\num{0.00896811067199} \numpmerr{0.00954876628835}{0.00857545109698} & \hphantom{+}\num{0.00960109088752} \numpmerr{0.00762497948713}{0.00761080414369} & \hphantom{+}\num{0.0101431880499} \numpmerr{0.00814944413581}{0.00807002264033}\\
 & $\Delta \alpha_\perp$ & \hphantom{+}\num{0.00288019795187} \numpmerr{0.00536981642058}{0.00628430554354} & \hphantom{+}\num{0.0022510739782} \numpmerr{0.00482577859531}{0.00461547906796} & \hphantom{+}\num{0.00227035311403} \numpmerr{0.00466550319812}{0.00465351652935}\\
 & $\Delta f \sigma_8$ & \num{-0.017364677906} \numpmerr{0.0148746073246}{0.0153758885962} & \num{-0.0042308242321} \numpmerr{0.0108046531677}{0.0100436508656} & \num{-0.00894894909859} \numpmerr{0.01157990098}{0.0107553005219}\\
\midrule
\multirow{3}{*}{$k_\mathrm{max} = 0.3 \ihMpc$} & $\Delta \alpha_\parallel$ & \hphantom{+}\num{0.00649154132276} \numpmerr{0.0077843768293}{0.00736347726873} & \hphantom{+}\num{0.00491874453564} \numpmerr{0.00717463874794}{0.00698691090136} & \hphantom{+}\num{0.00378223454153} \numpmerr{0.00689236000392}{0.00674700602438}\\
 & $\Delta \alpha_\perp$ & \hphantom{+}\num{0.0042100131777} \numpmerr{0.00492484808396}{0.00457219275207} & \hphantom{+}\num{0.00438216897989} \numpmerr{0.00426714965614}{0.00427298907051} & \hphantom{+}\num{0.0046475473907} \numpmerr{0.00445151453702}{0.00463204237357}\\
 & $\Delta f \sigma_8$ & \num{-0.00479761970043} \numpmerr{0.0122080594301}{0.0127761811018} & \hphantom{+}\num{0.00658949005604} \numpmerr{0.0099571198225}{0.0103905430606} & \hphantom{+}\num{0.00306755590439} \numpmerr{0.00929841399193}{0.0093460381031}\\
\midrule
\multirow{3}{*}{$k_\mathrm{max} = 0.4 \ihMpc$} & $\Delta \alpha_\parallel$ & \hphantom{+}\num{0.00889152541023} \numpmerr{0.00681966822276}{0.00768071320787} & \hphantom{+}\num{0.00298275783135} \numpmerr{0.00552401407104}{0.00540835152938} & \hphantom{+}\num{0.00119358278034} \numpmerr{0.00609093110544}{0.00677587624318}\\
 & $\Delta \alpha_\perp$ & \hphantom{+}\num{0.00495101701109} \numpmerr{0.00424141855313}{0.00404902872263} & \hphantom{+}\num{0.00285970376619} \numpmerr{0.00444283792695}{0.00388352871933} & \hphantom{+}\num{0.00119888029937} \numpmerr{0.00442587589345}{0.00409771663847}\\
 & $\Delta f \sigma_8$ & \num{-0.00308581900597} \numpmerr{0.00967335700989}{0.00907552242279} & \hphantom{+}\num{0.00309336471558} \numpmerr{0.00849893540503}{0.00756034255028} & \hphantom{+}\num{0.00589768874645} \numpmerr{0.00834973156452}{0.00807194411755}\\
\bottomrule
\end{tabular}
}
\caption{
The best-fit $\fsig$, $\aperp$, and $\apar$ values
and $1\sigma$ uncertainties obtained
when fitting the monopole, quadrupole, and hexadecapole from the mean of 10
RunPB galaxy catalogs at $z=0.55$ when including all galaxies,
centrals only, and type A centrals only, which are isolated
with no satellites in the same halo.
}
\label{tab:runPB-cen-comparison}
\end{table}

\section{Conclusion}\label{sec:conclusion}

We present a new model for the redshift-space power spectrum of
galaxies and demonstrate its accuracy in modeling the
monopole, quadrupole, and hexadecapole of the galaxy density
field down to $k = 0.4 \ihMpc$ through a series of tests on high-fidelity
$N$-body simulations. The model describes the clustering of galaxies
in the context of a halo model, building upon the formalism
presented in \cite{Okumura:2015}. We decompose galaxies into four
subsamples: centrals with and without satellites and satellites with one
or more neighboring satellite. We then model the clustering
of the underlying halos in redshift space using a
combination of Eulerian perturbation theory
and $N$-body simulations. The modeling of
RSD via the mapping from real space to redshift space is done using
the so-called distribution function approach. In order to achieve sufficient
accuracy in the modeling down to $k = 0.4 \ihMpc$, we utilize a
set of simulations to calibrate the most important terms
of the model. To this end, we extend the Halo-Zel'dovich
Perturbation Theory of \cite{Seljak:2015}, which combines
Lagrangian perturbation theory with physically motivated corrections
calibrated from simulations. We improve the accuracy of this model
for the dark matter density power spectrum and
develop models for the dark matter velocity correlators $P_{01}$
and $P_{11}$. Our final model has 13 free parameters, each of
which is physically motivated, as described
in table~\ref{tab:model-params}.
The model accounts for the FoG effect
from each of our galaxy subsamples, rather than using a single velocity
dispersion to describe the combined effect. We account for the linear
bias of each of the subsamples and describe the shot noise contributions
to the power spectrum via the amplitude of the 1-halo galaxy correlations.

We fit our 13 parameter model to the monopole, quadrupole, and
hexadecapole measured from several sets of simulations to test the accuracy and
precision of the recovered parameters. These mock catalogs
cover a range of cosmologies and galaxy bias models, providing stringent tests
of our model. The test suite also includes realistic mock catalogs of the
BOSS DR12 CMASS sample, properly modeling the volume and selection effects
of this data set. We perform fits as a function
of the maximum wavenumber included in the fit, using
$\kmax = 0.2, 0.3$, and $0.4 \ihMpc$. The results of these tests can
be summarized as follows:

\begin{enumerate}
  \item[(i)] Given the measurement covariance and degrees of freedom in
  the model, we find excellent agreement between our model and the measured
  $\ell=0$, 2, and 4 multipoles from simulations down to scales of
  $k = 0.4 \ihMpc$.
  \item[(ii)] A systematic shift in the best-fitting value of $\apar$ is
  identified at the level of $0.01-0.02$, independent of the $\kmax$ value
  used when fitting. Such a systematic shift can be calibrated from a large
  set of simulations and a correction applied to the best-fitting value.
  \item[(iii)] The level of systematic bias in the parameters $\fsig$ and $\aperp$
  is found to be small, at the level of $\sim$0.005, which is similar
  to other published RSD models in the literature, i.e., \cite{Alam:2016}.
  However, considering that the Planck results essentially fix the background
  cosmology model, comparisons between this level of systematics and
  the error on $\fsig$ for fixed AP parameters indicate that RSD analyses
  are nearly systematics dominated today. This will certainly be the case
  for the next generation of galaxy surveys, unless analyses are limited
  to the largest scales or substantial modeling improvements are made.
  \item[(iv)] Using a set of BOSS DR12 CMASS mock catalogs as a benchmark for
  comparison, we estimate an uncertainty on $\fsig$ that is $\sim$10-20\%
  larger than when using the models of \cite{Beutler:2017, Grieb:2017},
  when fitting over similar wavenumber ranges. With 5-6 fewer parameters,
  these models likely have a too-limited parametrization and are
  underestimating the resulting uncertainty of $\fsig$.
  \item[(v)] Extending the fitting range to $\kmax=0.4 \ihMpc$ provides
  15-30\% gains in the statistical precision of the $\fsig$ constraint
  relative to $\kmax=0.2 \ihMpc$. The gains are more modest
  when our model is compared to published models, which use a more
  limited parametrization; the error on
  $\fsig$ is roughly 5-10\% smaller with our model using $\kmax=0.4\ihMpc$ than
  constraints found when using the models of \cite{Beutler:2017, Grieb:2017}
  (with $\kmax \simeq 0.2 \ihMpc$) for the BOSS DR12 CMASS sample.
  \item[(vi)] We find a $\sim$10-15\% improvement in the constraint on $\aperp$
  and only marginal gains for $\apar$ when
  extending from $\kmax=0.2 \ihMpc$ to $\kmax=0.4 \ihMpc$. The constraint
  on $\aperp$ represents a 20\% improvement relative to the results found when
  the published models of \cite{Beutler:2017, Grieb:2017}.
  This improvement will further help constrain and de-correlate the parameters
  $D_A(z)$ and $H(z)$ when combined with post-reconstruction BAO-only analyses.
\end{enumerate}

Extending full-shape RSD modeling of galaxy clustering to smaller scales in
both an accurate and precise manner remains a complicated endeavor. Applying
models such as the one presented here will be necessary at the minimum to
fully capitalize on the cosmological information contained in future
galaxy surveys, such as Hobby Eberly Telescope Dark Energy Experiment
\cite{Hill:2008}, the Dark Energy Spectroscopic Instrument (DESI)\cite{Levi:2013},
the Subaru Prime Focus Spectrograph \cite{Takada:2014}, and the
ESA space mission \textit{Euclid} \cite{Laureijs:2011}. For example, we expect
our model to perform well on the DESI Emisson Line Galaxy sample, which has
a lower bias and high satellite fraction as compared to the BOSS CMASS sample.
Conversely, the gains when applying our model to next generation quasar samples
would likely be more modest due to the high shot noise and lower impact
of one-halo correlations.

There are several intriguing modeling approaches that can go beyond our
analytic modeling approach and potentially improve the constraints further.
Simulation-based approaches can leverage advances in high-performance computing
to accurately model nonlinear clustering on small-scales. Complementary
approaches such as halo reconstruction can simplify modeling and reduce the
need to include small-scale information by mitigating the complicated effects
of satellite galaxies on the modeling procedure. It remains to be seen if
these further advances will improve cosmological constraints, or whether
with our model we have reached the limit due to the effects of nonlinear
evolution and poorly known small-scale physics.

\acknowledgments{
We thank Jan N. Grieb and Ariel Sanchez for providing their RSD modeling results
and for comments on the manuscript. We also are grateful to Martin White
for providing his RunPB simulations on which large portions of the modeling in
this work are based. NH is supported by the National Science Foundation
Graduate Research Fellowship under grant number DGE-1106400. FB acknowledges
support from the UK Space Agency through grant ST/N00180X/1.
US is supported by NASA grant NNX15AL17G.
}

\appendix

\section{Velocity correlators in the Zel'dovich approximation}\label{app:zeldovich}

In this section, we use Lagrangian perturbation theory (LPT) to compute
two dark matter velocity correlators that enter into the DF model:
the density -- radial momentum cross spectrum, $P_{01}$, and the
radial momentum auto spectrum, $P_{11}$. We closely follow the notation
of \cite{Vlah:2015}; see e.g, \cite{Vlah:2015,Carlson:2013,Matsubara:2008}
and references therein for further review of Lagrangian perturbation theory.

\subsection{$P_{01}$ and $P_{11}$ using LPT}

Following the definitions of \cite{Vlah:2013}, the velocity correlators that
we wish to compute are given by

\begin{eqnarray}
(2 \pi)^3 P_{01}(k) \delta^D(\vk + \vk') & = &
			\langle \delta(\vk) | p_\parallel(\vk') \rangle, \nn \\
(2 \pi)^3 P_{11}(k) \delta^D(\vk + \vk') & = &
			\langle p_\parallel(\vk) | p_\parallel(\vk') \rangle,
\end{eqnarray}
where $p_\parallel$ is the momentum projected along the line-of-sight,
i.e., $p_\parallel = \vp \cdot \hat{z}$. The scalar component of the dark
matter momentum (which correlates with density) can be computed using
the continuity equation: $\ddelta(\vk) - i \vk \cdot \vp = 0$,
where the dot in $\dot{\delta}$ represents the derivative with respect
to conformal time $\tau$. Using this equation, we can express the
velocity correlators of interest as

\begin{eqnarray}
\label{eq:P01-to-Pdd}
P_{01}(k) &=& \frac{i \mu}{k} P_{\delta \ddelta}(k), \\
\label{eq:P11-to-Pdd}
P_{11, s}(k) &=& \frac{\mu^2}{k^2} P_{\ddelta \ddelta}(k),
\end{eqnarray}
where $\mu$ is defined as $k_\parallel / k$. Here, we explicitly note
that $P_{11, s}$ only includes scalar contributions, as only the
scalar component of momentum enters into the continuity equation.
The total contribution from these terms to the redshift-space
power spectrum $P^S(k, \mu)$ is given by:

\begin{eqnarray}
\label{eq:P01_ss}
P_{01}^S(\vk) &=&  2 \mathrm{Re}\left(\frac{-ik\mu}{\mathcal{H}}\right) P_{01}(k)
 			= 2\frac{\mu^2}{\mathcal{H}} P_{\delta \ddelta}(k), \\
\label{eq:P11_ss}
P_{11,s}^S(\vk) &=&  \left(\frac{k\mu}{\mathcal{H}}\right)^2 P_{11,s}(k)
 			= \frac{\mu^4}{\mathcal{H}^2} P_{\ddelta \ddelta}(k).
\end{eqnarray}
These are the spectra that we wish to compute in the Zel'dovich approximation.
In linear theory, these spectra are the anisotropic terms of the well-known
Kaiser formula: $P_{01}^S(\vk) = 2 f \mu^2 P_L(k)$ and
$P_{11,s}^S(\vk) = f^2 \mu^4 P_L(k)$ \cite{Kaiser:1987}.

We can compute $\delta$ and $\ddelta$ using Lagrangian
perturbation theory. In the Lagrangian clustering description,
the overdensity field is given by

\begin{equation}\label{eq:lpt-delta}
(2 \pi)^3 \delta^D(\vk) + \delta(\vk) = \int d^3q \ e^{i \vk \cdot \vq}
			\exp[i \vk \cdot \Psi(\vq)],
\end{equation}
where $\Psi(\vq)$ is the Lagrangian displacement field. The derivative
of this equation with respect to conformal time is given by

\begin{equation}\label{eq:lpt-deltadot}
\ddelta(\vk) = \int d^3 q e^{i \vk \cdot \vq}
			\left (i \vk \cdot \dot{\Psi} \right) \exp[i \vk \cdot \Psi(\vq)].
\end{equation}
The quantity of interest for $P_{01}$ is

\begin{eqnarray}
(2\pi)^3 P_{\delta \ddelta}(k) \delta^D(\vk + \vk')
			& = & \langle \delta(\vk) | \ddelta(\vk') \rangle, \nn \\
        &=& \int d^3 q d^3q' e^{i\vk\cdot\vq + i\vk'\cdot\vq'}
        	\left \langle \left (i \vk' \cdot \dot{\Psi}' \right)
            	e^{i \vk\cdot\Psi + i\vk'\cdot\Psi'} \right \rangle, \label{eq:Pddot}
\end{eqnarray}
where we have used the definition $\Psi' \equiv \Psi(\vq')$. Similarly,
for $P_{11,s}$ we need to compute

\begin{eqnarray}
(2\pi)^3 P_{\ddelta \ddelta}(k) \delta^D(\vk + \vk')
			& = & \langle \ddelta(\vk) | \ddelta(\vk') \rangle, \nn \\
        &=& \int d^3 q d^3q' e^{i\vk\cdot\vq + i\vk'\cdot\vq'}
        	\left \langle
            	\left (i \vk \cdot \dot{\Psi} \right)
            	\left (i \vk' \cdot \dot{\Psi}' \right)
            	e^{i \vk\cdot\Psi + i\vk'\cdot\Psi'}
             \right \rangle. \label{eq:Pdotdot}
\end{eqnarray}

\subsection{A generalized velocity generating function}

To facilitate the calculation of equations \ref{eq:lpt-delta} and
\ref{eq:lpt-deltadot}, we introduce a generalized velocity generating
function in this section. First, let us define the sum and difference
of the displacement field $\Psi(\vq)$ defined at points
$\vq_1$ and $\vq_2$ in space:

\begin{equation}
\Delta_i^- = \Psi_i(\vq_2) - \Psi_i(\vq_1), \ \ \ \
\Delta_i^+ = \Psi_i(\vq_2) + \Psi_i(\vq_1).
\end{equation}
Now we can define the generalized velocity generating function
$\mathcal{G}$ as

\begin{equation}
\label{eq:G-def}
(2\pi)^3\delta^D(\vk) = \G =
	\int d^3 q e^{i \vk \cdot \vq}
    \left \langle
    		e^{-i\vk\cdot\Delta^-
    		- i\gamma\vk\cdot\Dminusdot
     	    - i\lambda\vk\cdot\Dplusdot}
            \right \rangle.
\end{equation}
Note that the case of $\gamma = \lambda = 0$ gives the well-known
matter power spectrum $P_{\delta \delta}$ in the LPT formalism
\cite{Schneider:1995}. We define the following moments of
$\mathcal{G}$:

\begin{eqnarray}
\label{eq:G10}
G_{10}(k) &=& \frac{d}{d\gamma}\G\Big|_{\gamma=0,\lambda=0} =
	\int d^3q \edot{i\vk}{\vq}
    \la
    \left( i\vk\cdot\Dminusdot\right) \edot{-i\vk}{\Dminusdot}
    \ra,\\
\label{eq:G20}
G_{20}(k) &=& \frac{d^2}{d\gamma^2}\G\Big|_{\gamma=0,\lambda=0} =
	\int d^3q \edot{i\vk}{\vq}
    \la
    \left( i\vk\cdot\Dminusdot\right)^2 \edot{-i\vk}{\Dminusdot},
    \ra, \\
\label{eq:G02}
G_{02}(k) &=& \frac{d^2}{d\lambda^2}\G\Big|_{\gamma=0,\lambda=0} =
	\int d^3q \edot{i\vk}{\vq}
    \la
    \left( i\vk\cdot\Dplusdot\right)^2 \edot{-i\vk}{\Dminusdot},
    \ra .
\end{eqnarray}
Substituting the definitions of $\Dplus$ and $\Dminus$ into
these equations yields the following relations,

\begin{eqnarray}
\label{eq:P01-to-G}
G_{10}(k) = 2 P_{\delta \ddelta}(k), \\
\label{eq:P11-to-G}
G_{20}(k) + G_{02}(k) = 4 P_{\ddelta \ddelta}(k).
\end{eqnarray}
We can evaluate $\mathcal{G}$ using the cumulant expansion theorem,

\begin{equation}
\label{eq:cumulant-expansion}
\la e^{iX} \ra = \exp \left[ \sum_{N=0}^{\infty} \frac{(-i)^N}{N!}
			\la X^N \ra \right],
\end{equation}
where $X = \vk\cdot\Dminus + \gam\vk\cdot\Dminusdot + \lam\vk\cdot\Dplusdot$.
In the Zel'dovich approximation (tree-level LPT), the displacement field
remains Gaussian, so only the $N=2$ term is non-zero in the above expansion.
Thus, the quantity of interest is

\begin{eqnarray}
\label{eq:G-exponent}
\la \left ( \vk\cdot\Dminus + \gam\vk\cdot\Dminusdot +
	\lam\vk\cdot\Dplusdot \right)^2 \ra &=&
    k_i k_j \left [A_{ij} + \gam \dot{A}_{ij} + \gam^2 B_{ij}^-
    + \lam^2 B_{ij}^+ + ... \right] \nn \\
    &=& k_i k_j \left[\mathcal{A}_{ij} + ... \right],
\end{eqnarray}
where we have explicitly ignored terms that vanish upon
taking the derivatives in the expressions for
$G_{10}$, $G_{20}$, and $G_{02}$. The relevant definitions are

\begin{eqnarray}
A_{ij}(k) &=& \la \Dminus_i \Dminus_j \ra_c, \\
B_{ij}^-(k) &=& \la \Dminusdot_i \Dminusdot_j \ra_c, \\
B_{ij}^+(k) &=& \la \Dplusdot_i \Dplusdot_j \ra_c.
\end{eqnarray}
Note that this definition of $A_{ij}$ matches the notation used in
the recent LPT work of \cite{Vlah:2015,Carlson:2013}. Finally,
using equations \ref{eq:G-exponent}, \ref{eq:cumulant-expansion}, and
\ref{eq:G-def}, the velocity generating function becomes

\begin{equation}
(2\pi)^3 \delta^D (\vk) + \G = \int d^3 q \edot{i\vk}{\vq}
		\exp \left[ -\frac{1}{2} k_i k_j \mathcal{A}_{ij} \right].
\end{equation}

\subsection{The Zel'dovich approximation}

In the Zel'dovich approximation, the displacement field
and its time derivative are given

\begin{eqnarray}
\Psi(\vk) &=& -i \vk \delta_L(\vk) / k^2, \\
\dot{\Psi}(\vk) &=& f \mathcal{H} \Psi(\vk),
\end{eqnarray}
where $\delta_L$ is the linear overdensity field which
scales in time as the linear growth function $D$,
$\mathcal{H} = d\ln a / d\tau$ is the conformal Hubble
parameter, and $f = d\ln D/d\ln a$ is the logarithmic growth rate.

With these relations, we can now compute the expressions for
term of \ref{eq:G-exponent} (for a more in-depth discussion
of this procedure, see Appendix B of \cite{Carlson:2013}). The
relevant expressions are:

\begin{eqnarray}
A_{ij}(\vq) &=& I_{ij}^-(\vq), \\
\dot{A}_{ij}(\vq) &=& 2 f \mathcal{H} I_{ij}^- (\vq), \\
\dot{B}_{ij}^-(\vq) &=& (f \mathcal{H})^2 I_{ij}^-(\vq), \\
\dot{B}_{ij}^+(\vq) &=& (f \mathcal{H})^2 I_{ij}^+(\vq),
\end{eqnarray}
where we have defined the integral

\begin{equation}
\label{eq:Iij}
I_{ij}^{\pm}(\vq) = 2 \int \frac{d^3k}{(2\pi)^3}
				\left [ 1 \pm \cos(\vk\cdot\vq)\right] \frac{k_ik_j}{k^4} P_L(k),
\end{equation}
where $P_L(q)$ is the linear power spectrum. Here, $I_{ij}^{-}$ is
the same quantity that enters into the LPT calculation of
the density auto power spectrum; for example, our expression is
the same as equation A6 of \cite{Vlah:2015} (restricting to tree-level).

Equation \ref{eq:Iij} can be expressed in terms of two scalar
functions as

\begin{equation}
X_{ij}^\pm(\vq) = X^\pm(q) \delta^K_{ij} + Y^\pm(q) \hat{q}_i \hat{q}_j,
\end{equation}
We can compute $X_{ij}^\pm$ and $Y_{ij}^\pm$ by
performing the angular integration in equation \ref{eq:Iij}.
To facilitate comparisons with previous work
(e.g., \cite{Vlah:2015,Carlson:2013}), we define

\begin{eqnarray}
X_{ij}^{\pm}(q) &=& 2 \sigma^2 \pm \frac{1}{\pi^2}\int dk P_L(k) \frac{j_1(kq)}{kq}
			\equiv 2 \sigma^2 \pm X_0(q), \nn \\
X_{ij}^-(q) &=& X(q), \nn \\
Y_{ij}^\pm(q) &=& \mp Y(q),
\end{eqnarray}
where the $\sigma^2 = 1/(6\pi^2) \int dq P(q)$ is the square
of the linear displacement field dispersion, and the well-known
Zel'dovich integrals $X(q)$ and $Y(q)$ are

\begin{eqnarray}
X(q) &=& \int \frac{dk}{2\pi^2} P_L(k)
		\left[\frac{2}{3} - 2 \frac{j_1(kq)}{kq} \right], \nn \\
Y(q) &=& \int \frac{dk}{2\pi^2} P_L(k)
		\left [-2 j_0(kq) + 6\frac{j_1(kq)}{kq} \right],
\end{eqnarray}
where $j_n$ is the spherical Bessel function of order $n$.

With these integral expressions, we can now compute the relevant
moments of $\mathcal{G}$ in order to evaluate $P_{01}^S$
and $P_{11,s}^S$. First, for $P_{01}^S$, we have

\begin{align}
P_{01}^S(k, \mu) &= \frac{\mu^2}{\mathcal{H}} G_{10}(k), \nn \\
  &= 2 f\mu^2 \int d^3q e^{ikq\bar{\mu}}
  	\left[-\frac{1}{2}k^2 \left(X+ \bar{\mu}^2 Y \right) \right]
    e^{-\frac{1}{2}k^2 (X + \bar{\mu}^2 Y)}, \label{eq:P01_ss-final}
\end{align}
where we have introduced the angle between the
given $k$-mode and separation vector $\bar{\mu} = \hat{q}\cdot\hat{k}$.
Similarly, for $P_{11,s}^S$, we have

\begin{align}
P_{11,s}^S(k, \mu) &= \frac{\mu^4}{4\mathcal{H}^2}
		\left[ G_{20}(k) - G_{02}(k) \right], \nn \\
  &= \frac{1}{4}f^2\mu^4 \int d^3q e^{ikq\bar{\mu}}
  	k^2 \left[-2X_0 + k^2X^2 + 2(k^2X-1)Y\bar{\mu}^2 + k^2 Y^2\bar{\mu}^4 \right]
    e^{-\frac{1}{2}k^2 (X + \bar{\mu}^2 Y)}. \label{eq:P11_ss-final}
\end{align}
Equations \ref{eq:P01_ss-final} and \ref{eq:P11_ss-final} represent
the desired solution for $P_{01}^S$ and $P_{11,s}^S$ in the
Zel'dovich approximation. The angular integration over $\bar{\mu}$
in these expressions can be performed using the following
expression \cite{Schneider:1995}

\begin{equation}
\int_{-1}^{1} d\mu e^{iA\mu}e^{B\mu^2} = 2 e^B \sum_{n=0}^{\infty}
	\left ( -\frac{2B}{A}\right)^n j_n(A), \label{eq:mu0-integral-sum}
\end{equation}

and the subsequent derivatives of this expression with respect to $B$
yields

\begin{align}
\int_{-1}^{1} d\mu \mu^2 e^{iA\mu}e^{B\mu^2} &= 2 e^B \sum_{n=0}^{\infty}
	\left ( -\frac{2B}{A}\right)^n j_n(A)
    \left[1 + \frac{n}{B} \right], \label{eq:mu2-integral-sum} \\
\int_{-1}^{1} d\mu \mu^4 e^{iA\mu}e^{B\mu^2} &= 2 e^B \sum_{n=0}^{\infty}
	\left ( -\frac{2B}{A}\right)^n j_n(A)
    \left[1 + \frac{n}{B^2}(n + 2B - 1) \right]. \label{eq:mu4-integral-sum}
\end{align}
With equations \ref{eq:mu0-integral-sum}, \ref{eq:mu2-integral-sum},
and \ref{eq:mu4-integral-sum}, we can compute the desired quantities
in equations \ref{eq:P01_ss-final} and \ref{eq:P11_ss-final} as
a quickly-converging sum of one-dimensional integrals, where the
one-dimensional integrals can be computed rapidly with the aid
of software such as FFTLog \cite{Hamilton:2000}. Typically, the
sum over $n$ can be truncated at $n < 15$ for $k < 1 \ihMpc$.

\section{Improved HZPT modeling}\label{app:hzpt}

In this section, we give the best-fit parameters for the updated
HZPT modeling used in this work (as described in section \ref{sec:hzpt}).

\subsection{Dark matter correlators $P_{00}$ and $P_{01}$}\label{app:hzpt-P00}

For the dark matter power spectrum $P_{00}$, we follow the parameterization
of \cite{Seljak:2015} and provide updated best-fit parameters. We use
a Pad\'e expansion with $n_{\mathrm{max}} = 2$, such that the broadband
term is given by

\begin{equation}\label{eq:pade_power}
P^{BB}_{00}(k) = A_0 \left( 1 - \frac{1}{1 + k^2 R^2} \right)
		\frac{1 + (k R_1)^2}{1 + (k R_{1h})^2 + (k R_{2h})^4},
\end{equation}
where the free parameters of the model are given by:
$\{A_0, R, R_1, R_{1h}, R_{2h}\}$. For these parameters, we find the
best-fit parameters to be:

\begin{align}
\label{eq:hzpt-A0}
A_0    &= 708 \left( \frac{\sigma_8(z)}{0.8} \right)^{3.65} \ (h/\Mpc)^3, \\
\label{eq:hzpt-R}
R      &= 31.8 \left( \frac{\sigma_8(z)}{0.8} \right)^{0.13} (\Mpc/h), \\
\label{eq:hzpt-R1}
R_1    &= 3.24 \left( \frac{\sigma_8(z)}{0.8} \right)^{0.37} (\Mpc/h), \\
\label{eq:hzpt-R1h}
R_{1h} &= 3.77 \left( \frac{\sigma_8(z)}{0.8} \right)^{-0.10} (\Mpc/h), \\
\label{eq:hzpt-R2h}
R_{2h} &= 1.70 \left( \frac{\sigma_8(z)}{0.8} \right)^{0.42} (\Mpc/h).
\end{align}
As first shown in \cite{Seljak:2011} and discussed in
Appendix \ref{app:zeldovich} (see equation \ref{eq:P01-to-Pdd}),
$P_{01}$ is fully predicted from $P_{00}$ through the relation

\begin{equation}
P_{01}^S(\vk, a) = \mu^2 \frac{d P_{00}(k, a)}{d\ln a},
\end{equation}
where $a$ is the scale factor. Thus, the appropriate time derivative
of equation \ref{eq:pade_power}, combined with the Zel'dovich expression
for $P_{01}$ discussed in detail
in Appendix \ref{app:zeldovich} amounts to a full model for $P_{01}^S(\vk)$,
using the same 5 parameters defined in equations \ref{eq:hzpt-A0} -
\ref{eq:hzpt-R2h}.

We also include measurements of the small-scale dark matter correlation
function when finding the best-fit parameters discussed in this section.
For reference, we provide the full relation for $\xi_{BB}(r)$,
the Fourier transform of equation \ref{eq:pade_power},

\begin{align}
\xi_{BB}(r) &= -\frac{A_0 e^{-r/R}}{4\pi r R^2
			(1 - R^2_{1h}/R^2 + R^4_{2h}/R^4)} \nn \\
			&\times \Bigg[ 1 - R^2_1/R^2  \nn \\
			&+ A \exp \left[r
            	\left\{ R^{-1} - R_{2h}^{-2} \sqrt{(R_{1h}^2 - S)/2} \right\} \right] \nn \\
            &+ B \exp \left[r
            	\left\{ R^{-1} - R_{2h}^{-2} \sqrt{(R_{1h}^2 + S)/2} \right\} \right] \Bigg],
\end{align}
where we have defined the following quantities:

\begin{align}
S &\equiv \sqrt{R_{1h}^4 - 4 R_{2h}^4}, \\
A &\equiv (2 R_{2h}^4 S)^{-1} \Big[ R^2 \left(-2 R_{2h}^4 + R_1^2 (R_{1h}^2-S)\right)
		+ R_{2h}^4 (R_{1h}^2-S) \nn \\
		&+ R_1^2 (-R_{1h}^4 + 2 R_{2h}^4 + R_{1h}^2 S) \Big], \\
B &\equiv -(2 R_{2h}^4 S)^{-1} \Big[ R_{2h}^4 (R_{1h}^2 + S) - R_1^2 (R_{1h}^4 -
		2 R_{2h}^4 + R_{1h}^2 S) \nn \\
	    &+ R^2 \left(-2 R_{2h}^4 + R_1^2 (R_{1h}^2 + S) \right)\Big].
\end{align}

\subsection{Dark matter radial momentum power spectrum, $P_{11}$}\label{app:hzpt-P11}

We model the $\mu^4$ term of the scalar component of the radial
momentum auto power spectrum, $P_{11}[\mu^4]$, with a HPZT model,
as the sum of a Zel'dovich term and a Pad\'e sum

\begin{equation}
P_{11,s}^S[\mu^4](k) = P_{11,s}^\mathrm{zel}(k) + P^{BB}_{11}(k),
\end{equation}
where $P_{11,s}^\mathrm{zel}$ is the
Zel'dovich approximation expression for the radial momentum
power spectrum discussed in detail in Appendix \ref{app:zeldovich}.
For $P^{BB}_{11}(k)$, we use a Pad\'e sum of the form

\begin{equation}\label{eq:pade_power1}
P^{BB}_{11}(k) = A_0 \left( 1 - \frac{1}{1 + k^2 R^2} \right)
		\frac{1}{1 + (k R_{1h})^2}.
\end{equation}
The redshift-dependence of the parameters enters into the model through both
$\sigma_8(z)$ and $f(z)$, where $f$ is the logarithmic growth rate.
The best-fit parameters used in this work are given by

\begin{align}
\label{eq:P11-A0}
A_0    &= 659 \left( \frac{\sigma_8(z)}{0.8} \right)^{3.91}
				\left( \frac{f(z)}{0.5} \right)^{1.92} \ (h/\Mpc)^3, \\
\label{eq:P11-R}
R      &= 19.0 \left( \frac{\sigma_8(z)}{0.8} \right)^{-0.37}
				\left( \frac{f(z)}{0.5} \right)^{-0.25} \ (\Mpc/h), \\
\label{eq:P11-R1h}
R_{1h} &= 0.85 \left( \frac{\sigma_8(z)}{0.8} \right)^{-0.15}
				\left( \frac{f(z)}{0.5} \right)^{0.77}\ (\Mpc/h).
\end{align}
Note that in the large-scale, linear perturbation regime, we have
$P_{11,s}^S[\mu^4](k) = f^2 P_{\mathrm{lin}}$. As discussed in
\cite{Seljak:2015}, the density auto spectrum in both SPT and the
Zel'dovich approximation scales as the square of the linear power
spectrum. Noting the additional factor of $f^2$ in the case of
$P_{11,s}$, the low-$k$ amplitude scalings predict
$A_0 \propto f^2 \sigma_8^4$; this result is close to the best-fit
values found in equation \ref{eq:P11-A0}.

\subsection{Halo-matter power spectrum, $P^{hm}$}\label{app:hzpt-Phm}

The HZPT model for the halo-matter power spectrum, as discussed
in Section \ref{sec:hzpt-halo-matter}, is

\begin{equation}
P^{hm}(k) = b_1 P_{00}^{\mathrm{zel}}(k) + P_{00}^{\mathrm{BB}}(k, A_0, R, R_1, R_{1h}, R_{2h}),
\end{equation}
where $P_{00}^{\mathrm{BB}}$ is the broadband
Pad\'e term, as given by equation \ref{eq:pade_power}. The best-fit
parameters for the Pad\'e term used in this work are

\begin{align}
\label{eq:hzpt-hm-A0}
A_0    &= 752 \ b_1^{1.66} \left( \frac{\sigma_8(z)}{3.75} \right)^{3.65}
			   \ (h/\Mpc)^3, \\
\label{eq:hzpt-hm-R}
R      &= 16.9 \ b_1^{-0.12} \left( \frac{\sigma_8(z)}{0.8} \right)^{-1.07}
				\ (\Mpc/h), \\
\label{eq:hzpt-hm-R1}
R_1    &= 5.19 \ b_1^{-0.57} \left( \frac{\sigma_8(z)}{0.8} \right)^{0.16}
				\ (\Mpc/h), \\
\label{eq:hzpt-hm-R1h}
R_{1h} &= 8.25 \ b_1^{-0.84} \left( \frac{\sigma_8(z)}{0.8} \right)^{-0.13}
				\ (\Mpc/h), \\
\label{eq:hzpt-hm-R2h}
R_{2h} &= 3.05 \ b_1^{-1.03} \left( \frac{\sigma_8(z)}{0.8} \right)^{-0.36}
				 \ (\Mpc/h).
\end{align}

\section{Relation between model parameters in the halo model}
\label{app:model-constraints}

In this section, we describe the relations between parameters
of our model in the context of the halo model, as discussed in
section~\ref{sec:galaxy-model}. We apply previous analyses
of clustering in the halo model, i.e.,
\cite{Berlind:2002, Zheng:2004, Hikage:2013,Abramo:2015}, to the
specific notation used in our model. In particular, we are able
to constrain the linear bias (section~\ref{app:b1cB-constraint}) and
the relative fraction (section~\ref{app:fcB-constraint}) for
the sample of centrals with satellites in the same halo.
We also derive expressions for the 1-halo amplitudes,
$N_{c_B s}$ and $N_{s_B s_B}$, in terms of other model parameters
using the halo model in section~\ref{app:1halo-constraint}.

\subsection{The fraction of centrals with satellites}
\label{app:fcB-constraint}

The relative fraction for the $c_B$ sample $f_{c_B}$, which gives the
fraction of central galaxies that live in halos with
at least one satellite galaxy, can be related to the other
galaxy sample fractions. The number of galaxies
in the $c_B$ sample is equal to the number of centrals with only
one satellite plus the number of centrals with greater than
one satellite. Assuming each halo has exactly one central galaxy,
we can express this as

\begin{equation}
\label{eq:fcB-start}
f_{c_B} = \frac{N_{c_B}}{N_c} = \frac{N_{s_A}}{N_c}
		+ \frac{1}{\Nsatmult} \frac{N_{s_B}}{N_c},
\end{equation}
where we have defined $\Nsatmult$ to be the mean number
of satellites galaxies in halos with greater than one satellite. This
parameter normalizes the number of $s_B$ galaxies to the number
of centrals, such that $\Nsatmult^{-1} N_{s_B}$ gives
the number of centrals with greater than one satellite in the same halo.
For a HOD similar to the BOSS CMASS galaxy sample, we typically have
$\Nsatmult \sim 2.4$.

Using the definitions $f_s = N_s / N_g$ and $f_{s_B} = N_{s_B}/N_s$,
and noting that $N_s = N_{s_A} + N_{s_B}$ and $N_g = N_c + N_s$,
we can simplify equation~\ref{eq:fcB-start} as

\begin{align}
\label{eq:fcB-constraint}
f_{c_B} = \frac{f_s}{1-f_s} \left [ 1 +
				f_{s_B} \left ( \Nsatmult^{-1} - 1 \right)\right].
\end{align}

\subsection{The linear bias of centrals with satellites}
\label{app:b1cB-constraint}

Using the halo model, we can express the bias of a specific galaxy sample as
an integral over the halo mass function, weighted by bias

\begin{equation}
\label{eq:halo-model-bias}
b_{X} = \frac{1}{\bar{n}_X} \int d \ln M \frac{d \bar{n}_h}{d\ln M}\bar{N}_x(M) b(M) u(k|M),
\end{equation}
where $\bar{n}_x$ is the mean number density of the sample,
$d\bar{n}_h / d\ln M$ is the halo mass function, $\bar{N}_x$ gives the
mean halo occupation for the sample as a function of halo mass, $b(M)$ is the
halo bias -- mass relation, and $u(k|M)$ describes the halo profile in Fourier space.

For the sample of central galaxies with satellites in the same halo (denoted as $c_B$),
we are able to express the mean occupation $\bar{N}_{c_B}$ in terms of quantities
defined for the two satellite samples, $s_A$ and $s_B$. In particular, we can write

\begin{equation}
\label{eq:NcB-occupation}
\bar{N}_{c_B} = N_{s_A} + \Nsatmult^{-1} N_{s_B},
\end{equation}
where $N_{s_A}$ is the occupation of satellites with only a single satellite in a halo,
and $N_{s_B}$ is the occupation of satellites with multiple satellites in the same
halo. Here, we have defined $\Nsatmult$ to be the mean number
of satellites galaxies in halos with greater than one satellite.
Using equations~\ref{eq:halo-model-bias} and \ref{eq:NcB-occupation}, we can relate
the linear biases as

\begin{equation}
\bar{n}_{c_B}b_{1, c_B} = \bar{b}_{s_A} + \Nsatmult^{-1} \bar{n}_{s_B} b_{1,s_B}.
\end{equation}
We can relate the number density of individual samples to the total
galaxy number density $\bar{n}_g$ as

\begin{align}
\bar{n}_{c_B} &= f_{c_B}(1-f_s)\bar{n}_g \nn \\
\bar{n}_{s_A} &= f_s(1-f_{s_B}) \bar{n}_g \nn \\
\bar{n}_{s_B} &= f_s f_{s_B} \bar{n}_g. \nn
\end{align}
Finally, we obtain the expression for $b_{1,c_B}$

\begin{equation}
b_{1, c_B} = \frac{(1 - f_{s_B}) f_s}{f_{c_B}(1-f_s)} b_{1,s_A}
				+ \frac{f_{s_B} f_s}{\Nsatmult f_{c_B} (1- f_s)} b_{1, s_B}.
\end{equation}
Using the expression for $f_{c_B}$ from equation~\ref{eq:fcB-constraint},
we can simplify this equation as

\begin{equation}
\label{eq:b1cB-constraint}
b_{1, c_B} = \frac{1 - f_{s_B}}{1 + f_{s_B}(\Nsatmult^{-1}-1)} b_{1,s_A} +
					\frac{f_{s_B}}{\Nsatmult(1-f_{s_B}) + f_{s_B}} b_{1,s_B}.
\end{equation}
Note that, as expected, the weights in this linear combination,
$b_{1, c_B} = w_1 b_{1, s_A} + w_2 b_{1, s_B}$, sum to unity such
that $w_1 + w_2 = 1$.

\subsection{1-halo term amplitudes}
\label{app:1halo-constraint}

In this section, we express the 1-halo amplitudes $N_{c_Bs}$ and
$N_{s_Bs_B}$ in terms of other model parameters using a description
of the shot noise in terms of pair counts of galaxies. Generically,
we can write the shot noise of galaxies as

\begin{equation}
\label{eq:Pshot-start}
P^\mathrm{shot} = V \frac{\sum_{\halos} N_i^2}
						{\left(\sum_{\halos} N_i \right)^2}
                 = V \frac{\sum_{\halos} N_i^2}{N_g^2},
\end{equation}
where $V$ is the volume of the survey, $N_i$ represents the number
of galaxies in the $i^\mathrm{th}$ halo, $N_g$ is the total number of
halos, and we sum over all halos. Note that in the limit of a single
object per halo, this simplifies to the usual expression for the
Poisson shot noise, $P^\mathrm{shot} = V N_g / N_g^2 = \bar{n}_g^{-1}$,
where $\bar{n}_g = V / N_g$ is the number density of the galaxy sample.

We can decompose the sum in the numerator of
equation~\ref{eq:Pshot-start} as

\begin{align}
\label{eq:Pshot-numerator}
\sum_{\mathrm{halos}} N_i^2 &= N_g + \sum_{\halos}N_i(N_i-1), \nn \\
	&= N_g  + \sum_{\halos,N=2} N_i(N_i-1)
    		+ \sum_{\halos,N=3} N_i(N_i-1) + \dots, \nn \\
    &= N_g + 2 N_{N=2}^\halos + 6 N_{N=3}^\halos + \dots, \nn \\
    &= N_g + \sum_{\halos, j=2}^{\halos, j=\infty}j(j-1) N_{N=j}^\halos,
\end{align}
where $N_{N=j}^\halos$ is the total number of halos with exactly
$j$ galaxies in the halo.

To mirror our definitions of galaxy subsamples, we can decompose the
sum over halos with greater than one galaxy member in
equation~\ref{eq:Pshot-numerator} into the contributions
from central - satellite pairs and those between only satellites.
For the former case, we can consider the number of pairs between
centrals and satellites as

\begin{equation}
N_{cs}^\mathrm{pairs} = 2 \sum_{\halos} N_{s,i} = 2N_s = 2 f_s N_g,
\end{equation}
where $N_{s,i}$ is the number of satellite galaxies in the $i^\mathrm{th}$
halo. And then using using equation~\ref{eq:Pshot-start}, the total
contribution of this term to the shot noise is

\begin{equation}
P_{c_Bs}^{1h} = \frac{V}{N_g^2}N_{cs}^\mathrm{pairs} = \frac{2f_s}{\bar{n}_g},
\end{equation}
and using the fact that $P_{c_Bs}^{1h} = 2f_s(1-f_s)f_{c_B}N_{c_Bs}$, we have

\begin{equation}
\label{eq:NcBs-constraint}
N_{c_Bs} = \frac{f^{1h}_{c_Bs}}{\bar{n}_g} \left [ (1-f_s)f_{c_B} \right]^{-1},
\end{equation}
where $\bar{n}_g$ is the number density of the full galaxy sample, and
we have introduced an order-unity, normalization nuisance parameter to
allow for possible variations in the 1-halo amplitude.

Similarly, we can consider the contribution to equation~\ref{eq:Pshot-numerator}
from the correlations between satellites. The contribution to the shot noise
from satellite-satellite pairs is

\begin{align}
P_{ss}^\mathrm{shot} &= V \frac{\sum_{N_{s,i}>1} N_{s,i}(N_{s,i}-1)}{N_g^2}, \\
       &= \frac{V}{N_g^2} \la N_{s,i} (N_{s,i}-1) \ra_{>1,s} N_{>1,s}^\halos,
\end{align}
where we have defined  $\la N_{s,i} (N_{s,i}-1) \ra_{>1,s}$ as the mean
number of satellites in a halo, averaging over halos with greater than
one satellite, and $N_{>1,s}^\halos$ is the total number of halos that
have more than one satellite. We can express the latter quantity as

\begin{equation}
N_{>1,s}^\halos = N_g \left [ f_{c_B} (1-f_s) - f_s (1-f_{s_B}) \right],
\end{equation}
where the first term represents the total number of halos with at least one
satellite, and the second term is the number of halos with exactly one satellite.
Here, we have explicitly assumed that every halo has exactly one central galaxy.

Using the fact that $P_{s_Bs_B}^{1h} = f_s^2 f_{s_B}^2 N_{s_Bs_B}$, the
1-halo amplitude becomes

\begin{equation}
\label{eq:NsBsB-constraint}
N_{s_Bs_B} = \frac{f^{1h}_{s_B s_B}}{\bar{n}_g f_s^2 f_{s_B}^2}
			\left [ f_{c_B} (1-f_s) - f_s (1-f_{s_B}) \right],
\end{equation}
where we have defined a normalization nuisance parameter
$f^{1h}_{s_Bs_B}$, which allows for variations in the unknown
quantity $\la N_{s,i} (N_{s,i}-1) \ra_{>1,s}$. Typically, for a CMASS-like
galaxy sample, we find $f^{1h}_{s_Bs_B} \sim 4$. For comparison,
if $N_{s,i} = 2 \ (3)$ for all halos with greater than one satellite, then
$f^{1h}_{s_Bs_B} = 2 \ (6)$.

\bibliographystyle{JHEP}
\bibliography{main}

\end{document}